%% file: main_edit.tex
\documentclass[9pt,twoside]{extbook}
\pdfoutput=1
\usepackage{amsthm,amsmath,amsrefs}
\usepackage{amssymb}
\usepackage[pdftex]{graphicx}
\usepackage{listings}
\usepackage{pgfplots}
\usepackage[labelfont=bf,margin = 0.5in, font=footnotesize]{caption}%
\usepackage[top=1in, bottom=1in, left=1.75in, right=1.75in]{geometry}
\usepackage{fancyhdr}
\usepackage{forloop}
\usepackage[all]{xy}
\usepackage[pdf]{psfrag}
\usepackage{enumitem}
\usepackage{setspace}
\usepackage{rotating}
\usepackage{hyperref}
\usepackage[utf8]{inputenc}
\hypersetup{
    colorlinks,
    linkcolor={red!70!black},
    citecolor={blue!70!black},
    urlcolor={blue!50!black}
}
\usepackage{mdframed}
\usepackage{xcolor}

\definecolor{darkred}{RGB}{139,0,0}
\definecolor{chartreuse}{RGB}{127,255,0}
\definecolor{goldenrod}{RGB}{218,165,32}
\definecolor{gray}{RGB}{127,127,127}

\definecolor{Magenta}{RGB}{255, 0,255}
\definecolor{Orange}{RGB}{255,165, 0}
\definecolor{Gray}{RGB}{127,127,127}
\DeclareMathOperator*{\trace}{\mathsf{tr}}
\newcommand{\hess}{\mathsf{Hess}}
\renewcommand{\theequation}{\arabic{equation}}%

\include{preamble3}

\title{\Huge \bf On the Origins and Control of Community Types in the Human Microbiome } 
\author
{Travis~E.~Gibson$^{1,*}$, Amir~Bashan$^{1,*}$, Hong-Tai Cao$^{2}$,\\ Scott
  T. Weiss$^1$,  and Yang-Yu~Liu$^{1,3,\dag}$\\
\\
\begin{enumerate}
\item[1] sdsdf
\end{enumerate}
\\
\normalsize{$^{1}$Channing Division of Network Medicine, Brigham and Women's Hospital,}\\ 
\normalsize{Harvard Medical School, Boston, Massachusetts 02115, USA}\\
\normalsize{$^{2}$Chu Kochen Honors College, College of Electrical Engineering,}\\ 
\normalsize{Zhejiang University, Hangzhou, Zhejiang 310027, China}\\
\normalsize{$^{3}$Center for Cancer Systems Biology, Dana-Farber Cancer Institute,}\\
\normalsize{ Boston, Massachusetts 02115, USA}\\
\normalsize{$^\ast$These authors contributed equally}\\
\normalsize{$^\dag$Corresponding Author: yyl@channing.harvard.edu}
}

  \renewcommand{\thefigure}{\arabic{figure}}%

\begin{document}

\thispagestyle{empty}

\vspace*{1in}

\begin{centering}
{\Huge \bf On the Origins and Control of Community Types in the Human Microbiome } 

\vspace{.5in}

\Large Travis~E.~Gibson$^{1,*}$, Amir~Bashan$^{1,*}$, Hong-Tai Cao$^{2}$,\\ Scott
  T. Weiss$^1$,  and Yang-Yu~Liu$^{1,3,\dag}$ \\
  
\end{centering}

\vspace{.5in}

\begin{itemize}
\item[1] Channing Division of Network Medicine, Brigham and Women's Hospital,\\
Harvard Medical School, Boston, Massachusetts 02115, USA
\item[2] Department of Electrical Engineering, University of Southern California, \\
 Los Angeles, California 90089, USA
\item[3] Center for Cancer Systems Biology, Dana-Farber Cancer Institute,\\
 Boston, Massachusetts 02115, USA
\item[$\ast$] These authors contributed equally
\item[$\dag$] Corresponding Author: yyl@channing.harvard.edu
\end{itemize}

\vspace{.5in}

\textbf{
Microbiome-based stratification of healthy individuals into compositional categories, referred to as  ``community types'', holds promise for drastically improving personalized medicine. Despite this potential, the  existence of community types and the degree of their distinctness have been highly debated. Here we adopted a dynamic systems approach and found that heterogeneity in the interspecific interactions or the presence of strongly interacting species is sufficient to explain community types, independent of the topology of the underlying ecological network.  By controlling the presence or absence of these strongly interacting species we can steer the microbial ecosystem to any desired community type.  This open-loop control strategy still holds even when the community types are not distinct but appear as dense regions within a continuous gradient. This finding can be used to develop viable therapeutic strategies for shifting the microbial composition to a healthy configuration.}
\vfill

\tableofcontents

\chapter{Main Text}

\section{Introduction}
Rather than simple passengers in and on our bodies, commensal microorganisms have been shown to play key roles in our physiology and in the evolution of several chronic diseases \cite{DuPont:2011aa,Round:2009aa}.
Many scientific advances have been made through the work of
large-scale, consortium-driven metagenomic projects, such as the
{\em Metagenomics of the Human Intestinal Tract} (MetaHIT) \cite{Nelson:2011aa} and the {\em Human
Microbiome Project} (HMP) \cite{:2012aa,:2012ab}.
In particular, the HMP has analyzed the largest cohort and set of
distinct, clinically relevant body habitats to date, in order to characterize the
ecology of human-associated microbial
communities \cite{:2012aa}.   
These results thus delineate the range of structural and functional
configurations normal in the microbial communities of a healthy
population, enabling future characterization of the  translational applications of the human microbiome.  

A recent study proposed that a healthy gut microbiome falls
within one of three distinct community types, which the authors
coined as ``enterotypes''  \cite{Arumugam:2011aa}. More specifically, the authors calculated the relative abundance profiles of microbiota at the genus level and then performed standard cluster analysis, finding three distinct clusters (enterotypes).
 Each enterotype is a dominated by a particular genus ({\em Bacteroides}, {\em Prevotella}, or {\em Ruminococcus}) but not affected by gender, age, body mass index, or nationality of the host. 
These results suggest that enterotyping could be an efficient way to stratify healthy human individuals. The development of personalized microbiome-based therapies would then simplify to shifting an unhealthy microbiome to one of the distinct healthy configurations.

A meta-analysis, however, suggested that enterotypes, or in general community types, could be an artifact of the small sample size in \cite{Arumugam:2011aa} and what one should expect is a continuous gradient with  {dense regions} rather than distinct clusters \cite{Koren:2013aa}. 
The level of discreteness or continuity of the community types remains unclear.
 Interestingly, samples in the dense regions of this gradient are either highly abundant or deficient in {\em Bacteroides} \cite{Koren:2013aa}, %
 indicating that community types could still emerge as the dense regions within a continuous gradient.  {Indeed, some recent work actually supports the notion of distinct community types \cite{Wu07102011,hildebrand2013inflammation,zhou2014exploration,Ding:2014aa,Ravel15032011}.}

 {
 We still lack consensus on the nature and origins of community types \cite{Jeffery:2012aa,Claesson:2012aa,Faust:2012aa,Knights2014433,Arumugam:2014aa}. In principle the presence of community types can be explained by several different mechanisms. First, there may be true multi-stability, i.e. multiple stable states with all microbial species present in the same environment \cite{lewontin1969meaning}. Those stable states are interior equilibrium points of the corresponding ecological dynamic system. Although this type of multi-stability has been well discussed in macro-ecological systems \cite{connell1983evidence}, its detection in host-associated microbial communities is rather difficult and has not been demonstrated experimentally, partially because any subtle differences in the environment can drive those microbial communities \cite{Faust:2012aa}.  Second, there may be strong host heterogeneity, leading to host-specific microbial dynamics (parameterized by host-specific intra- and inter-species interactions). If those interactions, which serve as parameters of the host-associated microbial ecosystems, can be classified into distinct groups, then we can numerically demonstrate that distinct community types will naturally emerge (Supplementary Text Sec. 6.2 and 7.1). Yet, the presence of classifiable microbial dynamics has not been experimentally detected, presumably due to the lack of high-quality time-series data for a large number of subjects. Moreover, the overwhelming success of {\em Fecal Microbiota Transplantation} (FMT) in treating {\em recurrent Clostridium Difficile Infection} (rCDI) actually implies that difference in conditions between individuals are unlikely the cause of community types \cite{Youngster01062014,weingarden2015dynamic,doi:10.1001/jama.2014.13875}.}

 {Here we proposed a simple mechanism, without assuming multi-stability or host heterogeneity, to explain the origins of community types. In particular,}  using a dynamic systems approach, we studied compositional shift as a function of species collection and demonstrated that
 with heterogeneous interspecific  interactions, a phenomenon often observed in macroecology \cite{Paine:1992aa,JANE:JANE1041,JANE:JANE818},  community types can naturally emerge. Interestingly, this result is  independent of the topology of the underlying ecological network.
To our knowledge, this is the first  {quantitative} attempt to explore the analytical basis of community types. Furthermore, community types, even when they weakly exist, can be manipulated efficiently by controlling the  {\em Strongly Interacting Species} (SISs) only.\footnote{In this paper we use the term  species in the general context of ecology, i.e. a set of organisms adapted to a particular set of resources in the environment, rather than the lowest taxonomic rank. One could think of organizing microbes by genus or operational taxonomical units as well.} This provides theoretical justification for translational applications of the human microbiome.

\section{Dynamic Model}
The human microbiome is  a complex and dynamic ecosystem \cite{Gerber:2014aa}.  When modeling a dynamic system we should first decide how complex the model needs to be so as to capture the phenomenon of interest.
A detailed model of the intestinal microbiome would include mechanistic interactions among cells, spatial structure of the human intestinal tract, as well as host-microbiome interactions %
\cite{Munoz-Tamayo:2010aa,Shoaie:2013aa,waldram2009top,Bucci:aa}. That level of detail however is not necessary for this study, because we are primarily interested in exploring the impact that any given species has on the abundance of other species. To achieve that, a population dynamics model such as the canonical {\em Generalized Lotka-Volterra} (GLV) model is sufficient \cite{Pepper:2012aa,Faust:2012aa}. 
Indeed, GLV dynamics leveraging current metagenome data has already been used for predictive modeling of the intestinal microbiota \cite{stein:2013,Marino07012014,10.1371/journal.pone.0102451}.
Consider a
collection of $n$ species in a habitat with the population of species $i$ at time $t$ denoted as $x_i(t)$. The GLV model assumes
that the species populations follow a set of ordinary
differential equations
\be \label{eq:glv1}
\dot x_i(t) = r_i 
x_i(t) +  x_i(t)\sum_{j=1}^n a_{ij} \, x_j(t), \quad i=1,\ldots,n \ee 
where ${\dot{(\, )}}$ = ${\frac{\mathrm d}{\mathrm d t}(\, )}$. Here $r_i$ is the growth rate of species $i$, $a_{ij}$ (when $i\neq j$) accounts for the impact that species $j$ has on the population change of species $i$, and the
terms $a_{ii}x_i^2$ are adopted from Verhulst's logistic growth model \cite{Goel:1971aa}. 
By collecting
the individual populations $x_i(t)$ into a state vector $x(t)=[x_1(t),
\ \cdots \ ,  x_n(t) ]^\mathsf T$, Equation \eqref{eq:glv1}
can be represented in the compact form \be \label{eq:lvd2}
\dot x(t) = \diag(x(t))\left (r+  A x(t)\right),
\ee
where $r=[ r_1, \ \cdots \ ,  r_n]^{\mathsf T}$ is a column vector of the growth rates,  $A=(a_{ij})$ is
the interspecific interaction matrix, and $\diag$ generates a diagonal matrix from a vector. Hereafter we drop the explicit time dependence of $x$.

Next we discuss the notion of fixed point, %
or equivalently steady state, in the GLV dynamics. This notion is important in the context of the
human microbiome, as the
measurements taken of the relative abundance of intestinal microbiota in the
aforementioned studies typically represent steady behavior  \cite{:2012aa,Arumugam:2011aa}. In other words, the intestinal microbiota is a relatively resilient ecosystem \cite{Lozupone:2012aa,relman2012human}, and until the
next  large perturbation (e.g. antibiotic administration or dramatic change in diet) is introduced,  the system remains stable for months and possibly even years
 \cite{david2014host,caporaso2011moving,faith2013long}. The fixed points of system \eqref{eq:lvd2} are those
solutions $x$ that satisfy ${\dot x=0}$.  The solution $x=0$  (i.e. all species have zero abundance) is a trivial steady state.
The set of non-trivial steady states contains those solutions
$x^*$ such that $r+Ax^*=0$.  
When the matrix $A$ is invertible, it follows that the non-trivial steady state $x^* =
-A^{-1}r$ is unique %
\cite{goh1977global}.

%
%
%
%
%
%
%
%
%

Our study ultimately investigated the impact that different collections of microbial species have on their steady state abundances. In Box \hyperref[box:1]{1} we presented a detailed analysis showing that if we  introduce a  new species into the ecosystem in \eqref{eq:lvd2}, the shift of the steady state is proportional to the interaction strengths between the newly introduced species and the previously existing ones. Similarly, if two communities with the same dynamics differ by only one species, then it is the interaction strength of that species with regard to the rest of the community that dictates how far apart the steady states of the two communities will be. This analytical result indicates that heterogeneity of interspecific interactions could lead to the clustering of steady states, and hence the emergence of community types.

To systematically investigate how changes in species collection affect the steady state shift in the GLV dynamics, we assumed that two microbial species will interact in the same fashion regardless of the host. Otherwise, if the interactions are host specific and the dynamics are classifiable, we can show that distinct community types will emerge almost trivially (Supplementary Text Sec. \ref{sec:mm} and \ref{sec:mmodelsim}).

\section{Metacommunity and Local Communities}
Consider a universal species pool, also referred to as a metacommunity \cite{Costello08062012}, indexed by a set of integers $\mathbf S = \{1,\ \ldots, \ n\}$, an ${n\times n}$ matrix $\mathbf A$ representing all possible pairwise interactions between species, and a vector $\mathbf r$ of size $n$ containing the growth rates for all the  $n$ species. 
The global parameters for the metacommunity are completely defined by the triple $(\mathbf S, \mathbf A, \mathbf r)$.
We consider $q$ {\em Local Communities} (LCs),   defined by sets $S^{[\nu]}$ that are subsets of $\mathbf S$, denoting the species present in  $\mathrm{LC}_\nu$ with $\nu=1,\ldots, q$. This modeling procedure is inspired by the fact that alternative community assembly scenarios could give rise to the compositional variations observed in the human microbiome \cite{Costello08062012}. These LCs represent microbial communities in the same body site across different subjects.
For simplicity, we assume that each LC contains only $p$ species ($p\leq n$), randomly selected from the metacommunity. 

The GLV dynamics for each LC is given by 
\be
\label{eq:glv_environment}
\mathrm{LC}_\nu: \quad \dot x^{[\nu]}(t) = \diag\left(x^{[\nu]}(t)\right) \left(r^{[\nu]}+A^{[\nu]}x^{[\nu]}(t)\right),
\ee
where the LC specific interaction matrix and growth vector are defined as $\aa = \mathbf A_{S^{[\nu]},S^{[\nu]}}$ and $\rr = \mathbf r_{S^{[\nu]}}$, respectively.
That is, $\aa$ is obtained from $\mathbf A$ by only taking the rows and columns of $\mathbf A$ that are contained in the set $S^{[\nu]}$. A similar procedure is performed in order to obtain $\rr$. Finally for each $x^{[\nu]}$ there is a corresponding $\mathbf x^{[\nu]}\in\Re^n$ that has the abundances for species $S^{[\nu]}$ of  LC$_{\nu}$ in the context of the metacommunity species pool $\mathbf S$.

To reveal the origins of community types in the human microbiome, we decomposed the universal interaction matrix as \be\label{eq:A} \mathbf A =  \mathbf N  \mathbf H \circ  \mathbf G s,\ee
which contains four components. (i) $ \mathbf N\in\Re^{n\times n}$ is the nominal interspecific interaction matrix where each element is sampled from a normal distribution with mean 0 and variance $\sigma^2$, i.e. $[\mathbf N]_{ij}\sim \mathcal N(0,\sigma^2)$. (ii) $\mathbf H\in\Re^{n\times n}$ is a diagonal matrix that captures the overall interaction strength heterogeneity of different species.  When studying the impact of interaction strength heterogeneity the diagonal elements of $ \mathbf H$ will be drawn from a power-law distribution with exponent $-\alpha$, i.e. $[ \mathbf H]_{ii}\sim \mathcal P(\alpha)$, which are subsequently normalized so that the mean of the diagonal elements is equal to 1. This is to ensure that the average interaction strength is bounded. For studies that do not involve interaction strength heterogeneity $\mathbf H$ is simply the identity matrix. (iii) $ \mathbf G\in\Re^{n\times n}$ is the adjacency matrix  of the underlying ecological network: $[\mathbf G]_{ij}=1$ if species $i$ is affected by the presence of species $j$ and $0$ otherwise. For details on the construction of $\mathbf G$ for different network topologies see Supplementary Text Sec. \ref{sec:powerdigraph}. Note that the {\em Hadamard product} ($\circ$) between $\mathbf H$ and $\mathbf G$ represents element-wise matrix multiplication. (iv) The last component $s$ is simply a scaling factor between 0 and 1. Finally, we set $ [\mathbf A]_{ii} = -1$. The presence of the scaling factor $s$ and setting the diagonal elements of $\mathbf A$ to $-1$ are to ensure an asymptotic stability condition for the GLV dynamics (Supplementary Text Sec. \ref{sec:lstab}, \ref{sec:remove}, and \ref{sec:diag_random}). The elements in the global growth rate vector $\mathbf r$ are taken from the uniform distribution, $[\mathbf r]_i \sim \mathcal U(0,1)$. Details concerning the distribution $\mathcal N, \mathcal P$ and $\mathcal U$ can be found in Supplementary Text Sec. \ref{sec:dibin}.

\section{Origins of Community Types}
We first studied the role of interspecific interaction strength heterogeneity on the emergence of community types. %
In order to achieve this, we chose the complete graph topology, i.e. each species interacts with all other species. 
This eliminates any {\em structural} heterogeneity. The nominal interaction strengths were taken from a normal distribution $\mathcal{N}(0,1)$, the scaling component was set to $s=0.7$, and the interaction strength heterogeneity was varied from low heterogeneity ($\alpha = 7$) to a high level of heterogeneity ($\alpha=1.01$). Figure \ref{fig:2} displays the distributions of the diagonal elements of the interaction heterogeneity matrix $\mathbf H$ at various heterogeneity levels. 
For each level of heterogeneity we constructed ${500}$ LCs, each with ${80}$ species randomly drawn from a metacommunity of ${100}$ species. Figure \ref{fig:2}b illustrates the global interaction matrix $\mathbf A$ as a weighted network. With low heterogeneity all the link weights are of the same order of magnitude. As the heterogeneity increases fewer nodes contain highly weighted links, until there is only one node with highly weighted links when $\alpha=1.01$. These nodes with highly weighted links correspond to SISs.

Figure \ref{fig:2}c presents the results of {\em Principle Coordinates Analysis} (PCoA) of the steady states associated with the $500$ different LCs as a function of $\alpha$. For low interaction heterogeneity ($\alpha=7$) the classical clustering measure, Silhouette Index, is less than 0.1, suggesting a lack of clustering in the data. As the heterogeneity increases the steady states can be seen to separate in the first two principle coordinate axes. At one point ($\alpha=2.0$) three clusters is the optimal number of clusters. Then as $\alpha$ continues to decrease the optimal number of clusters becomes two.  The fact that there are three clusters when $\alpha=2.0$ is not special, as a different number of optimal clusters can be observed with different model parameters or different clustering measures (see Supplementary Text Sec. \ref{sec:smodel}) \cite{Koren:2013aa}. 
While the precise number of clusters is not important here, what is important is the fact that the degree of interaction strength heterogeneity controls the degree to which the clusters appear to be distinct. For low levels of interaction strength heterogeneity the clusters appear to be more like dense regions within a continuous gradient. As the heterogeneity increases, the clusters become more distinct.
Indeed, having two clusters for $\alpha=1.01$ is to be expected, because one of the clusters is associated with all the LCs that contain the single SIS, and the other LCs that do not contain the single SIS constitute the other cluster.

The overall trend observed in Figure \ref{fig:2}c is unaffected if the complete graph is replaced by an {\em Erd\H{o}s-R\'enyi} (ER) random graph, or if the total number of LCs is increased (Supplementary Figures \ref{fig:ext1} and \ref{fig:ext2}). The result is also generally unaffected by the specifics of the nominal distribution (Supplementary Text Sec. \ref{sec:hetstudy}), the mean degree of the ER graph (Supplementary Text Sec. \ref{sec:sparse_study}), or the number of species in the LCs (Supplementary Text Sec. \ref{sec:com_study}). %
{ Of course, each LC can be invaded by other species that are currently absent. If this migration occurs relatively fast, then all LCs will converge to roughly the same species collection and the clustering will disappear. Hence in our modeling approach we have to assume that the migration occurs at a relatively slow time scale, and the time interval between species invasions is too long to disrupt the clustering. We also note that if heterogeneous interactions are placed at random in the network the clustering of steady states does not arise (Supplementary Figure \ref{fig:ext31}). Our results are also robust (in the control theoretical sense) to stochasticity and the migration of existing species \cite{Fisher09092014}. Robustness to migration is illustrated in Supplementary Figures \ref{fig:ext41} and \ref{fig:ext51}, and robustness to stochastic disturbances is illustrated in Supplementary Figures \ref{fig:ext61}-\ref{fig:ext81} (see Supplementary Text Sec. \ref{sec:a:robustness} for analytical robustness results).}

We can explain the above results as follows: for low interaction strength heterogeneity all of the matrices $\aa$ are very similar. In other words, despite containing different sets of species, all the LCs have very similar dynamics.
Thus, clustering of steady states is not to be expected. As the heterogeneity of interaction strength increases, however, some of the LCs will have species that are associated with the highly weighted columns in $\mathbf A$, i.e. the SISs. 
Figure \ref{fig:3} presents a detailed analysis of the most abundant (dominating) species in each of the three clusters (community types) in Figure \ref{fig:2}c for $\alpha=2$ and $\alpha=1.6$, along with the abundances of the SISs within each cluster. It is clear that for different clusters their dominating species  are different, consistent with the empirical finding that each enterotype is dominated by a different genus \cite{Arumugam:2011aa}. The SISs that are present in each cluster also vary. For instance with $\alpha=1.6$ all LCs in the blue cluster contain SISs number 23 and 81, and none have species 60 or 51. For the orange cluster it is the opposite scenario. All of the LCs in the orange cluster contain SISs 60 and 51, and do not contain species 23 or 81. Most of the LCs in the yellow cluster contain SISs 23 and 51. Hence, each community type is well characterized by a unique combination of SISs. Note that none of the SISs are dominating species. These findings, along with the analysis in Box \hyperref[box:1]{1} , suggest that heterogeneity in interaction strengths or the presence of SISs leads to the clustering of steady states, i.e. the emergence of community types.

We then studied the impact of structural heterogeneity on community types. Four different scenarios are illustrated in Figure \ref{fig:4}: (a) a complete graph topology as in Figure \ref{fig:2}; (b) an ER random graph as in Supplementary Figure 1; (c)  a power-law out-degree network; (d) a power-law out-degree network with {\em no} interaction strength heterogeneity. Figures \ref{fig:4}a, \ref{fig:4}b and \ref{fig:4}c support the main result shown in Figure \ref{fig:2}, %
i.e. increasing interaction strength heterogeneity leads to the emergence of distinct community types.
Figure \ref{fig:4}d displays rather unexpected results as it suggests that structural heterogeneity alone does not lead to distinct community types. It is only with the inclusion of interaction strength heterogeneity that structurally heterogeneous microbial ecosystems can display strong clustering in their steady states as shown in Figure \ref{fig:4}c. This result is rather surprising, because structural heterogeneity is observed in many real-world complex networks \cite{bar,li_imath_04,ald_cyber_10} and has been shown to affect many dynamical processes over complex networks \cite{pastor2001epidemic,nishikawa2003heterogeneity,liu11nature}.

Note that in the preparation of Figure \ref{fig:4} the steady state  abundances were normalized to get relative abundances of the species and the Jensen-Shannon distance metric was used for clustering analysis \cite{Goodrich:2014aa}. The trends discussed above also hold when, instead of the Silhouette Index, the Variance Ratio Criterion is used as the clustering measure, or the Euclidean distance is used for clustering, or when absolute abundances are analyzed  along with the Euclidean distance being used  (Supplementary Figures \ref{fig:ext3}, \ref{fig:ext4}, and \ref{fig:ext5}). Supplementary Figure \ref{fig:ext5} correlates to the analytical results in Box \hyperref[box:1]{1} , where absolute abundances  and the Euclidean distance are implicitly  used.

\section*{Control of Community Types}

With the knowledge that each community type can be associated with a specific collection of SISs, we tested the hypothesis that a local community could be steered to a desired community type by controlling the combination of SISs only. Our results for three different scenarios are shown in Figure \ref{fig:5}a for $\alpha=1.6$. The local community that was controlled in each scenario is shown in magenta and is denoted LC$^*$, which initially belongs to the blue cluster. For Scenario 1, LC$^*$ had the SISs 23 and 81 removed, with species 60 and 51 simultaneously introduced with random initial abundances drawn from $\mathcal U(0,1)$. Recall that species 60 and 51 are the SISs present in the orange cluster. This swap of SISs shifts LC$^*$ to a slightly different state (green dot) within the blue cluster.
The GLV dynamics were then simulated and the trajectory goes from the blue cluster to the orange cluster. This result was independent of the initial condition of species 60 and 51 (Fig. \ref{fig:5}b). This open-loop control of the community type by manipulating a set of SISs also works at lower levels of heterogeneity (Fig. \ref{fig:5}c and \ref{fig:5}d). {Here we use the term open-loop to contrast closed-loop control where inputs are designed with feedback so as to continuously correct the system of interest.} These findings imply that the SISs, despite their low abundances, can be used to effectively control a microbial community to a desired community type.

In Scenario 2 we tested if the same result could be obtained by removing the six most abundant species from LC$^*$ and introducing the six most abundant species from the orange cluster at exactly the same abundance level as an arbitrary  local community in the orange cluster. The state  after this dominating species swap (red dot) starts close to the orange cluster, because the six most abundant species from a local community in that cluster were copied. The trajectory does not ultimately converge near the orange cluster, but goes toward the blue cluster instead. The trajectory, however, does not ultimately converge in the blue cluster because it does not contain any of the most abundant species present in the blue cluster.

In scenario 3  we explored how the open-loop control methodology just presented could also be used to conceptually justify the success of FMT in treating patients with rCDI \cite{Youngster01062014,weingarden2015dynamic,doi:10.1001/jama.2014.13875}.
This scenario begins by removing 20 species from  LC$^*$ (the top two SISs and 18 of the most abundant spaces) so as to emulate the effect of broad-spectrum antibiotics, resulting in an altered community (blue dot). Then the GLV dynamics were simulated and the local community converged to a new steady state (black dot), representing the CDI state. To emulate an oral capsule FMT 1\% of the species abundances from an arbitrary LC in the orange cluster, i.e. the donor, was added to the CDI state, resulting in a slightly altered community (gray dot). The GLV dynamics were then simulated until the final steady state was reached (white dot). As expected the post-FMT steady state is in the orange cluster, the same cluster that is associated with the donor's LC. Note that if during the FMT the SISs in the donor's LC were not transplanted then the patient's post-FMT steady state does not converge in the orange cluster (Supplementary Figure \ref{fig:ext6}).

The above results indicate that the presence of SISs simplifies the open-loop control design. However, the existence of community types is not a prerequisite for deploying this control methodology. The possibility for open-loop control of the human microbiome will likely be body site specific. Our work focused on the gut specifically because of the fact that this microbial community is very likely dominated by microbe-microbe and/or host-microbe interactions, rather than external disturbances. It is yet to be determined what factors drive the dynamics in other body sites.

\section{Discussion}

\begin{figure}[h!] \label{box:1}
\begin{mdframed}[backgroundcolor=blue!30!gray!20,hidealllines=true] \footnotesize
{\bf Box 1. Steady State Shift in the Generalized Lotka-Volterra Model} 

\rule{\textwidth}{0.4pt}

{\em Impact of one new species}: Consider the addition of one new species to the system described by \eqref{eq:lvd2}, so that now we have ${m=n+1}$ species, with the
abundance of the $i$-th species denoted as $z_i(t)$,
$i=1,\cdots,m$. The new $m$-dimensional state vector $z(t)=[ z_1(t),\, z_2(t),\,\ldots,\,  z_m(t) ]^{\mathsf T}$  evolves as  
\be \label{eq:lvd3}
\dot z(t) = \diag (z(t)) (g+  F z(t)), \tag{B1}
\ee
where 
\ben
g = \bb r \\ s \eb, \quad  \text{and} \quad   F= \bb A & b \\ c^{\mathsf T} & d\eb. %
\een
Note here that $A$ and $r$ are as defined in \eqref{eq:lvd2} and we
have only introduced four new elements $s,b,c,d$. The scalar element
$s$ represents the growth rate of the additional
species $z_m$. The $n$-dimensional vector $b$ represents the impact that species $z_m$ has on the first $n$ species $z_{1:n}$, which also
correspond to the $n$ species in the original state vector $x$. The scalar
element $d$ is the Verhulst term for the new species $z_m$. Finally,
the $n$-dimensional vector $c$ represents the effect that  $z_{1:n}$ have on the dynamics of the $m$-th state $z_m$. 
\\

Given the dynamics 
in \eqref{eq:lvd2} and  \eqref{eq:lvd3}, with the assumption that the
matrix $A$ is full rank, we can show that the difference between the original steady state $x^*$ and the shifted steady state of the same $n$ species, denoted as $z^*_{1:n}$, satisfies the following
equality
\be\label{eq:ss1}
z^*_{1:n}-x^*=-A^{-1} b z^*_m, \tag{B2}
\ee
where $z^*_m$ is the steady state value of the newly added species.
Given that $A$ is the same in \eqref{eq:lvd2} and \eqref{eq:lvd3}, the
shift in the steady state of the original $n$ species is  bilinear  in
terms of the vector $b$ and the steady state value of the newly added
$m$-th species $z^*_m$. \\

\psfrag{v}[cc][cc]{$x^*$}
\psfrag{w}[cc][cc]{$z^*$}
\psfrag{u}[cc][cc]{$z_{1:2}^*$}
\psfrag{b}[cl][cl]{$x^*-z_{1:2}^*$}
\hspace{1in}\includegraphics[width=3.25in]{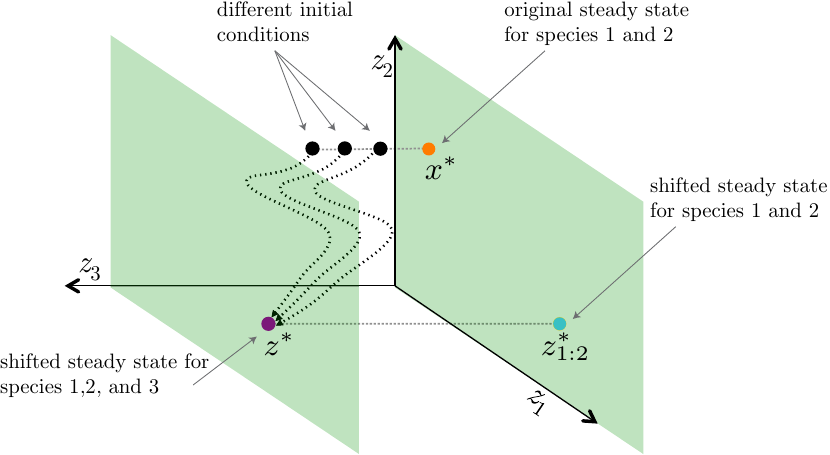} 

{\em Open-loop Control of Steady State}: If there exists a diagonal matrix $P>0$ such that $A^\mathsf{T}P+PA<0$ then for any $z^*$ there exists $b,c,d,s$ such that the original steady state $x^*$  of \eqref{eq:lvd2} can be steered to $z^*$ and furthermore this $z^*$ can be made to be uniformly asymptotically stable (Supplementary Text Sec. \ref{sec:add1}, Theorem \ref{thm3}).
\\

{\em Impact of multiple non-common species}:
Consider two systems with different collections of species. Let $\bar  z^*$ be the steady state of system 1 and $\hat z^*$ be the steady state of system 2. Assuming that the systems share $n$ species, then we can just apply the results from \eqref{eq:ss1} recursively to calculate the difference between the abundances of the common species as
\ben\label{eq:2}
\bar  z^*_{1:n} - \hat z^*_{1:n} = A^{-1}\left( \sum_{i\in\bar M} \bar b_i z^*_i  - \sum_{i\in \hat M} \hat b_i \hat z^*_i \right) %
\een
where $\bar M$ and $\hat M$ are the indices of the non-common species of system 1 and system 2 respectively. 
 \end{mdframed}
\end{figure}

In this work we studied compositional shift as a function of species collection using a dynamic systems approach, aiming to offer a possible mechanism for the origins  of community types. We found that the presence of interaction strength heterogeneity or SISs is sufficient to explain the emergence of community types in the human microbiome, independent of the topology of the  underlying ecological network.
The presence of heterogeneity in the interspecific interaction strengths in natural communities has been well studied in macroecology \cite{Paine:1992aa,JANE:JANE1041,JANE:JANE818,McCann:1998aa}. 
Extensive studies are still required to explore this interesting direction in the human microbiome. 
While preliminary analysis is promising, all existing temporal metagenomic datasets are simply not sufficiently rich to infer the interspecific interaction strengths among all of the microbes present in and on our bodies \cite{Faust:2012aa} even at the genus level, let alone the species level. Recent studies have tried to overcome this issue by only investigating the interactions between the most abundant species \cite{10.1371/journal.pone.0102451}. Our results, however, suggest that SISs need not be the most abundant ones and can still play an important role in shaping the steady states of microbial ecosystems. %
Ignoring the lack of sufficient richness, system identification analysis with regularization and cross-validation \cite{stein:2013,Buffie:2015aa}  of the largest temporal metagenomic dataset to date \cite{caporaso2011moving} does not disprove the existence of SISs. To the contrary, it supports this assertion (see Supplementary Figure \ref{fig:ext7}).  Permutation of the time series however also results in the identification of interaction strength heterogeneity (see Supplementary Figures \ref{fig:ext8} and \ref{fig:ext9}). Hence, the presence of SISs needs to be systematically studied with novel system identification methods and perhaps further validated with co-culture experiments \cite{Faust:2012aa}. {  For example, we could first use metabolic network models to predict levels of competition and complementarity among species \cite{Levy30072013}, which could then be used as prior information to further improve system identification \cite{angulo2015fundamental}.

Note that our notion of SIS is fundamentally different from that of keystone species, which are typically understood as species that have a disproportionate deleterious effect (relative to its abundance) on the community upon their removal \cite{paine1995conversation}. One can apply a brute-force leave-one-out strategy to evaluate the ``degree of keystoneness'' of any species in a given community network \cite{berry2014deciphering}. Even without any interaction strength heterogeneity, a given community may still have a few keystone species. The SISs defined in this work are those species that have very strong impacts (either positive or negative) on the species that they directly interact with. The presence of SISs requires the presence of interaction strength heterogeneity. We emphasize that an SIS is not necessarily a keystone species. In fact, without any special structure embedded in the interaction matrix (and hence the ecological network), there is no reason why the removal of any SIS would cause mass extinction. It does have a profound impact on the steady-state shift, which is exactly what we expected from our analytical results presented in Box \hyperref[box:1]{1}.
}

Our findings also have important implications as we move forward with developing microbiome-based therapies, whether it be through drastic diet changes, FMT, drugs, or even engineered microbes \cite{sa_microbes,yaung2014recent,tanouchi2012engineering,leonard2008engineering,yaung2015improving,mee2014syntrophic,Kotula01042014}. Indeed, our results suggest that a few strongly interacting microbes can determine the steady state landscape of the whole microbial community. Therefore, it may be possible to control the microbiome efficiently by controlling the collection of SISs present in a patient's gut. Finer control  may be possible through the engineering of microbes.  This will involve a detailed mechanistic understanding of the metabolic pathways associated with the microbes of interest.  As discussed in Box \hyperref[box:1]{1} ,  given a new steady state of interest, the parameters $b,c,d,s$ could be chosen such that the new steady state is feasible and stable (Supplementary Text Sec. \ref{sec:add1}). Then, with the knowledge of the appropriate parameters $b,c,d,s$ it would be possible to introduce a known microbe with those characteristics or engineer one to have the desired properties. We emphasize that the stability and control of the microbial ecosystem must be studied at the macroscopic scale using a systems and control theoretic approach. This is  similar to what is carried out in  aerospace applications. The design of wings and control surfaces for an aircraft incorporate sophisticated fluid dynamic models. The control algorithms for planes however are often derived from simple linearized reduced order dynamic models where linear control techniques can be easily deployed \cite{stevens2003aircraft}. Taken together, our results indicate that the origins and control of community types in the human microbiome can be explored analytically if we combine the tools of dynamic systems and control theory, opening new avenues to translational applications of the human microbiome.

\section*{Acknowledgements}
We thank George Weinstock, Curtis Huttenhower, Rob Knight, Domitilla Del Vecchio, and Doug Lauffenburger for helpful discussions. Special thanks to Aimee Milliken for a careful reading of the text. This work was partially supported by the John Templeton Foundation (award number 51977) and National Institutes of Health (R01 HL091528).

\section*{Contributions}
Y-YL conceived the project. All authors performed research. AB performed initial numerical calculations. TEG performed systematic analytical and numerical calculations. HTC and TEG performed the system identification. TEG and Y-YL wrote the manuscript. AB and STW  edited the manuscript.

\clearpage

\chapter{Figures}

\clearpage

\begin{figure}\centering
  \includegraphics[width=\textwidth]{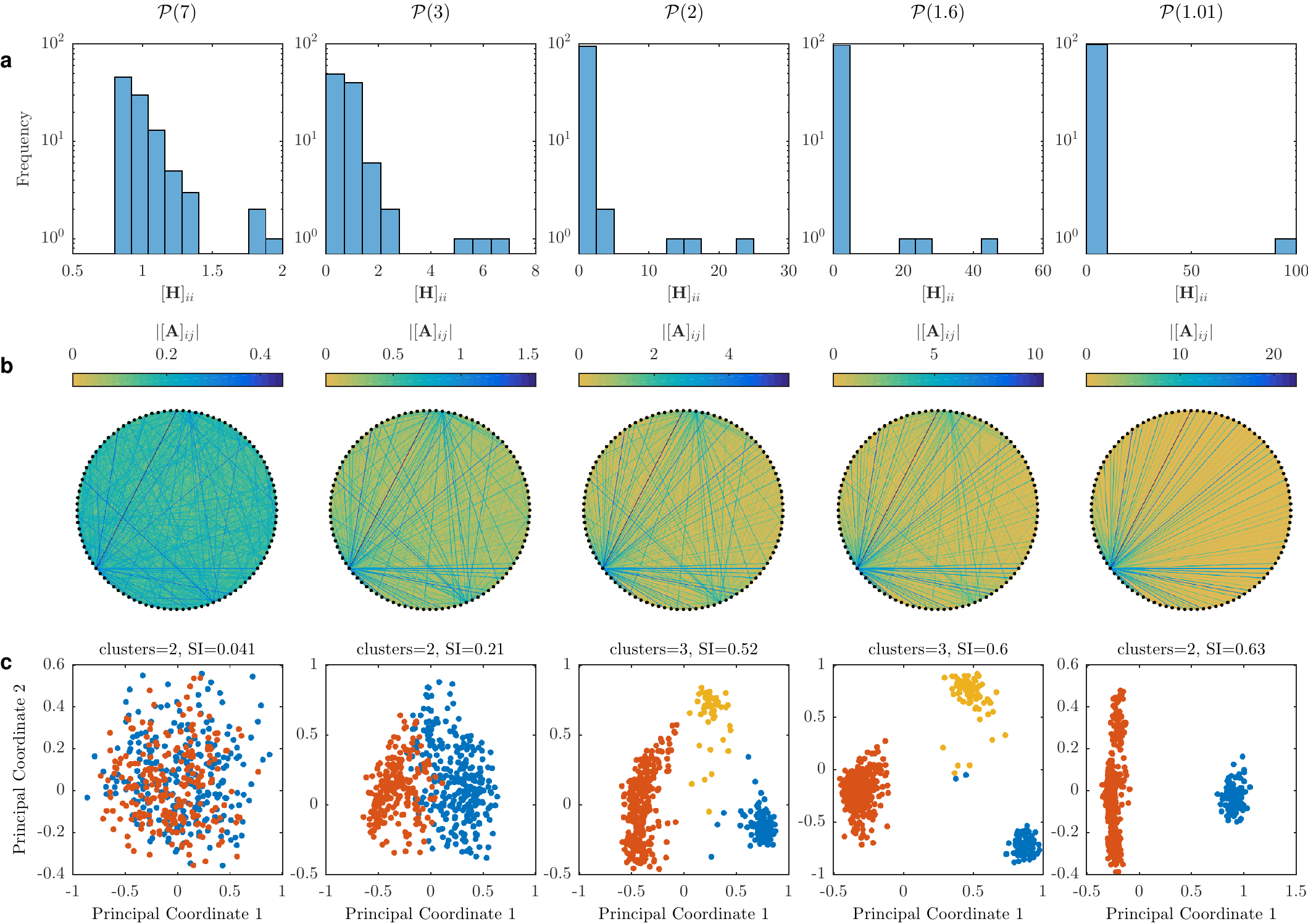}
  \caption{{\em Impact of interaction strength heterogeneity on the distinctness of community types}. A total of $q=500$ local communities, each with  $p=80$ species randomly drawn from a universal pool of $n=100$ species. The nominal components were drawn from $\mathcal N(0,1)$, the interaction heterogeneity matrix elements were taken from $\mathcal P(\alpha)$ and $\alpha$ is varied with the set of values $\{7,\ 3,\ 2,\ 1.6,\ 1.01\}$ for each column in the figure. The topology component $\mathbf G$ has all elements equal to 1, giving a complete graph. The scaling factor was set at $s=0.07$.  (a) Histogram of the diagonal elements of the heterogeneity matrix $\mathbf H$.     (b) Visualization of the universal interaction matrix $\mathbf A$ as a weighted adjacency matrix of a digraph. 
  (c) Principle coordinate analyses of the normalized steady state for each local community using the Jensen-Shannon distance. The Silhouette Index and optimal number of clusters are denoted. Further details can be found in the Methods Section.}\label{fig:2}
\end{figure}

\clearpage

\begin{figure}\centering
  \includegraphics[width=5in]{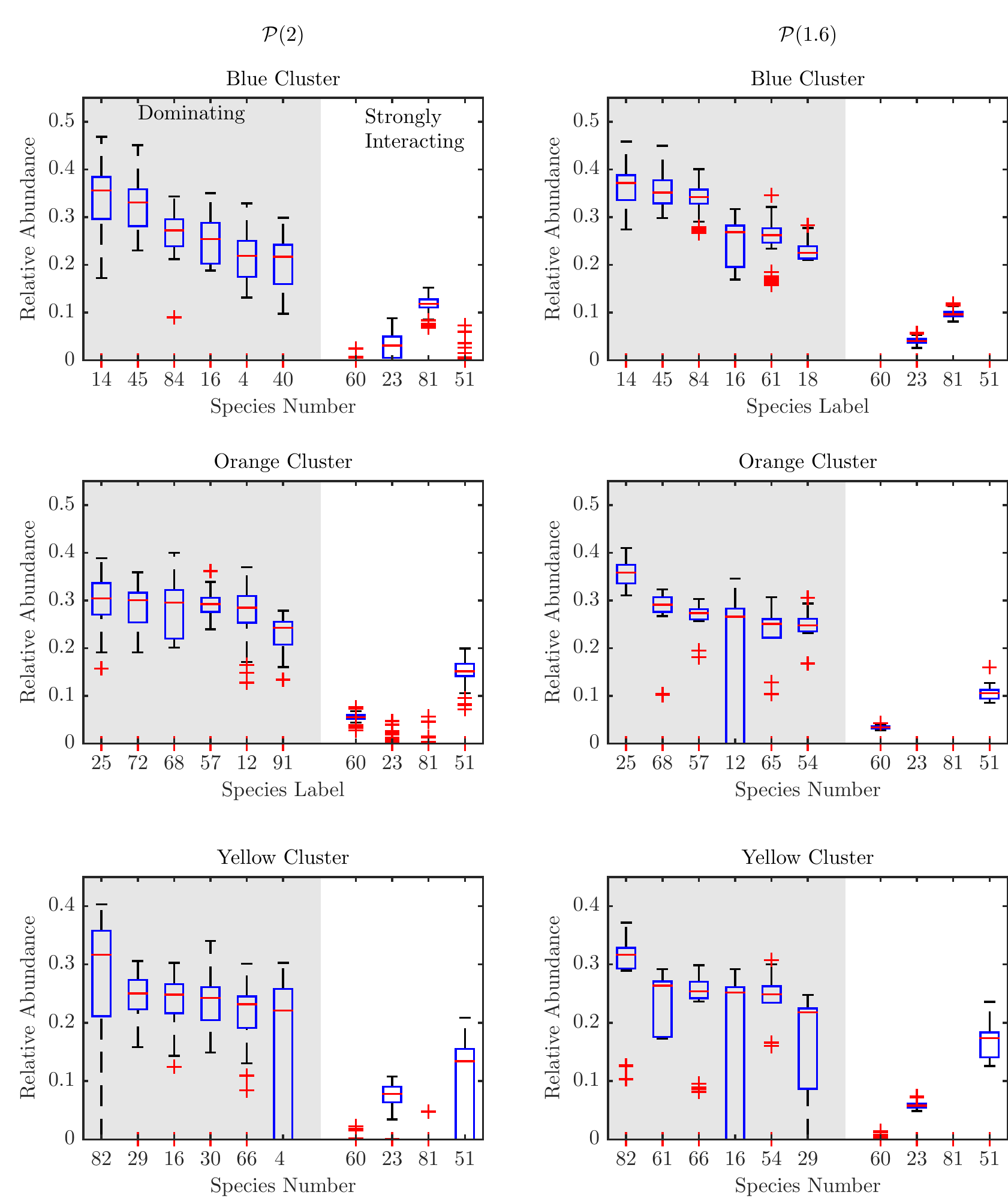}
  \caption{{\em Comparison of dominating species to SISs in different community types (clusters)}. The relative abundances of the six most abundant species from each of the three clusters in Figure \ref{fig:2}c for $\alpha=2$ and $\alpha=1.6$ are compared to that of the four species with the largest interaction strengths ($60$, $23$, $81$, and $51$).} \label{fig:3}
\end{figure}

\clearpage

\begin{figure}\centering
  \includegraphics[width=\textwidth]{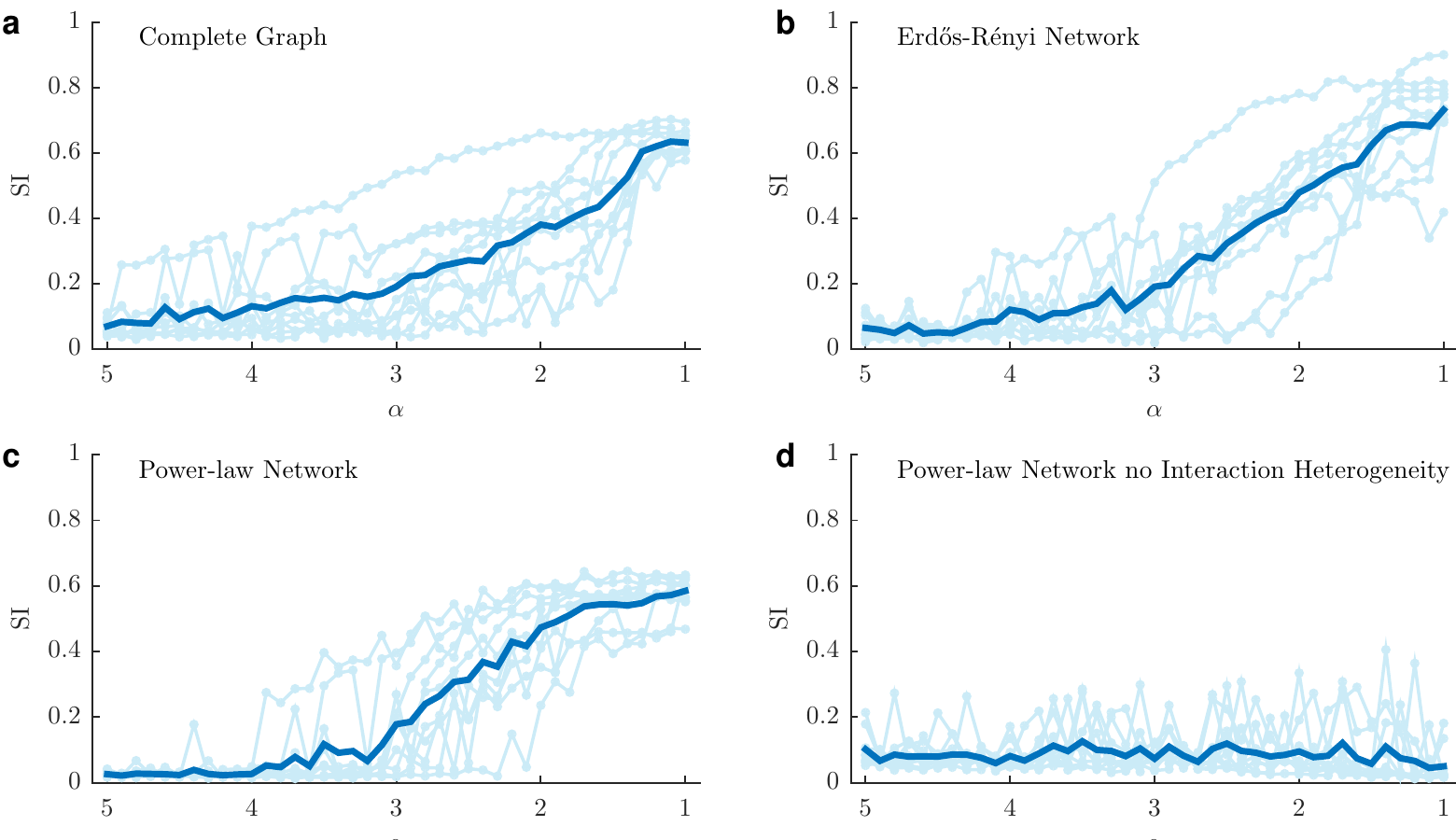}
  \caption{{\em Impact of network structure on the distinctness of community types.}  For each type of network structure 10 different Universal Triples $(\mathbf S, \mathbf A, \mathbf r)$ with $n=100$ species  and $q=500$ local communities of size $p=80$ were generated with results shown in the lighter color and averaged results shown in bold. (a) {\em Complete graph}. Same study as in Figure 2 with $\alpha\in [5,1)$. (b) {\em Erd\H{o}s-R\'enyi} network (digraph) $[\mathbf N]_{ij}\sim\mathcal N(0,1)$, $[\mathbf H]_{ii}\sim\mathcal P(\alpha)$ where $\alpha\in (1,5]$, Probability $[\mathbf G]_{ij}=1$ is $0.1$, i.e. a mean in(out)-degree of 10, and scaling factor $s=1/\sqrt{10}$. (c) {\em Power-law out-degree network} $[\mathbf N]_{ij}\sim\mathcal N(0,1)$, $[\mathbf H]_{ii}\sim\mathcal P(\alpha)$, $\mathbf G$ is the adjacency matrix for a digraph with out-degree having a power-law distribution $\mathcal P(\alpha)$.  The high-degree nodes have the largest interaction scaling. (d) {\em Power-law out-degree network, no interactions strength heterogeneity} $[\mathbf N]_{ij}\sim\mathcal N(0,1)$, $\mathbf H$ is the identity matrix, $\mathbf G$ is the adjacency matrix for a digraph with out-degree having a distribution $\mathcal P(\alpha)$. Further details can be found in the Methods Section.}\label{fig:4}
\end{figure}

\hfill
\clearpage

\begin{figure}\centering
\includegraphics[width=4.5 in]{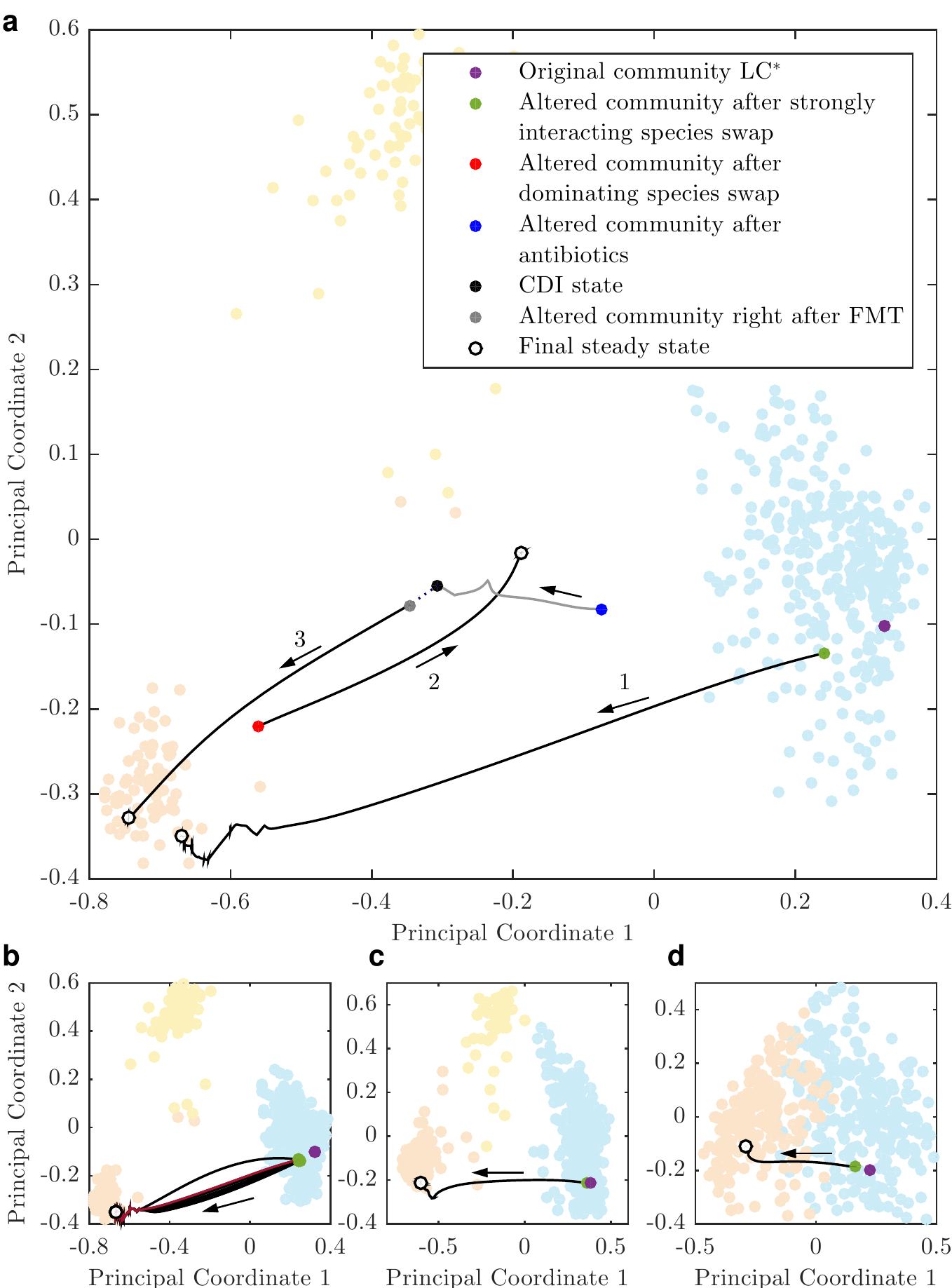}
  \caption{{\em Open-loop control of the human microbiome.}  
  (a) Background of clustering analysis for $\alpha=1.6$ from Figure \ref{fig:2}, but with Euclidean distance used so that a projection matrix could be found to show the trajectories in the 2D principle coordinate plane (Supplementary Text Sec. 5.6). We aim to steer a local community (denoted as LC$^*$, shown in magenta) in the blue cluster to the orange cluster. Three different scenarios are presented, per the three numbers above the arrows. Scenario 1:  SISs swap. The SISs (23 and 81) of LC$^*$ were replaced by the SISs present in the orange cluster (60 and 51). The initial abundances of species 60 and 51 were drawn from $\mathcal U(0,1)$, resulting in  the green dot, and the GLV dynamics were simulated. Scenario 2: dominating species swap.  The six most abundant species in LC$^*$ were removed and replaced by the six most abundant species from a local community in the orange cluster, with the initial condition after the switch of species shown as the red dot, and the dynamics were simulated until steady state was reached. Scenario 3: Fecal Microbiota Transplantation (FMT). The two SISs and 18 of the most abundant species (for a total of 20) were removed from LC$^*$ with the initial condition shown in blue (post-antibiotic state). Then the GLV dynamics were simulated (gray line) and the system converged to the black dot (CDI state). Then 1\% of the steady abundances from an arbitrary LC in the orange cluster were  added to the CDI state (gray dot, emulating oral capsule FMT) and the dynamics were then simulated until steady state was reached. 
  (b) The SISs swap process was repeated ten times, each time the initial abundances of species 60 and 51 were randomly drawn from $\mathcal U(0,1)$. Nine of the simulations are shown in black and the simulation that pertains to Figure 4a is shown in maroon.
  (c) The same analysis as for Figure \ref{fig:5}a, in terms of SISs swap, but for $\alpha=2$. 
  (d) The same analysis as for Figure \ref{fig:5}a, in terms of SISs swap, but for $\alpha=3$.}\label{fig:5}
\end{figure}

\chapter{Materials and Methods}

  \renewcommand{\theequation}{M\arabic{equation}}%

The methods section begins with a toy example to illustrate the construction of the  universal interaction matrix  $\mathbf A  = {\color{red}\mathbf  N} {\color{blue} \mathbf  H} \circ {\color{violet} \mathbf G} {\color{green} s}$ in \eqref{eq:A}, where
\begin{align*}
\text{steps:} \quad  & (i)&   {\color{red} \mathbf N } & {\color{red} = }{\color{red} \begin{bmatrix}
0 &0.2 &0.4& -0.1\\
0.7 &0& 0.3 & 0.4\\
-0.1 &0.7 &0& 0.1\\
-0.3 & -0.2 &0.4 & 0
\end{bmatrix}}\\
& (ii)&{\color{blue}\mathbf H } & {\color{blue} = }{\color{blue} \begin{bmatrix}
10 & 0 & 0 & 0 \\
0 & 0.2 & 0 & 0 \\
0 & 0 & 0.2 & 0 \\
0 & 0 & 0 & 0.4 
\end{bmatrix}}\\
& (iii)&\quad{\color{violet} \mathbf G }&{\color{violet} = } {\color{violet} \begin{bmatrix}
0 & 1 & 1 & 1\\
1 & 0 & 1 & 0 \\
1 & 0 & 0 & 0 \\
0 & 0 & 1 & 0 
\end{bmatrix}}\\
 & (iv)& {\color{green} s } & {\color{green}=}{\color{green} 1} \\
 & (v)& [\mathbf A]_{ii}&=-1\\
 \text{final result:} \quad & & \mathbf A&=\begin{bmatrix}
-1 & 0.04 & 0.08 & -0.04\\
7 & -1 & 0.06 & 0 \\
-1 & 0 & -1 & 0 \\
0 & 0 & 0.08 & -1 \end{bmatrix}
\end{align*}
\noindent Given that $\mathbf H$ is diagonal, it scales the columns of $\mathbf N$.  If one thinks of $\mathbf A$ as the adjacency matrix of a digraph, then $\mathbf H$ scales all of the edges leaving a node. Thus one can consider $\mathbf H$ as controlling the interaction strength heterogeneity of $\mathbf A$. Given the Hadamard product between  $\mathbf H$ and $\mathbf G$, the off-diagonal elements of $\mathbf G$ that are zero will result in the corresponding off-diagonal elements of $\mathbf A$ being zero as well.

In the first study (Figure \ref{fig:2}), to explore the impact of interaction heterogeneity on steady state shift, we varied the exponent $-\alpha$ of the power-law distribution of $[\mathbf H]_{ii}$ to generate five different universal interaction matrices $\mathbf A$ of dimension $100\times100$.  For each universal  interaction matrix $\mathbf A$, the nominal component $\mathbf N$ consists of independent and identically distributed elements sampled from a normal distribution $\mathcal N(0,1)$. The topology for this study was a complete graph and thus all the elements in $\mathbf G$ are equal to 1. The heterogeneity element $\mathbf H$ is constructed in two steps. First, five different   vectors $\bar h(\alpha)\in \Re^{100}$ are constructed where each element is sampled from a power-law distribution $\mathcal P(\alpha)$ for $\alpha \in \{7,3,1.6,1.2,1.01\}$. Then, each of the $\bar h(\alpha)$ is normalized to have a mean of 1, $h = \bar h / \mathrm{mean}(\bar h).$ Finally the heterogeneity matrix is defined as $\mathbf H=\diag\left( h\right)$. For this study $ s=0.07$, ensuring uniform asymptotic stability for the case of low heterogeneity (see Supplementary Text Theorem \ref{thm:asym_stab}). The final step in the construction of $\mathbf A$ is to set the diagonal elements to $-1$. 

For each $\alpha$ the following simulation steps were taken. There are a total of 100 species, $\mathbf S=\{1,\, 2,\, \ldots,\, 100\}$, in the metacommunity, and  each of the 500 local communities contains $80$ species, randomly chosen from  $\mathbf S$. The MATLAB command used to perform this step is \verb*:randperm:. The initial condition for each of the 500 local communities, $\xx(0)$, were sampled from $\mathcal U(0,1)$. The dynamics were then simulated for 10 seconds using the MATLAB command \verb*:ode45:. If any of the 500 simulations crashed due to instability or if the norm of the terminal discrete time derivative was greater than $0.01$ then that local community was excluded from the rest of the study. Those simulations that finished without crashing and with small terminal discrete time derivative were deemed steady. Less than $1\%$ of simulations were deemed unstable in the preparation of Figure \ref{fig:2}.

The networks presented in the second row of Figure \ref{fig:3} were constructed by considering $\mathbf A$ as the weighted adjacency matrix of the network. Note that arrows showing directionality and self loops were suppressed. The links were color coded in proportion to the absolute value of the entries in $\mathbf A$.

For the last row of Figure \ref{fig:2} a clustering analysis was performed. For each $\alpha$ the steady state abundances of the 500 local communities were normalized so that we have 500 synthetic microbial samples. Then $k$-medoids clustering was performed for $k\in\{1,\, 2,\,\ldots,\,10\}$ using the Jensen-Shannon distance metric (Supplementary Text Sec. \ref{sec:dm}). Silhouette analysis was performed to determine the optimal number of clusters and the clustering results were  illustrated in the 2-dimensional principle coordinates plot.  For Supplementary Figure \ref{fig:ext1} the same steps as for the preparation of Figure \ref{fig:2} were performed, but with $\mathbf G$ representing the adjacency matrix of an Erd\H{o}s-R\'enyi  digraph with mean degree of $20$ (mean in-degree of 10 and mean out-degree of 10) and $s=1/\sqrt{10}$. Details on the construction of an Erd\H{o}s-R\'enyi digraph can be found in Supplemental Information Section \ref{sec:ER}. For Supplementary Figure \ref{fig:ext2} the same steps as above were performed in Figure \ref{fig:2} but with $p=5,000$ local communities.

Figure \ref{fig:4} is a macroscopic analysis of how network structure plays a role in the steady state shift with values of ${\alpha\in(1,5]}$. For each topology ten different universal matrices $\mathbf A$ were generated. Figure \ref{fig:4}(a) shows the results of a complete graph and for each of the ten universal  $\mathbf A$  the same steps as in the preparation of Figure \ref{fig:2} were carried out. Figure \ref{fig:4}(b) shoes the result of an Erd\H{o}s-R\'enyi random digraph topology and for each of the ten  $\mathbf A$ matrices the same steps as in the preparation of Supplementary Figure \ref{fig:ext2} were carried out. Figure \ref{fig:4}(c) shows results for networks with a power-law out-degree distribution with a mean out-degree of 10, where the out-degree sequence uses the same $\bar h$ in the construction of $\mathbf H$. More information on the construction of  $\mathbf G$ for a power-law out-degree network can be found in Supplementary Text Sec. \ref{sec:powerdigraph}. Figure \ref{fig:4}(d) shows results for networks with a power-law out-degree distribution with mean out-degree of 10 and there is no interaction strength heterogeneity, i.e. $\mathbf H$ is the identity matrix. For this study the Silhouette Index was constructed from normalized steady state data using the Jensen-Shannon distance. Supplementary Figure  \ref{fig:ext3}  is the same as Figure \ref{fig:4}, but instead of the Silhouette Index, the variance ratio criterion is used with the Jensen-Shannon distance, from normalized steady state abundance  (Supplementary Text Sec. \ref{sec:vrc}). In Supplementary Figure  \ref{fig:ext4}  the Silhouette Index is determined from the Euclidean distance with normalized steady state abundance. Finally, in Supplementary Figure  \ref{fig:ext5}  the Silhouette Index is determined by the Euclidean norm with the absolute steady state abundance.

Figure \ref{fig:5} contains a PCoA analysis of the results from Figure \ref{fig:2}, but with the Euclidean distance being used instead of the Jensen-Shannon distance, making PCoA equivalent to principle component analysis. This enables us to project the open-loop control trajectories  into the principle coordinates (Supplementary Text Sec \ref{sec:pca}). This procedure was also used in the preparation of Supplementary Figure  \ref{fig:ext6} .

Supplementary Figures  \ref{fig:ext7}  to  \ref{fig:ext9}  contain system identification analyses for temporal gut microbiome data of two subjects \cite{caporaso2011moving}. The data is publicly available from the metagenomics analysis server MG-RAST:4457768.3-4459735.3 and can also be accessed (as we did) from Qiita (http://qiita.ucsd.edu) under study ID 550. The processed data was downloaded as biom file ``67\_otu\_table.biom'' (2014-11-17 13:18:50.591389). The {\em Operational Taxonomic Units} (OTUs) were then grouped from the genus level and up, depending on the availability of known classifications for OTUs, and converted to a txt file using MacQIIME version 1.9.0-20140227 with the command {\verb:summarize_taxa.py:} with the options \verb:-L 6 -a true:. Data was collected over 445 days with 336 fecal samples from Subject A and 131 fecal samples from Subject B. Details on the system identification algorithm are now given. 
The dynamics in \eqref{eq:lvd2} can be approximated in discrete time as \cite{stein:2013}
\be \label{eq:logdiff}e_i(k)+\log\left( x_i(t_{k+1})\right)- \log\left( x_i(t_k)\right)  =r_i+\sum_{j=1}^{n}a_{ij}x_j(t_k)\ee
for $i=1,2,\ldots, n$ where $k=1,2,\ldots, N-1$ is the sample index, $N$ is the total number of samples,  $t_k$ is the time stamp of sample $k$, and $e$ is an error term that arises because of the assumption that $x(t)$ is constant over each interval $t\in[t_k,t_{k+1})$. Equation \eqref{eq:logdiff} can be rewritten 
in terms of a regressor vector $$\phi(k)  =[1,x_1(t_k),\ x_2(t_k),\ \ldots,\ x_n(t_k)]^\mathsf T,$$ the parameter vector $\theta_{i}  =[r_i,\ a_{i1},\ a_{i2},\ \ldots,\ a_{in}]$ and the log difference $y_{i}(k) = \log\left( x_i(t_{k+1})\right)- \log\left( x_i(t_k)\right)$ as
\begin{equation*}
 e_i(k)+y_{i}(k) = \theta_{i}\phi(k).
 \end{equation*}
The identification problem can then be defined as finding the parameter matrix estimate ${\hat{\Theta}} = \left[\hat{\theta}_{1}^\mathsf T,\hat{\theta}_{2}^\mathsf T,\cdots,\hat{\theta}_{n}^\mathsf T \right]^\mathsf T$ of the true parameter matrix ${{\Theta}} = \left[{\theta}_{1}^\mathsf T,{\theta}_{2}^\mathsf T,\cdots,{\theta}_{n}^\mathsf T \right]^\mathsf T$. Letting
$$y(k)=[y_1(k),\ y_2(k),\ \ldots,y_n(k)]^\mathsf T$$ be the log difference vector for all species and ${Y} = \left[y(1),\ y(2),\ \ldots,\ y(N-1)\right]$ be the log difference matrix the system identification problem can be compactly presented as 
\begin{equation*}
\min_{{\hat{\Theta}}}  \|{Y}-{\hat{\Theta}}{\Phi}\|_F^2+\lambda\|{\hat{\Theta}}\|_F^2
\end{equation*}
where ${{\Phi} = \left[\phi(1),\ \phi(2),\ \ldots,\ \phi(N-1)\right]}$ is the regressor matrix, $\norms{\cdot}_F$ denotes the Frobenius norm, ${\lambda\geq 0}$ is the  Tikhonov regularization term  \cite{tikhonov1963solution}. The minimal solution to the above problem can be given directly as 
\ben
\argmin_{\hat\Theta}  \left(\|{Y}-{\hat{\Theta}}{\Phi}\|_F^2+\lambda\|{\hat{\Theta}}\|_F^2 \right)=  F Y^\mathsf T(YY^\mathsf T + \lambda I)^{-1}
\een
where $I$ is the identity matrix. 

Next we discuss how missing data, zero reads, and $\lambda$ were chosen. The difference equation in \eqref{eq:logdiff} only uses sample data over two consecutive time samples. Therefore, in the construction of $Y$ and $\Phi$ we only include samples that for which there is data from the next day as well. Also, given that the logarithms are used, when a sample has zero reads for a given taxa, a read value of one is inserted. Then relative abundances are computed before the logarithm is taken. Finally we discuss how the regularization parameter is chosen. For Supplementary Figures  \ref{fig:ext7}  and  \ref{fig:ext8}  the following cross-validation is performed.
For Subjects A and B two-thirds of data was used for training and one-third for testing. More precisely, for each $\lambda$ two-thirds of the data from Subject A and two-thirds of the data from Subject B were used to identify their corresponding dynamical constants. Then the combined error from the two test sets was used to find the optimal $\lambda$. The regularization value used in Supplementary Figure  \ref{fig:ext9}  is simply the same regularization value used in Supplementary Figure  \ref{fig:ext7} .

\clearpage

\chapter{Supplementary Figures}

\renewcommand{\figurename}{Supplementary Figure}
 \setcounter{figure}{0}
  \renewcommand{\thefigure}{S\arabic{figure}}%

\begin{figure}[h!]\centering
  \includegraphics[width=\textwidth]{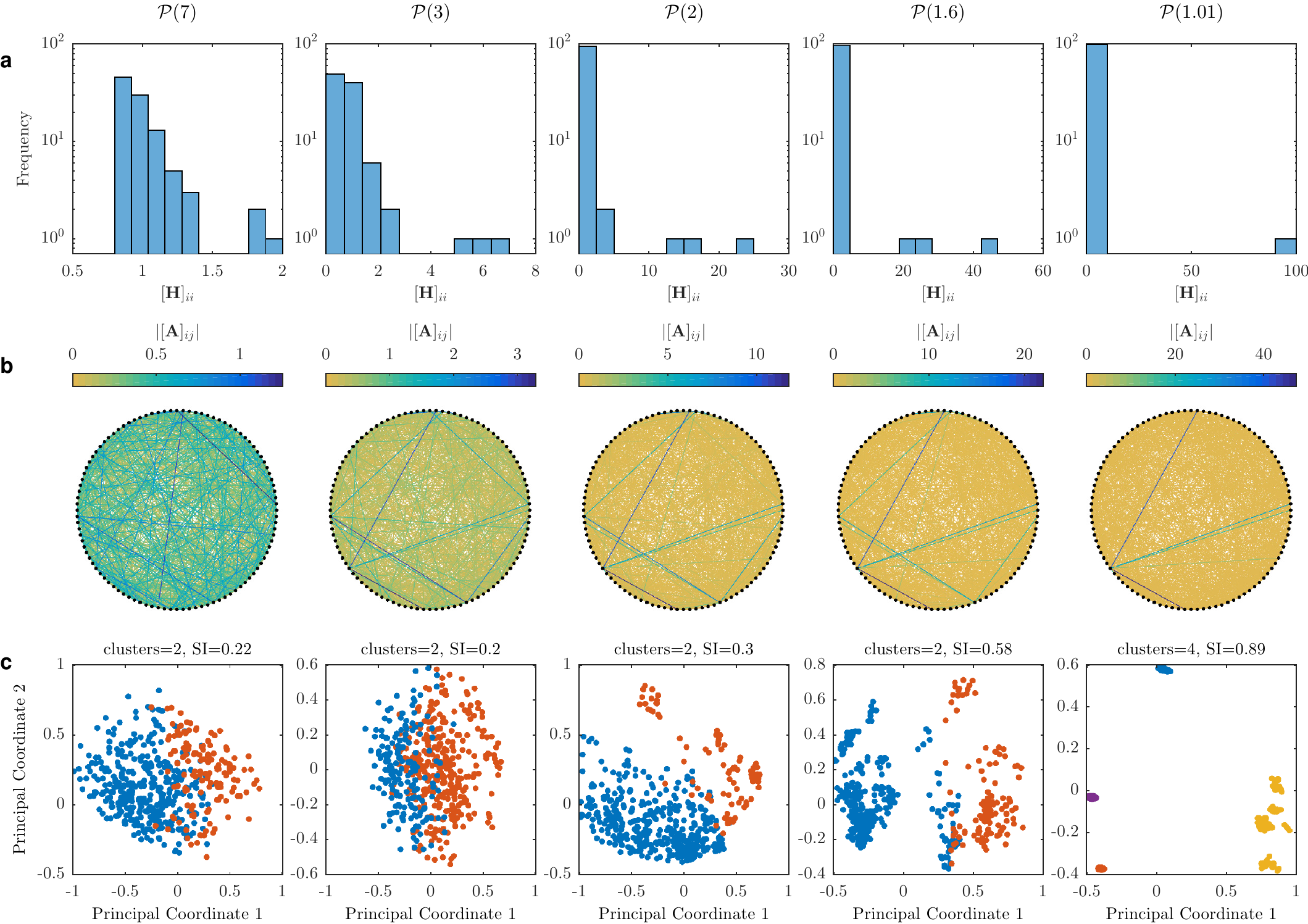}
  \caption{{\em  Impact of interaction strength heterogeneity on the distinctness of  community types}.  Same as Figure \ref{fig:2} but with
  the topology component $\mathbf G$ chosen to be an Erd\H{o}s-R\'enyi digraph with a link probability of 0.1 and the scaling factor was set at $s=1/\sqrt{10}$.} \label{fig:ext1} \end{figure}

\clearpage

\begin{figure}\centering
  \includegraphics[width=\textwidth]{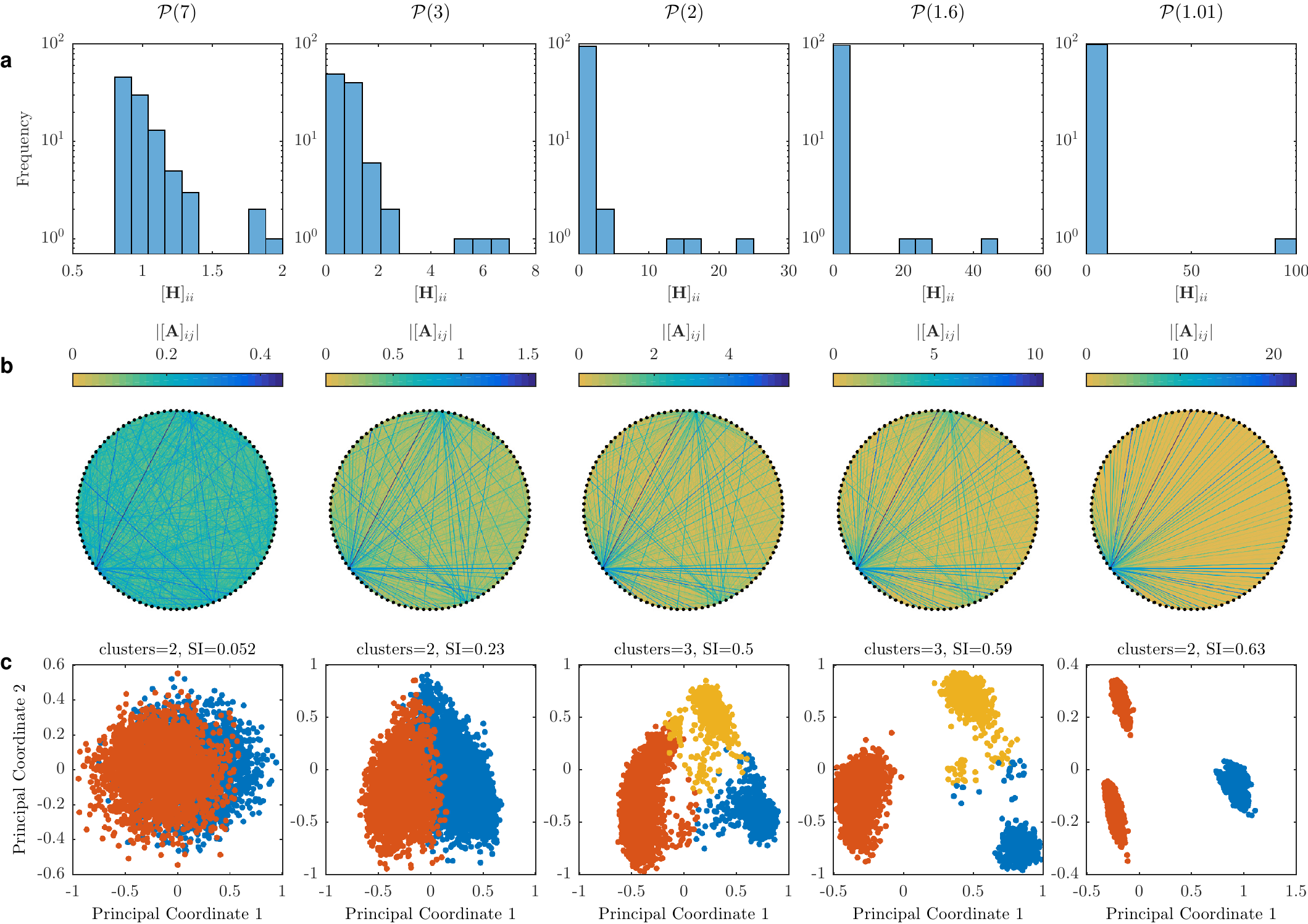}
  \caption{{\em  Impact of interaction strength heterogeneity on the distinctness of community types}.  Same as Figure \ref{fig:2} but with $p=$5,000 local communities. Note that it is rather counter-intuitive that for $\alpha=1.01$ the Silhouette Index suggests that there are two clusters, while PCoA suggests three clusters. We emphasize that as a typical ordination method, the PCoA just produces a spatial representation of the entities in the dataset, rather than the actual determination of cluster membership \cite{milligan1987methodology,Jain:1999:DCR:331499.331504}. Note that as compared to Figure \ref{fig:2}, because there are more samples in this figure, the distinctness of the clusters when $\alpha=2$ has shifted to more of a continuous gradient as apposed to distinct clusters.}  \label{fig:ext2}\end{figure}

\clearpage

\begin{figure}\centering
  \includegraphics[width=\textwidth]{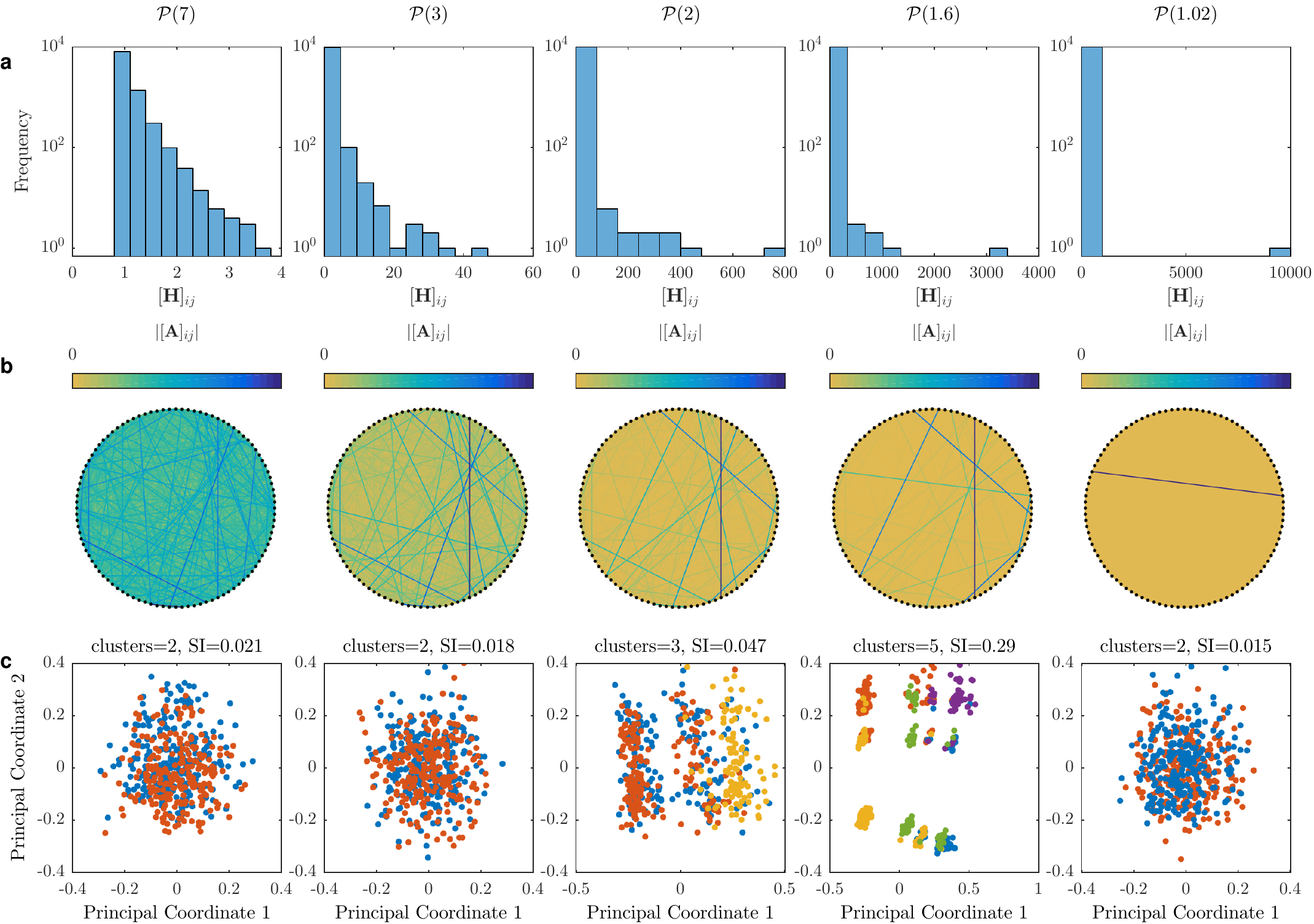}
  \caption{{\em  Impact of  interaction heterogeneity disbursed randomly throughout the network}. The set up is the same as that of Figure  \ref{fig:2} but instead of $\mathbf H$ being a diagonal matrix, it is a full matrix, so that individual interactions are scaled randomly from a power-law distribution. }  \label{fig:ext31}\end{figure}

\clearpage

\begin{figure}\centering
  \includegraphics[width=\textwidth]{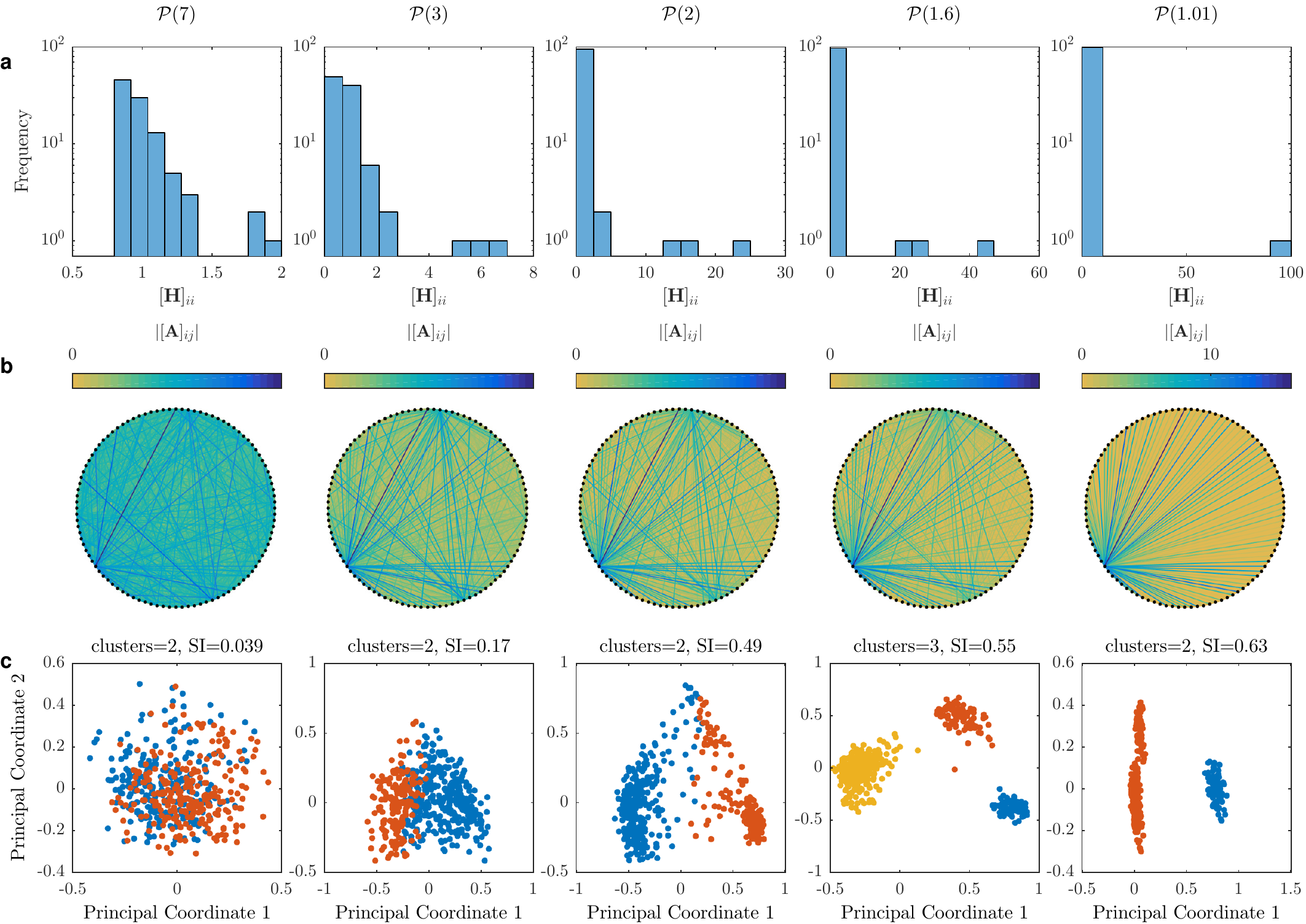}
  \caption{{\em  Impact of low levels of migration}. Same as Figure \ref{fig:2} but with a new term $\lambda(t)\in[0,1]^n$ added to the dynamics so that now
$\dot x = \lambda + \diag(x)(r+Ax)$. In this example $\lambda_i \sim \mathcal U(0,0.1)$. The disturbance is sampled every $0.01$ seconds and held constant until the next sample is taken..}  \label{fig:ext41}\end{figure}

\clearpage

\begin{figure}\centering
  \includegraphics[width=\textwidth]{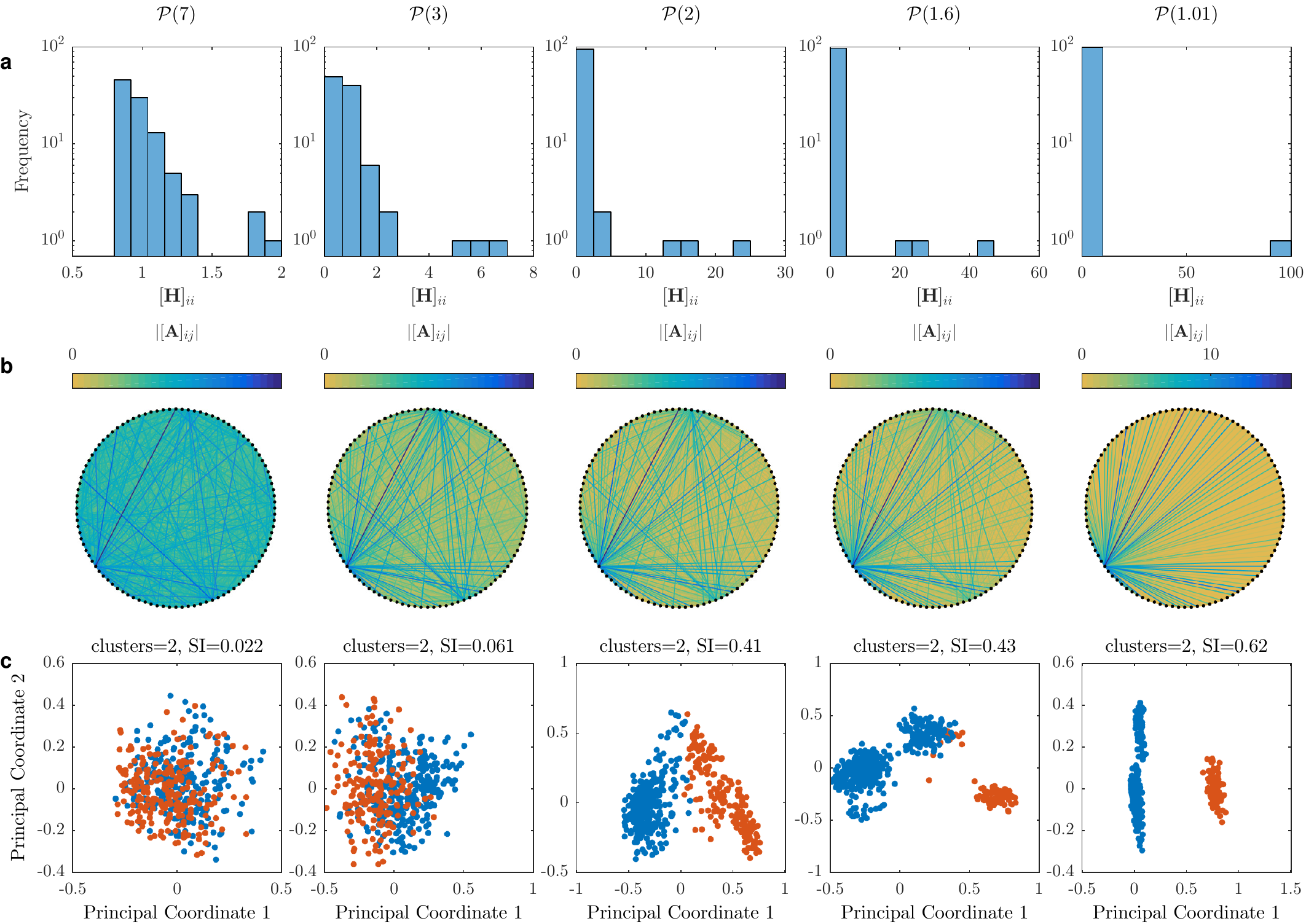}
  \caption{{\em  Impact of moderate levels of migration}. Same as Figure \ref{fig:2} but with a new term $\lambda(t)\in[0,1]^n$ added to the dynamics so that now
$\dot x = \lambda(t) + \diag(x)(r+Ax)$. In this example $\lambda_i \sim \mathcal U(0,1)$. The disturbance is sampled every $0.01$ seconds and held constant until the next sample is taken.}  \label{fig:ext51}\end{figure}

\clearpage

\begin{figure}\centering
  \includegraphics[width=\textwidth]{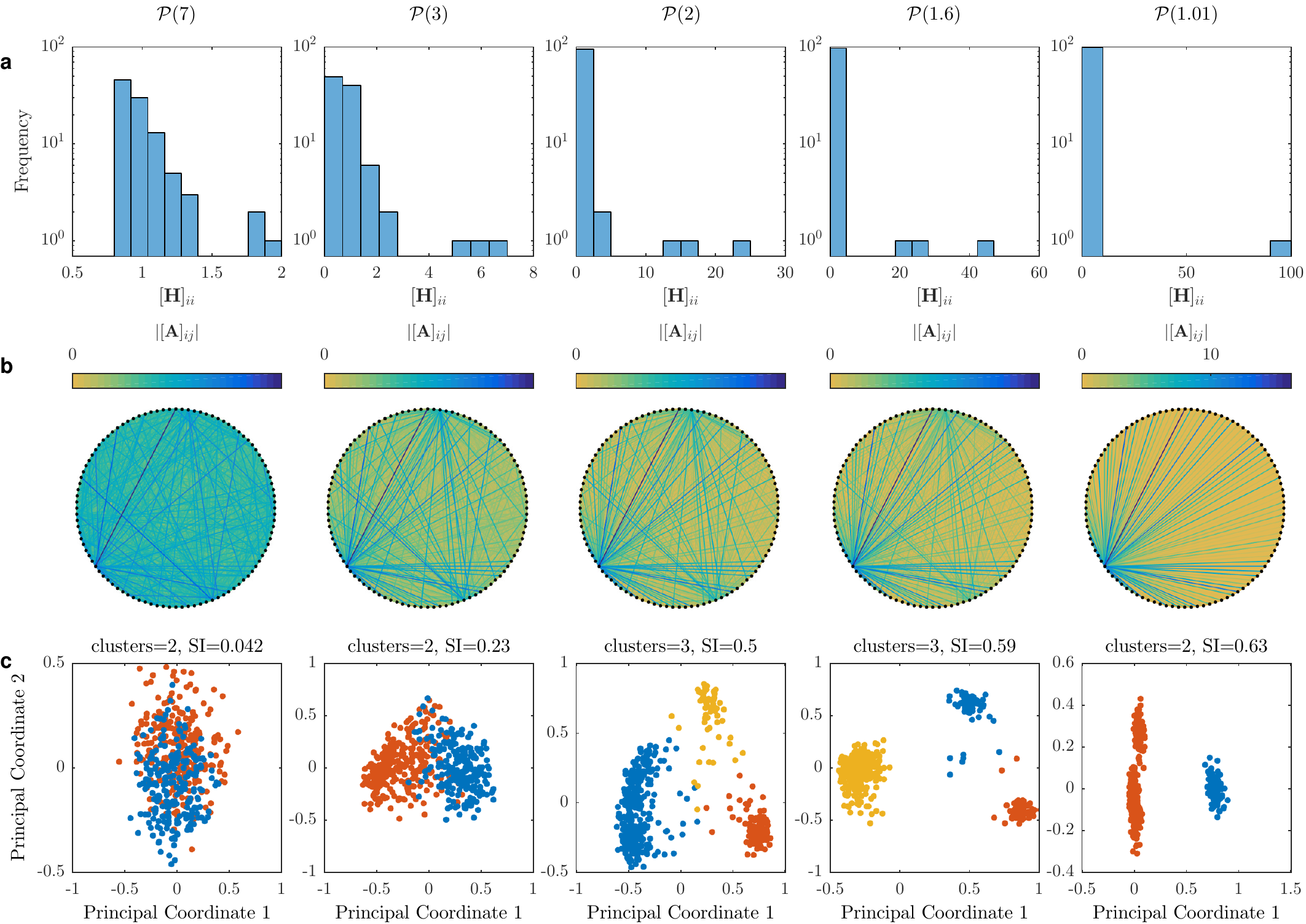}
  \caption{{\em  Impact of small stochastic disturbance}. Same as Figure \ref{fig:2} but with stochastic It\^o dynamics $\mathrm d x = \diag(x)(r\, \mathrm dt+Ax\,\mathrm dt + c\, \mathrm d w)$ where $w$ is a $n$-dimensional Brownian motion and $c$ represents the stochastic disturbance strength. Dynamics were simulated with a discrete time step of $0.01$ seconds and $c=0.1$. }  \label{fig:ext61}\end{figure}

\clearpage

\begin{figure}\centering
  \includegraphics[width=\textwidth]{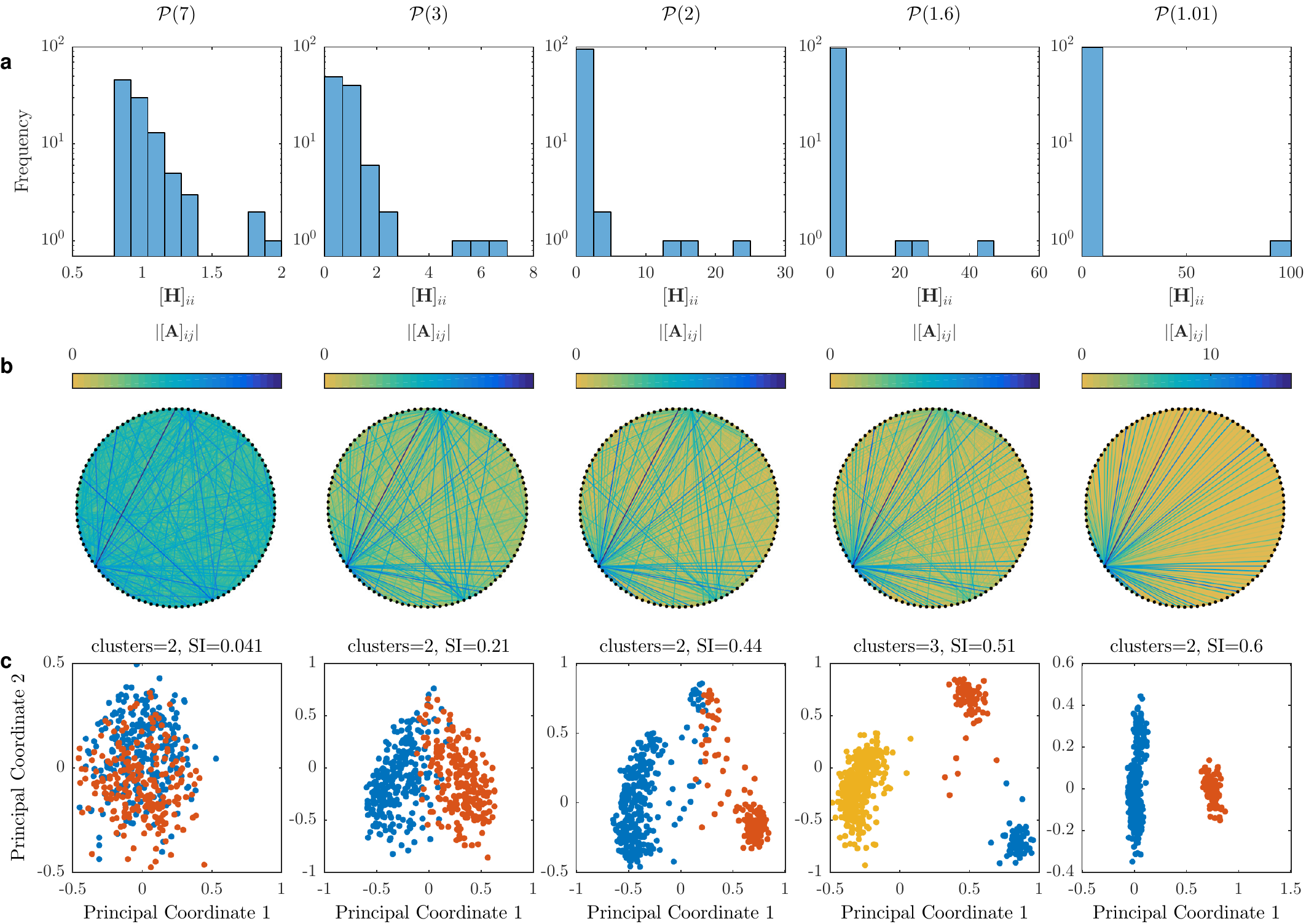}
  \caption{{\em  Impact of moderate stochastic disturbance}. Same as Figure \ref{fig:2} but with stochastic It\^o dynamics $\mathrm d x = \diag(x)(r\, \mathrm dt+Ax\,\mathrm dt + c\, \mathrm d w)$ where $w$ is a $n$-dimensional Brownian motion and $c$ represents the stochastic disturbance strength. Dynamics were simulated with a discrete time step of $0.01$ seconds and $c=0.5$. 
}  \label{fig:ext71}\end{figure}

\clearpage

\begin{figure}\centering
  \includegraphics[width=\textwidth]{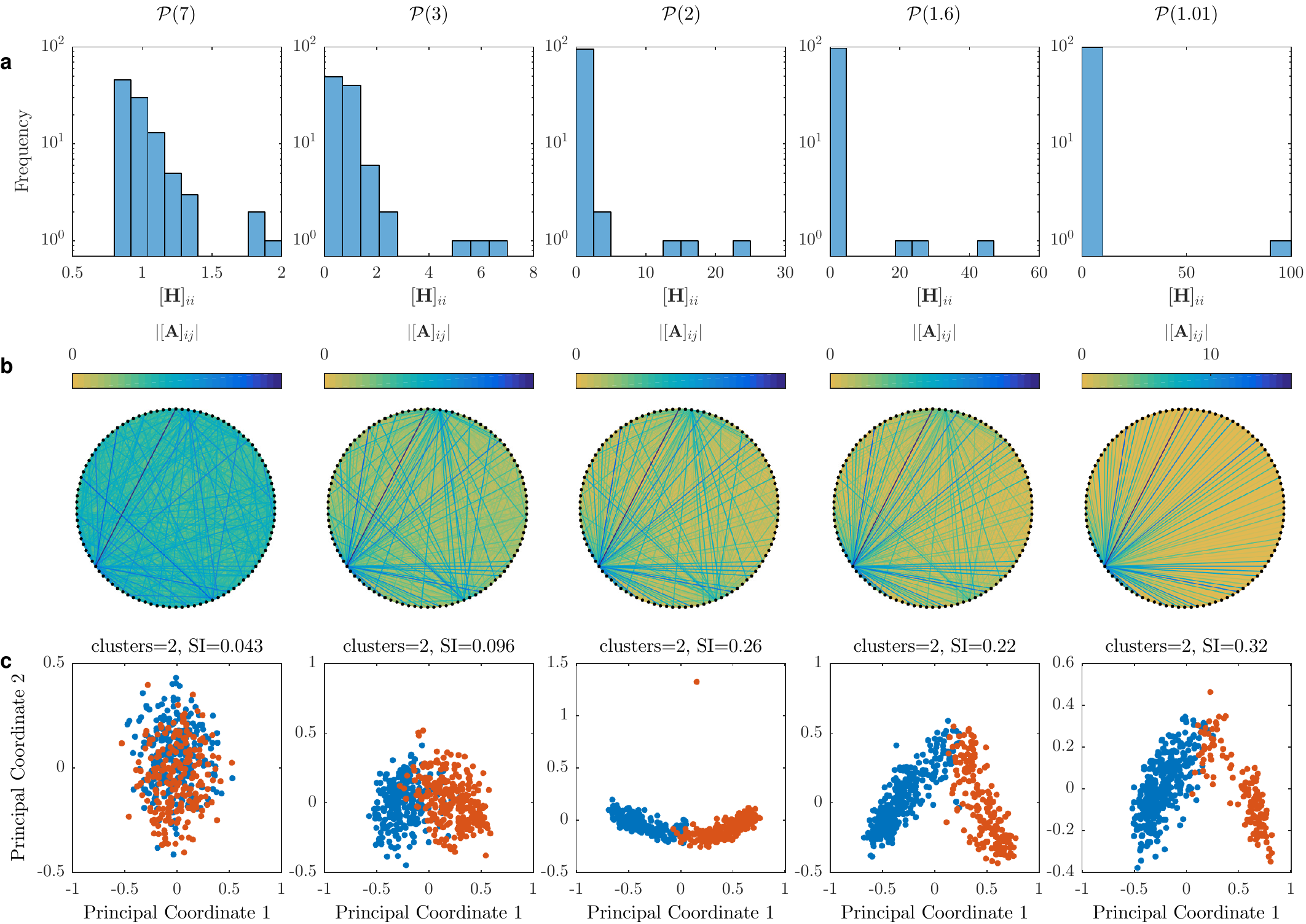}
  \caption{{\em  Impact of large stochastic disturbance}. Same as Figure \ref{fig:2} but with stochastic It\^o dynamics $\mathrm d x = \diag(x)(r\, \mathrm dt+Ax\,\mathrm dt + c\, \mathrm d w)$ where $w$ is a $n$-dimensional Brownian motion and $c$ represents the stochastic disturbance strength. Dynamics were simulated with a discrete time step of $0.01$ seconds and $c=1$. 
}  \label{fig:ext81}\end{figure}

\clearpage

\begin{figure}\centering
  \includegraphics[width=\textwidth]{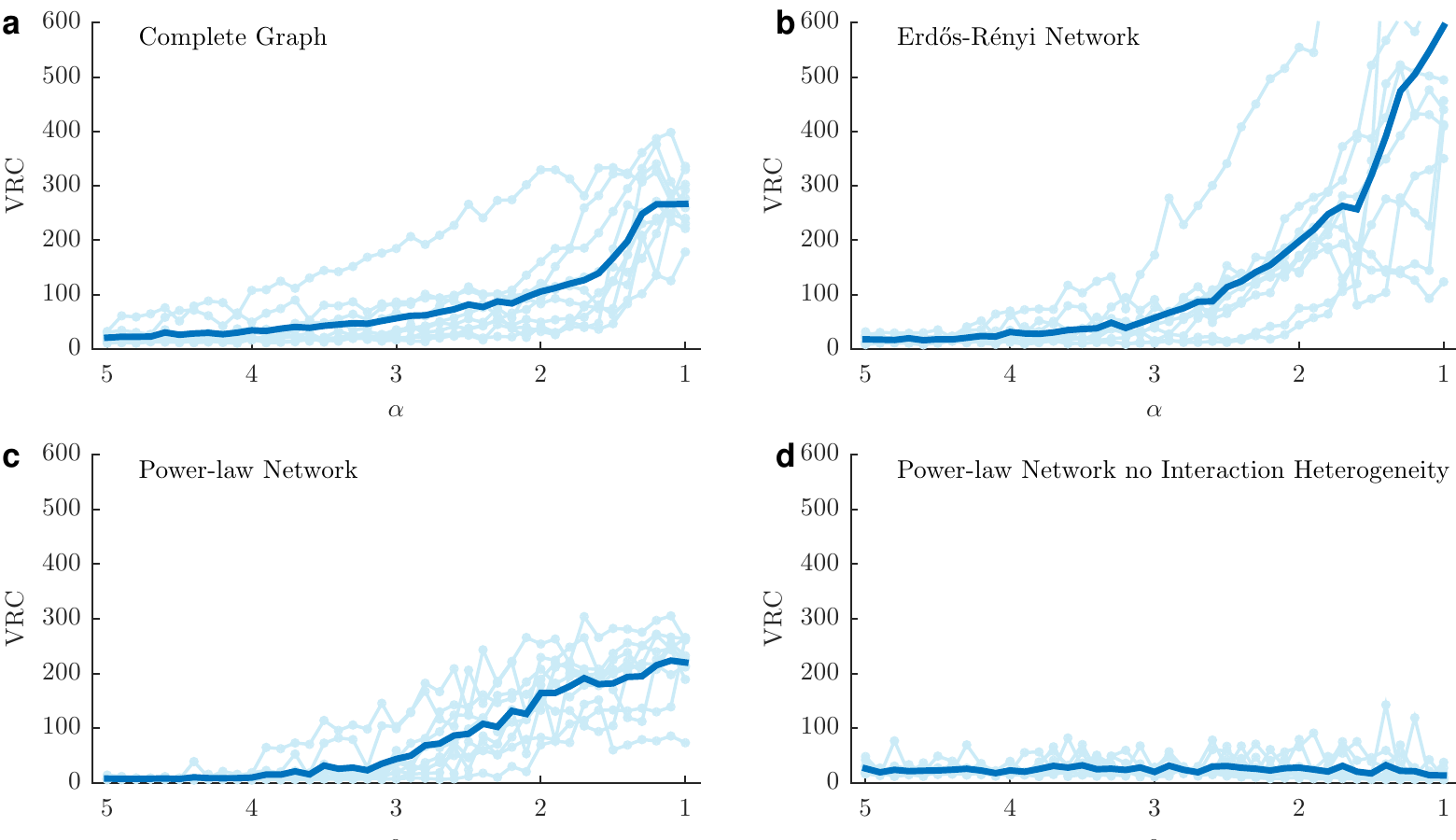}
  \caption{{\em Impact of network structure on the distinctness of community types.} The same as Figure \ref{fig:4} with the {\em Variance Ratio Criterion}  (VRC) used as apposed to the Silhouette Index for the clustering measure. See Supplemental Information \S5.4 for details on the VRC.}\label{fig:ext3}
\end{figure}
\clearpage
\begin{figure}\centering
  \includegraphics[width=\textwidth]{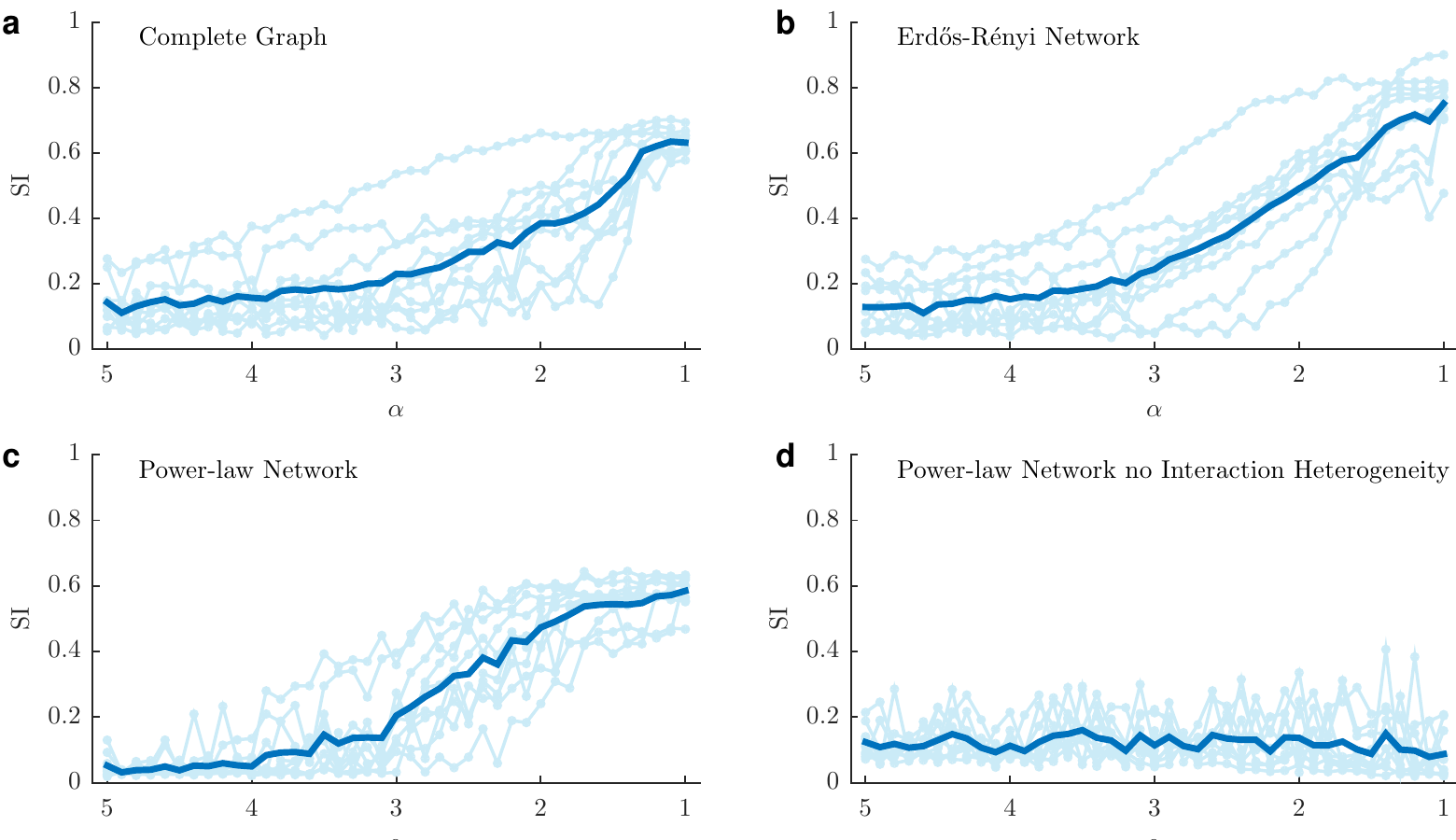}
   \caption{{\em Impact of network structure on the distinctness of community types.} The same as Figure \ref{fig:4} with the Euclidean distance metric used instead of the Jensen-Shannon distance metric.}\label{fig:ext4}
\end{figure}

\clearpage

\begin{figure}
\centering  \includegraphics[width=\textwidth]{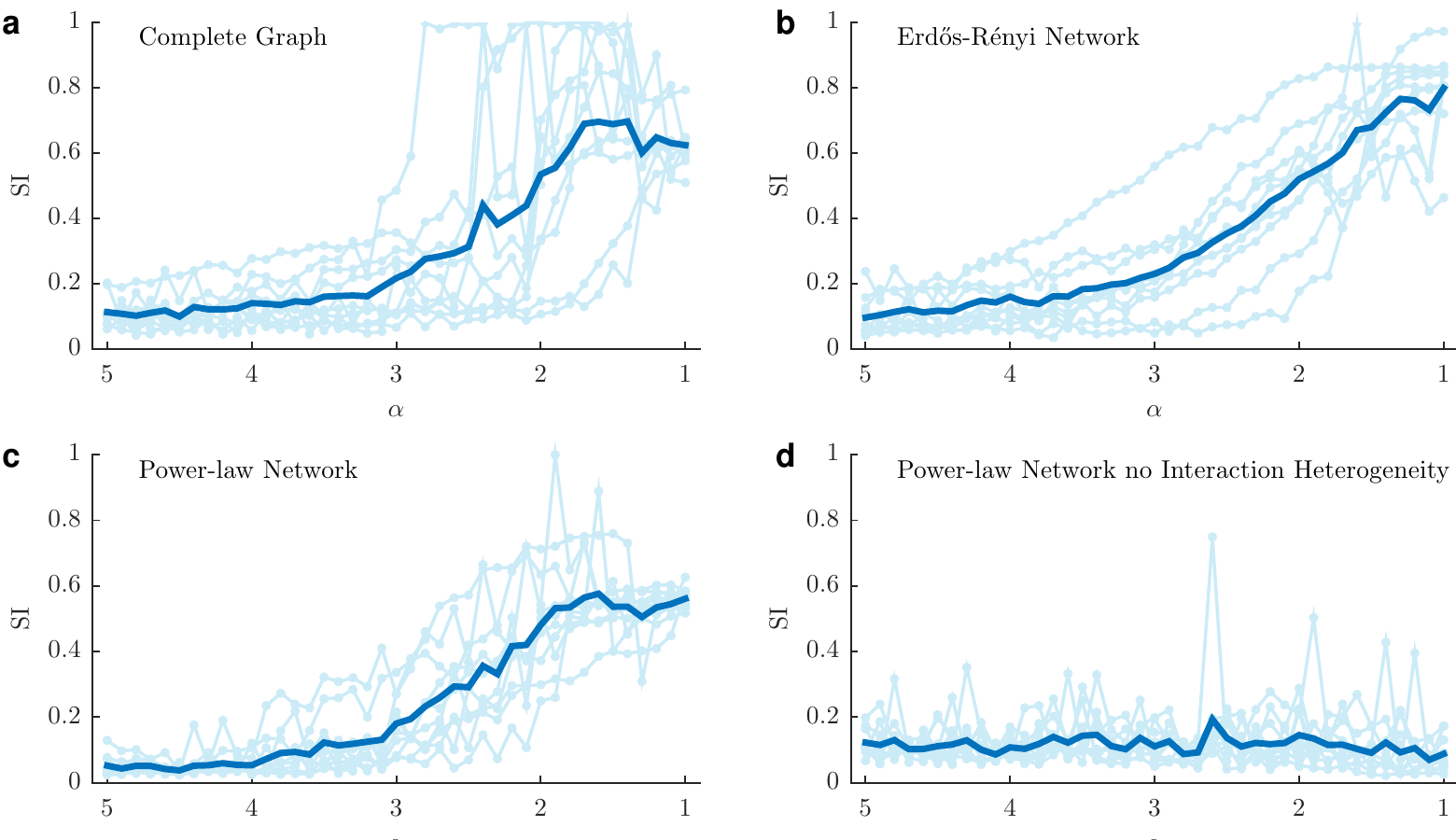}
   \caption{{\em Impact of network structure on the distinctness of  community types.} The same as Figure \ref{fig:4} with the Euclidean distance metric used instead of the Jensen-Shannon distance metric and absolute abundance used instead of relative abundace.}\label{fig:ext5}
\end{figure}

\clearpage

\begin{figure}\centering
  \includegraphics[width=4.5 in]{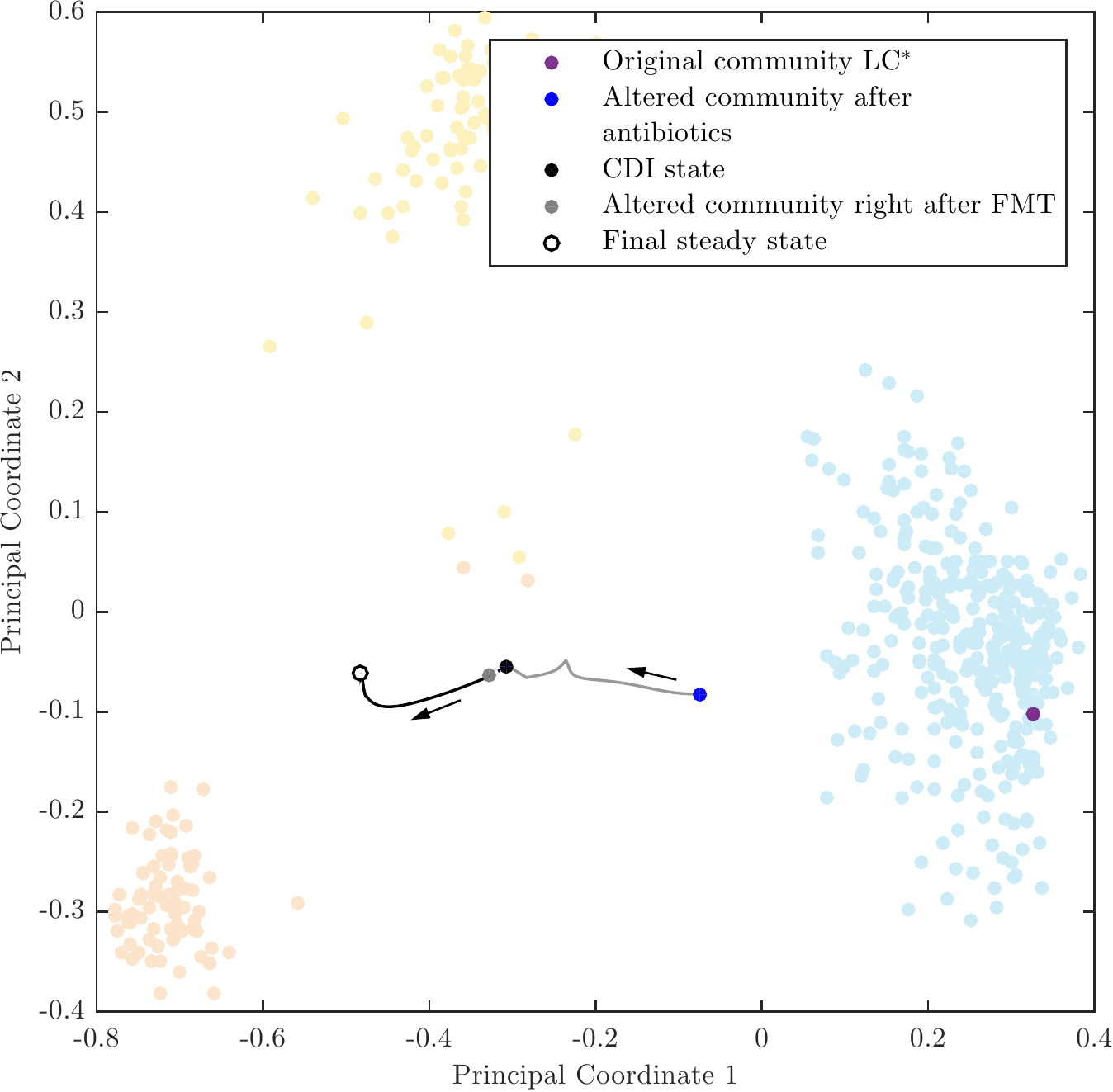}
  \caption{{\em Unsuccessful Fecal Microbiota Transplantation.}    Similar to Scenario 3 shown in Figure  \ref{fig:5}a, but during the FMT, the SISs (60 and 51) of the donor's local community in the orange cluster were not transplanted to the CDI state (black dot). This FMT resulted in a slightly altered community (gray dot) and the system eventually evolved to a steady state (white dot) which is not in the orange cluster. Hence the FMT failed.}\label{fig:ext6}
\end{figure}

\clearpage
\begin{figure}\centering
  \includegraphics[width=4.5 in]{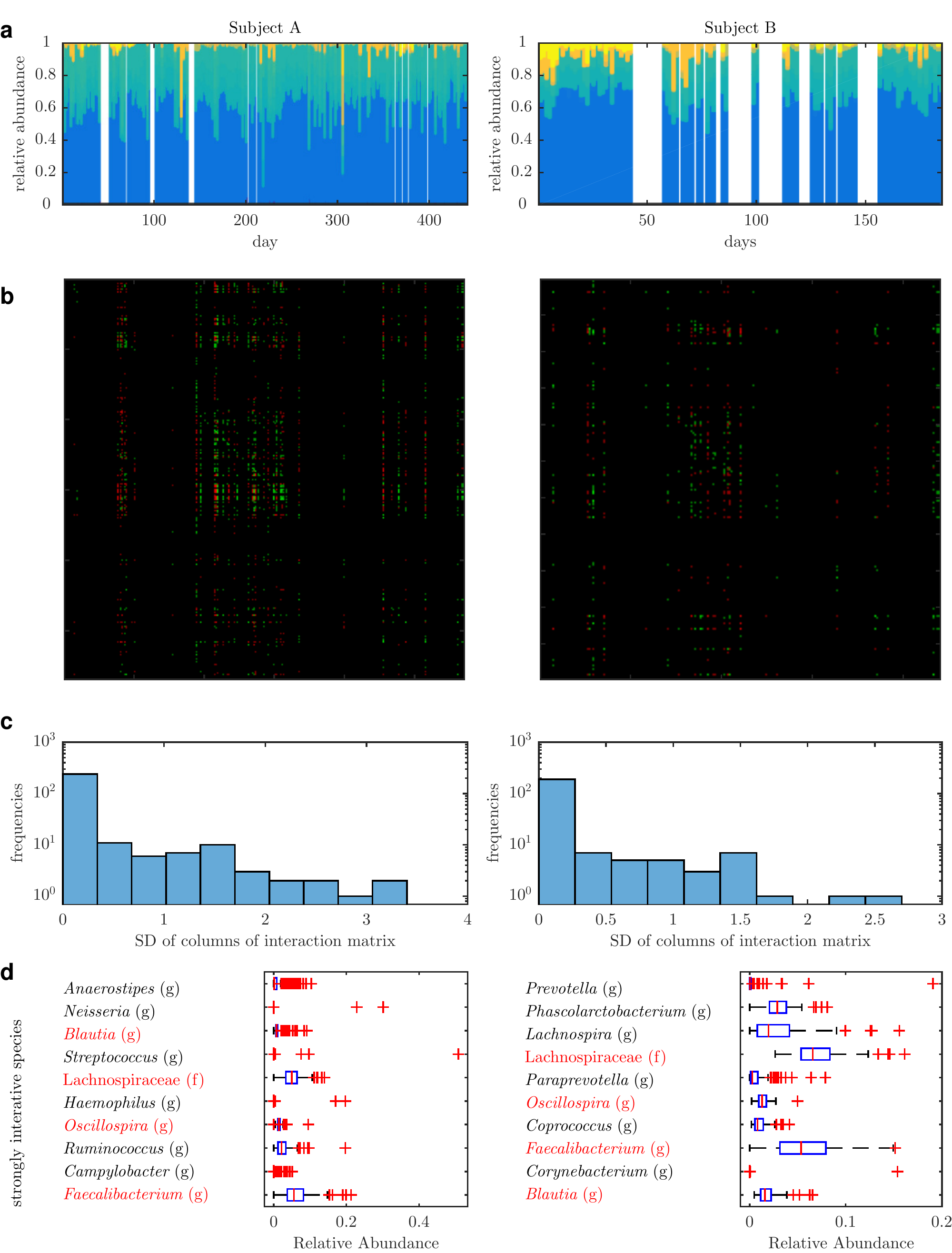}
  \caption{{\em System Identification, Tikhonov Regularization $\lambda=0.0423$.}  System identification was performed on the stool samples from the longitudinal data in \cite{caporaso2011moving} for two subjects as described in the Supplementary Methods where $\lambda$ was determined by cross-validation. (a) Visualization of microbial taxa in terms of relative abundances versus day sample was taken. (b) Heat map of the interaction matrix for top 100 SISs. (c) Histogram of {\em Standard Deviation} (SD) of the columns of the interaction matrix. (d) List of top ten SISs in descending interaction strength (defined by the SD of each column in the interaction matrix) with relative abundances over all samples shown as a box plot. The banded structure shown in the heat map supports the assertion that SISs do exist in the gut microbiome. However this banded structure is also seen when the dates of the sample collections are permuted, see Supplementary Figure \ref{fig:ext8} and \ref{fig:ext9}.}\label{fig:ext7}
\end{figure}

\clearpage
\begin{figure}\centering
  \includegraphics[width=4.5 in]{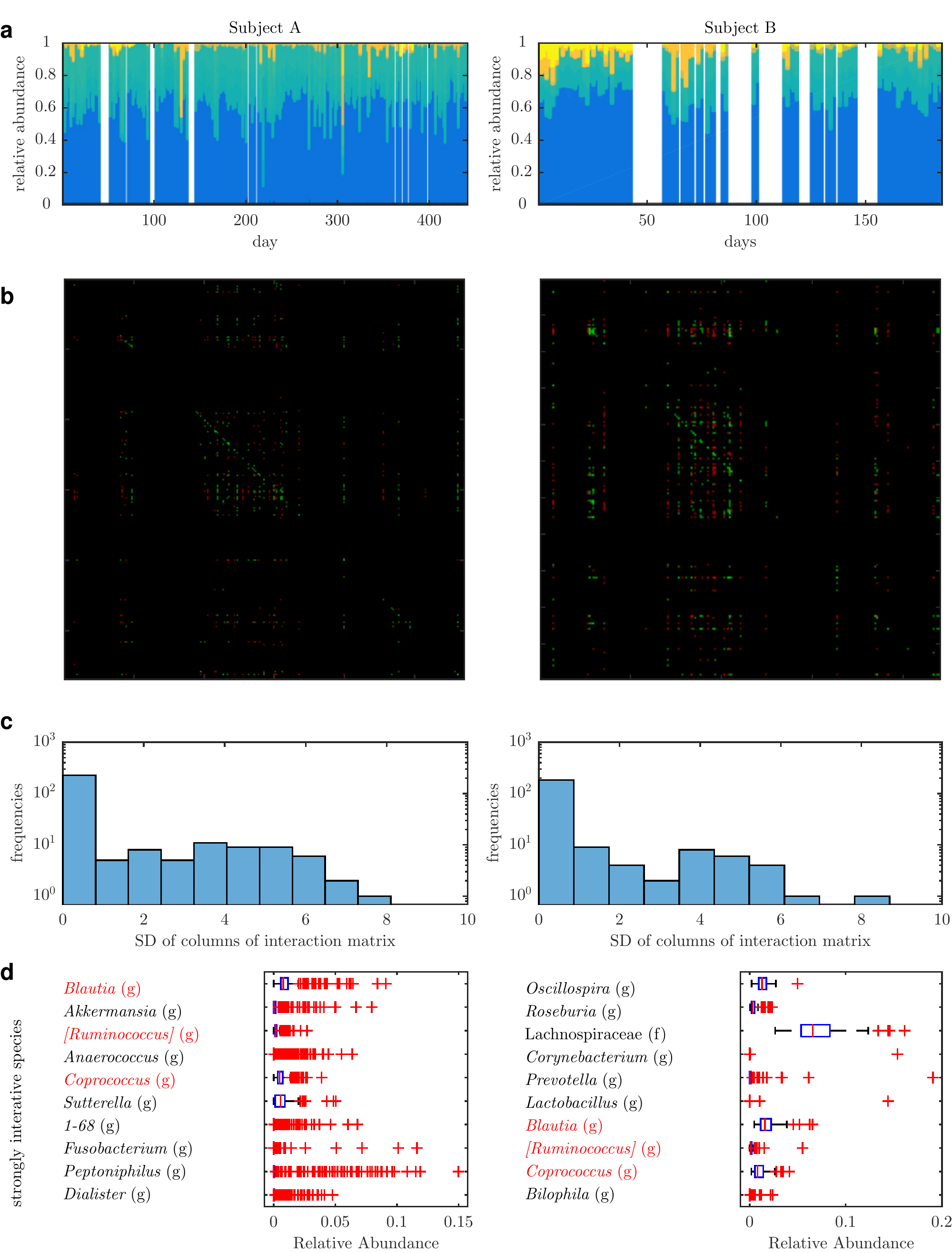}
  \caption{{\em System Identification, Day Swap, Tikhonov Regularization $\lambda=0.0057$.}   System identification was performed on the stool samples from the longitudinal data in \cite{caporaso2011moving}, but with the collection dates permuted, $\lambda$ was determined by cross-validation on the permuted data. (a) Visualization of microbial taxa in terms of relative abundances versus day sample was taken (not permuted samples). (b) Heat map of the interaction matrix for top 100 SISs. (c) Histogram of {\em Standard Deviation} (SD) of the columns of the interaction matrix. (d) List of top ten SISs in descending interaction strength (defined by the SD of each column in the interaction matrix) with relative abundances over all samples shown as a box plot. Even though the sample days have been permuted the banded structure still persists.}\label{fig:ext8}
\end{figure}

\clearpage
\begin{figure}\centering
  \includegraphics[width=4.5 in]{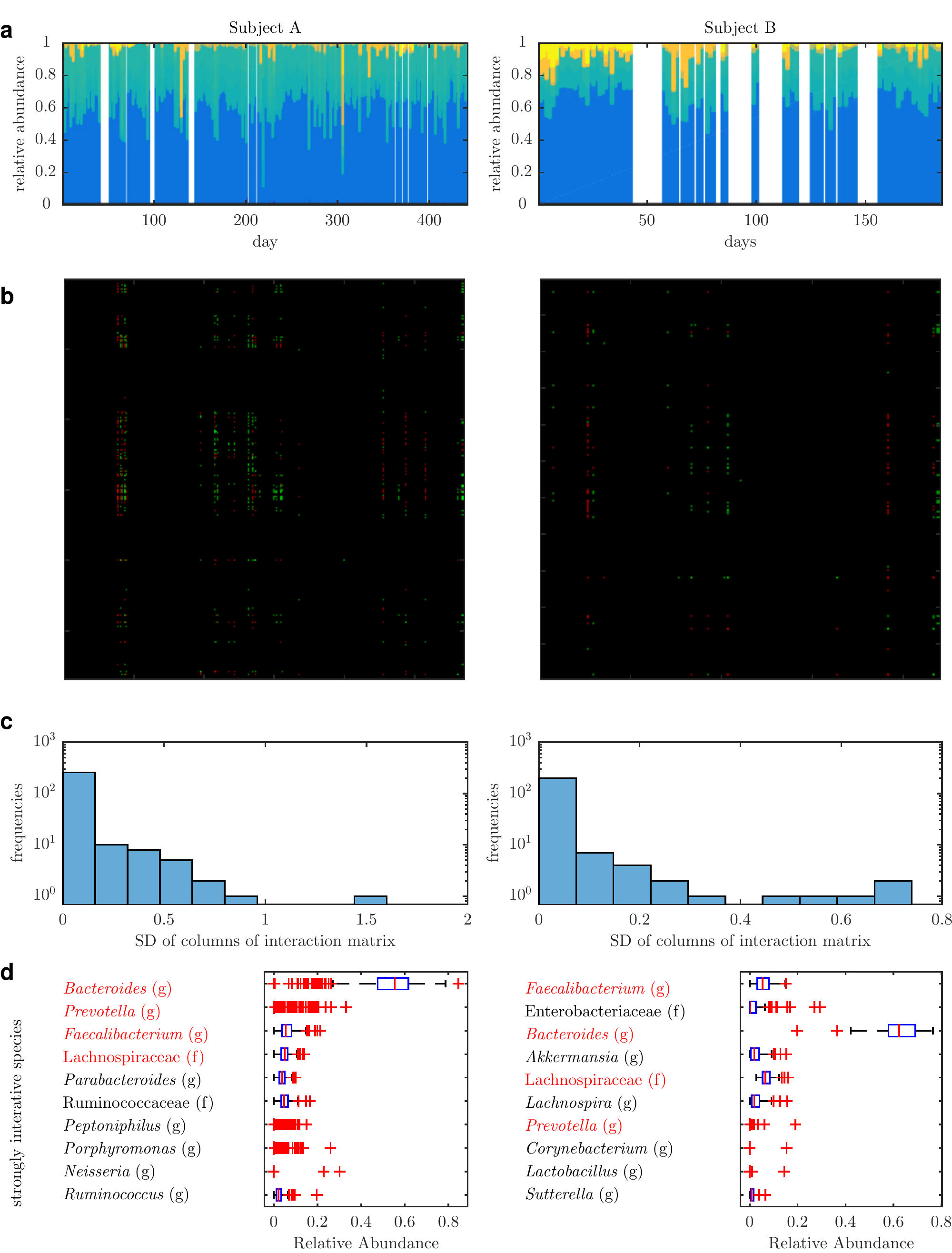}
  \caption{{\em System Identification, Day Swap, Tikhonov Regularization $\lambda=0.0423$.}  System identification was performed on the stool samples from the longitudinal data in \cite{caporaso2011moving}, but with the collection dates permuted, $\lambda$ was selected to be the same as in Supplementary Figure \ref{fig:ext7}. (a) Visualization of microbial taxa in terms of relative abundances versus day sample was taken (not permuted samples). (b) Heat map of the interaction matrix for top 100 SISs. (c) Histogram of {\em Standard Deviation} (SD) of the columns of the interaction matrix. (d) List of top ten SISs in descending interaction strength (defined by the SD of each column in the interaction matrix) with relative abundances over all samples shown as a box plot. For the permuted data when $\lambda$ is larger than the optimal value from the cross-validation the identification method biases towards making the most abundant species also the SISs.}\label{fig:ext9}
\end{figure}

\renewcommand{\figurename}{Supplementary Text Figure}

\chapter{Supplementary Text}
      \setcounter{table}{0}
        \renewcommand{\thetable}{T\arabic{table}}%
        \setcounter{figure}{0}
        \renewcommand{\thefigure}{T\arabic{figure}}%
        \setcounter{equation}{0}
        \renewcommand{\theequation}{T\arabic{equation}}%

\input{introduction3}

\input{notation3}

\input{random3}

\input{stability3}

\input{clustering3}

\input{model3}

\input{simulations3}

%

%
%
%

\clearpage

\input{tables3}

\input{figures3}

\clearpage
\addcontentsline{toc}{chapter}{Bibliography}
\bibliography{../../master}

\end{document}

%% file: preamble3.tex
\definecolor{mygreen}{rgb}{0,0.6,0}
\definecolor{mygray}{rgb}{0.5,0.5,0.5}
\definecolor{mymauve}{rgb}{0.58,0,0.82}

\pgfmathdeclarefunction{gauss}{2}{\pgfmathparse{1/(#2*sqrt(2*pi))*exp(-((x-#1)^2)/(2*#2^2))}}
\pgfmathdeclarefunction{power}{2}{\pgfmathparse{(#2-1)/(#1)*(x/#1)^(-#2)}}

 \theoremstyle{plain}
 \newtheorem{thm}{Theorem}
 \newtheorem{lem}{Lemma}
 
 \newtheorem{cor}[thm]{Corollary}
 
 \theoremstyle{definition}
 \newtheorem{defn}{Definition}
 \newtheorem{exmp}{Example}
 \theoremstyle{remark}
 \newtheorem{rem}{Remark}

 \newtheorem{assm}{Assumption}

\newcommand{\bb}[0]{\begin{bmatrix}}
\newcommand{\eb}[0]{\end{bmatrix}}
\newcommand{\be}[0]{\begin{equation}}
\newcommand{\ee}[0]{\end{equation}}
\newcommand{\ben}[0]{\begin{equation*}}
\newcommand{\een}[0]{\end{equation*}}

\renewcommand{\aa}[0]{A^{[\nu]}}
\newcommand{\rr}[0]{r^{[\nu]}}
\newcommand{\xx}[0]{x^{[\nu]}}
\renewcommand{\ss}[0]{S^{[\nu]}}

\newcommand{\norms}[1]{\lVert#1\rVert}

\let\norm=\normB

\newcommand{\abs}[1]{\left\lvert#1\right\rvert}

\renewcommand{\exp}[1]{{\bf e}^{#1}}
\renewcommand{\Re}[0]{\mathbb R}

\newcommand{\Co}[0]{\mathbb C}

\newcommand{\bea}{\begin{eqnarray}}
\newcommand{\eea}{\end{eqnarray}}
\newcommand{\bw}{\begin{widetext}}
\newcommand{\ew}{\end{widetext}}

\newcommand{\sa}{{$\sigma$-algebra}}

\newcommand{\bi}{\begin{itemize}}
\newcommand{\ei}{\end{itemize}}

\DeclareMathOperator*{\argmin}{arg\,min}
\DeclareMathOperator*{\argmax}{arg\,max}

\DeclareMathOperator*{\diag}{diag}

\newcommand{\eqnum}{\refstepcounter{equation}\textup{\tagform@{\theequation}}}

%% file: introduction3.tex
\section{Introduction}

The purpose of this supplement is to give a self-contained, rigorous account of the topics needed to fully understand the main text. The supplement is organized as follows. Section 2 contains some remarks on notation. Section 3 covers the basics of random variables, gives details on how one constructs a well defined random variable from a power-law distribution, and covers some basic results in random matrix theory, Wigner's circle and semi-circle laws. Section 4 contains details on the definition of asymptotic stability, how this relates to the generalized Lotka-Volterra dynamics,  some resent results on diagonal stability are discussed, and new results regarding the stochastic stability of Lotka-Volterra dynamics is discussed. Section 5 gives the details of the multi-dimensional scaling analysis and clustering algorithms employed in this study. Section 6 gives the details of our modeling approach. Section 7 contains simulation results that are intended to complement the figures in the main text and the supplementary figures. 

Much of the foundational material is well known in each of their respective areas. The diagonal stability result for random matrices has never been discussed in the context of Lotka-Volterra dynamics. The discussions regarding stable steady state shift for Lotka-Volterra dynamics are the first of their kind as well.

Those familiar with probability theory need only read \S\ref{sec:spec_random} for a refresher on Wigner random matrices and the rest of \S\ref{sec:rand} can be skipped. Those familiar with stability need only visit \S\ref{sec:stab_add} and \S\ref{sec:diag_random}. That being said, even the seasoned stability theorist may find new things in the stochastic subsection of \S\ref{sec:a:robustness}. Section \ref{sec:clust} can be skipped for those familiar with the commonly used tools in clustering analysis.

%% file: notation3.tex
\section{Notation}
Throughout we denote the real numbers as $\Re=(-\infty,\infty)$, and the complex numbers as $\Co$. A number $z\in \Co$ if $z=x+iy$ where $x,y\in\Re$ and $i\triangleq \sqrt{-1}$. The positive real numbers are denoted as ${\Re_{>0}\triangleq (0,\infty)}$ and the non-negative reals as ${\Re_{\geq 0}\triangleq [0,\infty)}$. The $n$-dimensional reals are denoted as $\Re^n$, $\Re_{>0}^n$ denotes the $n$-dimensional space of positive vectors which we will refer to as the {\em positive orthant}, and $\Re_{\geq0}^n$ as the {\em non-negative orthant}. An $m\times n$ matrix $A$ of values in the reals is denoted as ${A\in\Re^{m\times n}}$. The element of $A$ in the $i$-th row and the $j$-th column will often be denoted with lower case letters as $a_{ij}=[A]_{ij}$. If there is an issue with formatting and it is not clear that both elements $i$ and $j$ are subscripts in the previous notation then it is equivalent to denote $a_{ij}$ as $a_{i,j}$.  The dual notation for an $n\times n$ matrix constructed from elements $a_{ij}$ is denoted as $A=(a_{ij})_{1\leq i,j\leq n}$. 

The superscript ${(\cdot)^{\mathsf T}}$ is used to denote transpose. The components of an $n$ dimensional vector $x$ are defined as follows ${x=[x_1,\ x_2, \ \ldots, \ x_n]^\mathsf{T}}$. We will often make use of the following nonstandard subscript notation, $x_{j:k}$ to denote the vector obtained from taking the $j$-th element to the $k$-th element of $x$.\footnote{This notation was motivated by the index notation used in Matlab.} As an example, consider ${y=[1,\ 3, \ 5]^\mathsf{T}}$, then $y_{2:3}=[3, \ 5]^\mathsf{T}$. The Euclidean norm of a vector $x\in\Re^n$ is defined as ${\norm{x}\triangleq \left( \sum_{i=1}^n x_i^2\right)^{1/2}}$. When applied to a matrix $A\in \Re^{m\times n}$ the norm is an induced norm $\norm{A}\triangleq \sup_{\norm x=1} \norm{Ax} $. For a square symmetric matrix ${P=P^{\mathsf{T}}\in\Re^{n\times n}}$ the inequality $<\, (\leq)$ is used as $P<0\,(P\leq 0)$ if and only if ${x^{\mathsf{T}}Px<0}\, {(x^{\mathsf{T}}Px\leq 0)}$ for all ${x\in\Re^n}$.

The notation for union and intersection of sets is $\cup$ and $\cap$ respectively. For a set $A\subset U$ the complement of $A$ is defined as ${A^{\mathsf c}= \{y \in U\ | y\notin A \}}$. The set minus notation is defined as ${A\setminus B = A\cap B^{\mathsf c}}$. Let $A$ be a set, then $\abs{A}$ denotes the  cardinality of that set. If for instance $A=\{1,2,4\}$, then $\abs{A}=3$.

Given that $x$ will primarily be defined as the state variable, we will use $Y$ when discussing a generic random variable. The probability distribution for $Y$ is denoted as $\mu_Y$ for which we will generically denote its probability density function as $f(y)$ where $\mu_Y = \int{f(y)}\, dy$. When given a probability density function $f(y)$, the notation ${Y\sim f(y)}$ reads as the random variable $Y$ is drawn from the probability density function $f$. Also, if we generically denote the standard normal distribution as $\mathcal N(0,1)$, then with a slight abuse of notation we can write ${Y\sim \mathcal N(0,1)}$ to denote that $Y$ is drawn from the standard normal distribution. The notation $X\equiv Y$ is used to denote when two random variables are drawn from the same distribution.

We will also make use of the following asymptotic notation. $f(n)=O(g(n))$ if there exists a $C$ independent of $n$ such that $\norm{f(n)} \leq C \norm{g(n)}$ for $n$ sufficiently large. Another form of asymptotic notation that will be borrowed is the following, $f(n)=o(g(n))$ if there exists a $c(n)\geq 0$ where $\lim_{n\to\infty} c(n)=0$ and $\norm{f(n)}\leq c(n) \norm{g(n)}$. 

Results from dynamics and control, to probability and random matrix theory will be called upon in this work. If the notation of a particular section seems to overlap, it is assumed that the convention from the field of origin overrides. For instance, when discussing dynamics a capital letter will denote a matrix, however capital letters are used in probability theory when denoting generic random variables.

%% file: random3.tex
\section{Random Variables and Random Matrices}\label{sec:rand}

\subsection{Primer on Random Variables}

We will use the following three distributions in constructing the random interaction matrices for the microbial communities: the uniform distribution taking values in the interval $[0,1]$, the normal (Gaussian) distribution of mean $0$ and variance $\sigma^2$ and a power-law distribution with minimum value $1$ and exponent $-\alpha$. One can define a random variable just with a variable space and ignore the underlying sample space \cite{terence2012topics}*{\S 1.1.2}. We however introduce the full probabilistic machinery into the discussion. Formality in this section is two fold. First, to give the reader confidence that the random variables generated from the power-law distribution are indeed well defined random variables \cite{doi:10.1137/070710111}. Second, the Wigner circle law and semi-circle law can not be defined without this machinery, and these two laws are fundamental to understanding stability results that are presented in the next section. The following is a nearly verbatim presentation of probability spaces following \cite{terence2011introduction,terence2012topics}.

Let $\Omega$ be a sample space. When given a measure and a $\sigma$-algebra, $\Omega$ becomes a probability space  $\Omega=(\Omega, \mathcal F, \mathbf P)$, where $\mathcal F$ is a $\sigma$-algebra of subsets of $\Omega$ and $\mathbf P$ is a probability measure. The probability measure along with the sample space satisfy the following equality $\mathbf P(\Omega)=1$. Events $E$ are then taken from the $\sigma$-algebra and will have a probability of occurring, i.e. $E\in\mathcal F$ and $E\mapsto \mathbf P(E)$ where ${\mathbf P(E)\in [0,1]}$. A random variable $Y$  takes values in a measurable space ${R=(R,\mathcal R)}$ where $\mathcal R$ is a \sa{} of subsets of $R$. We will also refer to  $R$ as the variable space. The formal definition of a random variable is a map $Y:\Omega \to R$ where $Y$ is measurable. When one asks for the probability that event $Y$ is in $S\in \mathcal R$, we are interested in the probability of event $E=Y^{-1}(S)$ occurring. From the definition of our probability space $\Omega$, this is simply $\mathbf P(Y^{-1}(S))$. Note that this is equivalent to ${\mathbf P(\{\omega\in\Omega:Y(\omega)\in S\})}$ which we can unambiguously denote in shorthand as ${\mathbf P(Y\in S)}$ \cite{terence2012topics}*{\S 1.1.2}. This notation makes no reference to the original sample space and thus can be used without ambiguity when discussing the probability of an event occurring. All of our probability spaces and variable spaces will be subsets of $\Re$ or $\Co$. Thus we will implicitly use the Borel $\sigma$-algebra. Consequently, the specific $\sigma$-algebra of interest will no longer be denoted.

We now give rigorous definitions for the probability measure $\mu_Y$ of a random variable $Y\in R=(R,\mathcal R)$ and the corresponding probability density function $f$ associated with $\mu_Y$. The distribution of $Y$ is defined as 
\ben
\mu_Y (S) \triangleq \mathbf P(Y\in S).
\een 
Given the above definition we define the probability density function as 
\ben
\mu_Y(S) = \int_S f(y) \, dy.
\een
The cumulative distribution function is defined as
\ben
F_Y(y)\triangleq \mathbf P(Y\leq y ) = \mu_Y((-\infty,x]).
\een
The expected value of a random variable $Y\sim f(y)$ (taking values from the probability density function $f$) is defined as 
\be\label{eq:defE}
\begin{split}
\mathbf E Y& \triangleq  \int_R y\, d_Y(y) \\ & = \int_R y f(y)\, dy \\ & = \int_R \mathbf P (Y\geq \lambda)\, d\lambda.
\end{split}
\ee
Sometimes parentheses are used, i.e. $\mathbf E(Y)$, for clarity. The variance of a distribution is defined as 
\be\label{eq:var}
\mathbf{Var} (Y) \triangleq \mathbf E \abs{Y- \mathbf E(Y)}^2.
\ee

{
\begin{defn}[[Almost Surely]  An event $E$ occurs {\em almost surely} if $\mathbf P(E)=1$.
\end{defn}}

\subsubsection{Distributions of interest}\label{sec:dibin}
The uniform distribution generates random variables on the measure space $R=[0,1]$ with the following
 probability density function
\ben
u(y) = \begin{cases} 1 \text{ if } y\in[0,1] \\
0 \text{ otherwise }. \end{cases}
\een
A pseudorandom variable in $[0,1]$ can be generated uniformly using the MATLAB command \verb:random('unif',0,1):. We will use the shorthand notation $\mathcal U(0,1)$ to denote a uniform distribution taking values in $[0,1]$.

The normal distribution with mean $0$ and standard deviation $\sigma$ generates random variables on the measure space $\Re$ and satisfies the well known probability density function
\ben
n(y)= \frac{1}{\sigma \sqrt{2\pi} } e^{ -\frac{y^2}{2\sigma^2} }.
\een 
A pseudorandom variable with mean $0$ and variance $\sigma^2$ can be generated using the MATLAB command \verb:random('norm',0,sigma):.  We will use the shorthand notation $\mathcal N(0,\sigma^2)$ to denote a normal distribution with mean $0$ and standard deviation $\sigma$. A log-normal distributed random variable is simply one in which $Y= \exp{X}$ where $X\sim \mathcal N(0,\sigma^2)$. 

The generic power-law used in this work generates random variables on the measure space $[1,\infty)$ with the following probability density function
\be\label{eq:pldist}
p(y) = (\alpha-1) y^{-\alpha}
\ee
where $\alpha>1$ \cite{doi:10.1137/070710111}*{Equation (2.2)}. We will use the following short hand notation $\mathcal P(\alpha)$ to denote a power-law distribution with exponent $-\alpha$. 

Later we will generate power-law distributions from uniform distributions, and thus we have the following simple result. Let $U$ be a uniform random variable from the set $[0,1)$, then we can generate a random variable $P$ with a power-law distribution in $[1,\infty)$ using the following measurable monotonic function $r:[0,1)\to[1,\infty)$,
\be\label{eq:genP}
r(U) = ( {1-U} )^{\frac{1}{1-\alpha}}.
\ee
Now, letting $P\triangleq r(U)$ we can see that $P$ indeed satisfies all of the requirements to be a well defined random variable. $P$ is a measurable function defined from a probability space (the variable space of $U$ with a probability measure $\mathbf P$) to an event space by a measurable function $r$. This mapping for the generation of a random variable $P$ with powerlaw distribution is illustrated with the following commutative diagram.
\begin{displaymath}
\xymatrix{& [0,1) \ar[rd]^r &  \\  \Omega \ar[ru]^U \ar[rr]^P & & [1,\infty)}
\end{displaymath}
Random variables satisfying a power-law probability density function can be generated in MATLAB using the following line \verb*:(1-random('unif',0,1))^(1/(1-alpha)):.

The nice  trick in \eqref{eq:genP} is introduced in the literature without proof, and is simply denoted as following from the fundamental transformation law of probabilities \cite{press1987numerical}*{\S 7.3}. Given the rigorous introduction of probability theory at the beginning of this section, we can however directly derive this result by analyzing the probability that $P>y$ under the assumption that there exists a  monotonically increasing bijection $r:[0,1)\to[1,\infty)$, which yields
\be\label{eq:}
\begin{split}
\int_y^\infty p(x) dx & = \mathbf{P}(P>y) \\ 
                                       & = \mathbf{P}(r(U)>y) \\
                                       & = \mathbf{P}(U>r^{-1}({y})) \\
                                       & = \int_{r^{-1}({y})}^\infty u(x) dx. 
\end{split}
\ee
The transition from line 2 to line 3 in the above equality follows from the fact that $r$ is monotonically increasing and thus when the inverse is taken the sign of the inequality is preserved. Integrating the above equality it follows that 
\ben
y^{1-\alpha}= 1-r^{-1} (y). 
\een
Substituting $z=r^{-1}(y)$ and solving for $y$ in the above equality it follows that
\ben
y=(1-z)^ \frac{1}{1-\alpha}
\een
which proves the relation in \eqref{eq:genP}.

\begin{figure}
\centering
\begin{tikzpicture}[scale=0.7]
\begin{axis}[width=0.45\textwidth, 
  axis lines*=left, 
    title=Uniform Distribution,
    xmin=-1.2,
   xmax=2.2,
    ylabel near ticks,
  xlabel={$y$},
  ylabel={Probability Density},
  ] 
\addplot [mark=*,color=black,fill=white,]coordinates {(0,0)};
\addplot [mark=none,color=black,very thick]
coordinates {(-1.2,0)
(0,0)};
\addplot [mark=none,color=black,very thick]
coordinates {(0,1)
(1,1)};
\addplot [mark=*,color=black,fill=black]coordinates {(0,1)
(1,1)};
\addplot [mark=*,color=black,fill=white,]coordinates {(1,0)};
\addplot [mark=none,color=black,very thick]
coordinates {(1,0)
(2.2,0)};
\end{axis}
\end{tikzpicture}
\quad  
\begin{tikzpicture}[scale=0.7]
\begin{axis}[ 
  axis lines*=left,
  width=0.45\textwidth,
  ylabel near ticks,
  title=Normal Distribution,
  xmin=-2.2,
  xmax=2.2,
  xlabel={$y$},
  ylabel={Probability Density}]
  \addplot[very thick,smooth,samples=50] {gauss(0,0.5)};
\end{axis}
\end{tikzpicture}
\quad  
\begin{tikzpicture}[scale=0.7]
\begin{axis}[every axis plot post/.append style={samples=50,smooth},
    width=0.45\textwidth, 
    axis lines*=left,
    title={Power-law Distribution},
    ylabel near ticks,
      xmin=0.8,
  xmax=4.2,
   xlabel={$y$},
   ylabel={Probability Density}] 
  \addplot [mark=*,color=black,fill=black]coordinates {(1,2)};
    \addplot [mark=*,color=black,fill=white]coordinates {(1,0)};
\addplot [mark=none,color=black,very thick]
coordinates {(0,0)
(1,0)};
  \addplot [very thick,domain=1:5]{power(1,3)};
\end{axis}
\end{tikzpicture}
\caption[Comparison of uniform, normal and power-law distributions.]{Illustration of the distributions for the three types of probability functions used in this work, from left to right: uniform, normal, and power-law. }
\label{fig:distributions}
\end{figure}
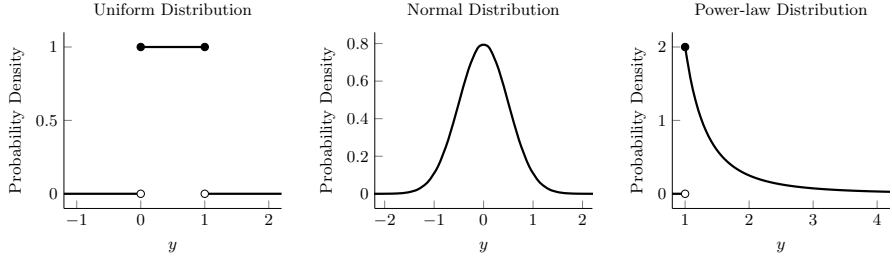

\begin{figure}
\centering
\begin{tikzpicture}[scale=0.7]
\begin{axis}[every axis plot post/.append style={samples=50,smooth},
    width=0.45\textwidth, 
    axis lines*=left,
    title={Power-law $\alpha=7$},
    ylabel near ticks,
      xmin=0.8,
  xmax=4.2,
   xlabel={$y$},
   ylabel={Probability Density}] 
  \addplot [mark=*,color=black,fill=black]coordinates {(1,6)};
    \addplot [mark=*,color=black,fill=white]coordinates {(1,0)};
\addplot [mark=none,color=black,very thick]
coordinates {(0,0)
(1,0)};
  \addplot [very thick,domain=1:5]{power(1,7)};
\end{axis}
\end{tikzpicture}
\quad
\begin{tikzpicture}[scale=0.7]
\begin{axis}[every axis plot post/.append style={samples=50,smooth},
    width=0.45\textwidth, 
    axis lines*=left,
    title={Power-law $\alpha=2$},
    ylabel near ticks,
      xmin=0.8,
  xmax=4.2,
   xlabel={$y$},
   ylabel={Probability Density}] 
  \addplot [mark=*,color=black,fill=black]coordinates {(1,1)};
    \addplot [mark=*,color=black,fill=white]coordinates {(1,0)};
\addplot [mark=none,color=black,very thick]
coordinates {(0,0)
(1,0)};
  \addplot [very thick,domain=1:5]{power(1,2)};
\end{axis}
\end{tikzpicture}
\quad
\begin{tikzpicture}[scale=0.7]
\begin{axis}[every axis plot post/.append style={samples=50,smooth},
    width=0.45\textwidth, 
    axis lines*=left,
    title={Power-law $\alpha=1.2$},
    ylabel near ticks,
      xmin=0.8,
  xmax=4.2,
   xlabel={$y$},
   ylabel={Probability Density}] 
  \addplot [mark=*,color=black,fill=black]coordinates {(1,.2)};
    \addplot [mark=*,color=black,fill=white]coordinates {(1,0)};
\addplot [mark=none,color=black,very thick]
coordinates {(0,0)
(1,0)};
  \addplot [very thick,domain=1:5]{power(1,1.2)};
\end{axis}
\end{tikzpicture}

\caption[Power-law distributions as a function of the exponent.]{Illustration of the power-law distribution from low heterogeneity to high heterogeneity. }
\label{fig:distributions}
\end{figure}
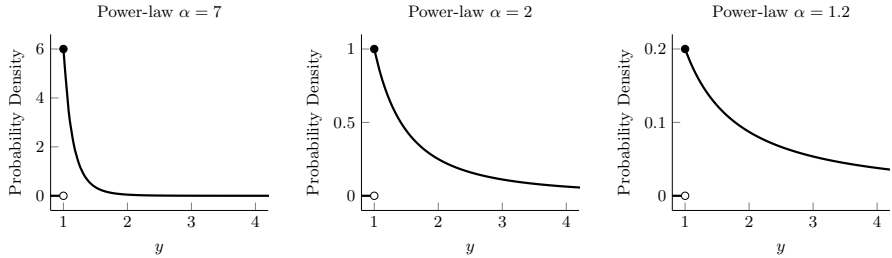

\subsection{Random Network Models}
Formally a digraph $\mathcal G$ is defined by the double $(\mathcal V,\mathcal E)$ where $\mathcal V=\{1, 2, \ldots, n\}$ is the vertex set  and  the directed edges are defined by the ordered pairs ${(i, j)\in \mathcal E \subset{\mathcal V}\times{ \mathcal V}}$. An element $(i,j)\in  \mathcal E$ if and only if there is a directed edge from vertex $i$ to vertex $j$. We are primarily interested in the adjacency matrix $\mathcal A$ of the digraph which is defined as
\ben
[\mathcal A]_{ij}=\begin{cases}  1 & \text{if }  (j,i) \in \mathcal E \\
0 & \text{otherwise}\end{cases}.
\een
When  $[\mathcal A]_{ij}=1$ for all $i$ and $j$ the digraph is said to be {\em complete}. Note that we allow for self-loops in our construction.  Two other digraph topologies to be discussed shortly are the Erd\H{o}s-R\'enyi (Gilbert) random digraph and the power-law degree digraph.
\subsubsection{Erd\H{o}s-R\'enyi (Gilbert) Digraph}\label{sec:ER}

An Erd\H{o}s-R\'enyi (Gilbert) digraph is a digraph $\mathcal G(n,p)$ of $n$ nodes, where the probability of a directed edge from node $i$ to node $j$ is $p$ for any $i,j\in \mathcal V$. The adjacency matrix for this model is constructed as follows. Let $G\in[0,1]^{n\times n}$ be a matrix with elements independently sampled from the uniform distribution between 0 and 1, $[G]_{ij}\sim \mathcal U(0,1)$. Then let $\mathcal A$ be defined as follows 
\ben
[\mathcal A]_{ij}=\begin{cases}  1 & \text{if }  [G]_{ij}<p \\
0 & \text{otherwise}\end{cases}.
\een
If one is interested in defining an Erd\H{o}s-R\'enyi model for a 100 node digraph with an expected mean in-degree (or out-degree) of 10, then simply set $p=10/100$ in this construction.

While the above model is often credited to Erd\H{o}s and R\'enyi  \cite{Erdos:1959aa,erdos}, it was actually first presented by Gilbert in \cite{gilbert1959}. We will follow convention however and simply refer to this random network model as the {\em Erd\H{o}s-R\'enyi} (ER) model. The adjacency matrix for this model can be generated in MATLAB using the following boolean expression \verb*:rand(n,n)<p:.

\subsubsection{Power-law Out-degree Digraph}\label{sec:powerdigraph}
In this section we outline how one can generate the adjacency matrix for a power-law out-degree digraph. Let ${h=[h_1,\ h_2, \ \ldots, \ h_n]^\mathsf{T}}$ be the column vector of out-degrees for nodes $\{1,\ 2, \ \ldots, \ n\}$. If one is interested in having a digraph with a power-law out-degree of exponent $-\alpha$ with the mean out-degree approximately $d$, then setting 
\be h_i = \min \left\{\left \lceil{d \frac{[\bar h]_i}{\mathrm{mean}(\bar h)}}\right\rceil, n \right\}  \ee 
is sufficient, where $[\bar h]_i\sim \mathcal P(\alpha)$, $i=1,2,\ldots,n$. Note that $\lceil \cdot \rceil$ is the ceiling operator. We also note that this will not guarantee that the mean out-degree is $d$ for any finite sized digraph. Finally the adjacency matrix is constructed by selecting $h_i$ random elements in column $i$ of $\mathcal A$ and setting them to 1.

This method of generating power-law degree distributions leaves much to be desired. It is not based on any known theory for power-law degree graphs \cite{bol_handbook_03,li_imath_04}. So as to be able to compare Erd\H{o}s-R\'enyi digraphs with the above power-law digraphs we have normalized by degree, which is not a common practice in the literature either. We note also that few authors have rigorously analyzed power-law digraphs with the exception of   \cite{bol_handbook_03}*{\S 11} and \cite{Bollobas:2003:DSG:644108.644133}. 

\subsection{Spectrum of Random Matrices}\label{sec:spec_random}

\begin{figure}
\centering
  \includegraphics[width=4in]{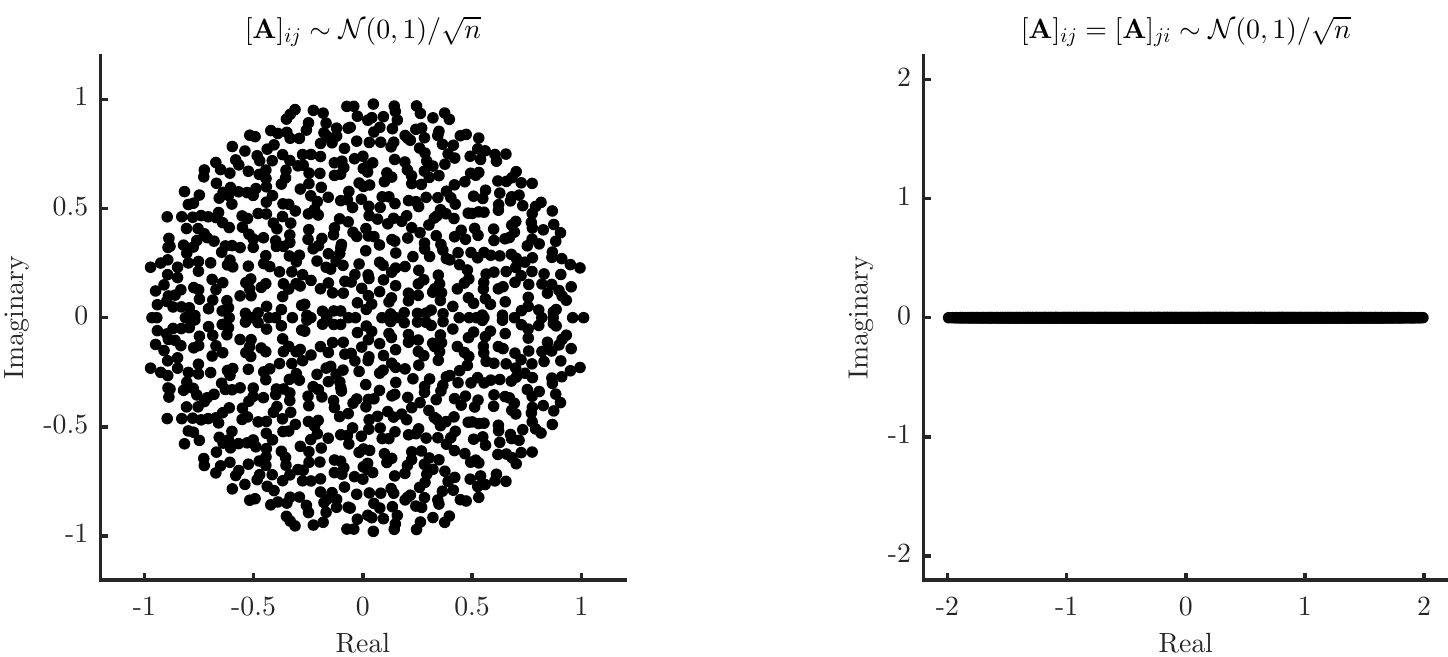}
  \caption[Spectrum of random matrices.]{Eigenvalues for (left) random matrix and (right) symmetric random matrix of dimension $n\times n$ with $n=1000$ where elements are drawn from the distribution $\mathcal N(0,1)/{\sqrt{n}}$.} 
  \label{fig:rand_matrix_example}
\end{figure}

Two classic results from Wigner \cite{Wigner:1955aa,Wigner:1957aa,Wigner:1958aa} are now discussed. 
First we define the {\em Empirical Spectral Distribution} (ESD) of an $n\times n$ real valued matrix $A$ as $\mu_A:\mathbb C \to \mathbb N$
\ben
\mu_A (z) \triangleq \frac{1}{n}\abs{\{1\leq i \leq n: \mathrm{Re}\,{\lambda_i(A)}\leq \mathrm{Re}\, z ,\; \mathrm{Im}\, {\lambda_i(A)\leq \mathrm{Im}\, z}\}}.
\een
Recall that  when applied to a finite set $\abs{\cdot}$ denotes the cardinality of that set. The ESD simply counts the number of eigenvalues of $A$ within radius $\abs{z}$ of the origin.
A weaker version of \cite{tao2010}*{Theorem 1.10} is now stated.
\begin{thm}\label{thm:circ}
Let $B_n$ be a real  $n\times n$ matrix whose elements are independently and identically distributed random variables with mean 0 and variance 1. Then it follows that $\mu_{B_n/\sqrt{n}}$ converges almost surely to the uniform disk in the complex plane, $\mathbf 1_{\abs z\leq 1}$, with probability 1. Let $R$ denote the variable space for the $i.i.d.$ random variables, then 
$$
\mathbf P\left(\limsup_{n\to\infty} \abs{\int_\Co h(z)\, d\mu_{B_n/\sqrt{n}}(z) -\int_\Co h(z) \mathbf 1_{\abs{z}\leq 1} \, dz} \leq \epsilon\right)=1
$$
for all $\epsilon>0$ and every bounded continuous function $h: R \to \Co$.
\end{thm}
%
\noindent The semi-circle distribution is defined as $\mu_{\mathrm{sc}}(S)\triangleq \int_S \rho_{\mathrm{sc}}(y) dy $ where the probability density function is defined as 
\ben
\rho_{sc}(y) = \begin{cases}  \frac{1}{2\pi} ({4-y^2 })^{1/2}_+, & \abs{y}\leq 2 \\ 0, & \abs{y}> 2\end{cases}
\een
We now state a more conservative version of \cite{2012arXiv1202.0068T}*{Theorem 5}.
\begin{thm}\label{thm:semicirc}
Let $M_n$ be a symmetric $n\times n$ matrix whose diagonal and upper right elements are independently and identically distributed random variables with mean 0 and variance 1. Then it follows that $\mu_{M_n/\sqrt{n}}$ converges in the sense of probability to $\mu_{\mathrm{sc}}$. Let $R$ denote the variable space for the $i.i.d.$ random variables, then 
$$
\liminf_{n\to\infty} \mathbf P\left( \abs{\int_{-2}^x h(y)\, d\mu_{B_n/\sqrt{n}}(y) -\int_{-2}^x h(y) d\mu_{\mathrm{sc}}(y)} \leq \epsilon\right)=1
$$
for all $\epsilon>0$ and every bounded continuous function $h: R \to \Re$.
\end{thm}
A conservative version of \cite{bai1988necessary}*{Theorem A} is now stated
\begin{thm}\label{thm:spec_bound}
Let $M_n$ be a symmetric $n\times n$ matrix whose upper right  elements are independently and identically distributed random variables with mean 0 and variance $\sigma^2$, and whose diagonal elements are  independently and identically distributed with mean 0 and finite variance. Then it follows that $\sup_{1\leq i\leq n} \abs{\lambda_i(M_n)}=2\sigma\sqrt{n}(1+o(1))$ asymptotically almost surely. Stated another way, 
$$
\liminf_{n\to\infty} \mathbf P\left(\sup_{1\leq i\leq n} \abs{\lambda_i(M_n)}=2\sigma\sqrt n (1+o(1))\right)=1.
$$
\end{thm}
Theorem \ref{thm:circ} states that the spectral distribution of a random matrix with elements drawn from a normal distribution with mean 0 and variance $1/n$  converges to the unit disk centered at the origin of the complex plain. For the same random matrix but with the added assumption of symmetry, the spectrum converges to the line segment $[-2,2]$ on the real line, see Figure \ref{fig:rand_matrix_example}. The spreading of the spectrum from a diameter of 2 to a diameter of 4 can be explained by the fact that all 2-cycles in the matrix now have a positive loop game. Positive feedback for 2-cycles always repels the eigenvalues along the real axis \cite{5060121}. 

%
%


%% file: stability3.tex
\section{Stability}
\label{sec:stab}

\subsection{Primer on Stability}

Consider the time-varying dynamical system defined by
\be\label{math:eq:sys}\begin{split}
\dot {x}(t) &  = f( x(t),t)\\
x(t_0)&= x_0 
\end{split}\ee
where ${x:\Re\to\Re^n}$ is the state vector, $t$ is time, $t_0$ is the initial time, and ${\dot{(\ )}\triangleq \frac{\mathrm d}{\mathrm d t}(\ )}$.
Let ${x^*\in\Re^n}$ be the equilibrium solution so that $f(x^*,t)=0$ for all $t$. The solution to the ordinary differential equation in \eqref{math:eq:sys} is a transition function $\phi(t;x_0,t_0)$ such that ${{\phi}(t_0;x_0,t_0)=x_0}$ and ${\dot {\phi}(t;x_0,t_0)   = f( \phi(t;x_0,t_0),t)}$. For existence and uniqueness conditions see \cite{coddington1955theory}.
Below we give the various definitions of stability as defined in \cite{hah67,mas56,kal60,annbook}.

\begin{defn}[Stability] \label{defn:a:stab}Let ${t_0\geq 0}$, the equilibrium $x^*$ is \label{math:def:stable}  \begin{enumerate}
\item[(i)] {\em Stable}, if for all ${\epsilon>0}$ there exists a ${\delta(\epsilon,t_0)>0}$ such that ${\norm{x_0-x^*}\leq \delta}$ implies ${\norm{\phi(t_0; t_0, x_0)-x^*} \leq \epsilon}$ for all ${t \geq t_0}$.
\item[(ii)] {\em Attracting}, if there exists a $\rho(t_0)>0$ such that for all $\eta>0$ there exists a  $T(\eta,x_0,t_0)$ such that ${\norm{x_0-x^*}\leq\rho}$ implies ${\norm{\phi(t;x_0,t_0)-x^*} \leq\eta}$ for all $t\geq t_0+T$.
\item[(iii)] {\em Uniformly Stable}, if the $\delta$ in (i) is uniform in $t_0$, thus taking the form $\delta(\epsilon)$.
\item[(iv)] {\em Uniformly Attracting}, if it is attracting where $\rho$ does not depend on $t_0$ and $T(\eta,\rho)$ does not depend on $x_0$ or $t_0$.
\item[(v)] {\em Uniformly Asymptotically Stable} (UAS), if it is uniformly stable and uniformly attracting.
\item[(vi)] {\em Uniformly Bounded}, if for all $r>0$ there exists a $B(r)$ such that $\norm{x_0-x^*}\leq r$ implies that $\norm{s(t; t_0, x_0)-x^*}\leq B$ for all $t\geq t_0$.
\item[(vii)] {\em Uniformly Attracting in the Large}, if for all $\rho>0$ and $\eta>0$ there exists a  $T(\eta,\rho)$ such that $\norm{x_0-x^*}\leq\rho$ implies $\norm{s(t;x_0,t_0)-x^*} \leq\eta$ for all $t\geq t_0+T$.
\item[(viii)] {\em Uniformly Asymptotically Stable in the Large} (UASL), if it is uniformly stable, uniformly bounded, and  uniformly attracting in the large.
\item[(ix)] {\em UAS in the Positive Orthant}, if it is uniformly stable, uniformly bounded, and  uniformly attracting in the positive orthant.

\end{enumerate}
\end{defn}

The precise definition  of UAS is important in the context of dynamical systems. If one is able to show that a given system is UAS, then it follows that the dynamics are stable in the presence of bounded disturbances as well. That is, if the dynamics in \eqref{math:eq:sys} are UAS then for $\norm{d(t)}\leq \gamma$ where $\gamma>0$ is sufficiently small, the dynamics
\ben
\dot x(t) = f(x(t)) + d(t)
\een
are uniformly stable. If the dynamics in \eqref{math:eq:sys} are UASL then $x(t)$ is bounded for all bounded $d(t)$  and $\gamma$ can be arbitrarily large.  A detailed discussion regarding this fact can be found in  \cite[Definition 56.1 and Theorem 56.4]{hah67} and a practical example in the context of adaptive systems can be found in \cite{narendra86}.

For a linear dynamical system $\dot x(t) = A x(t)$ the uniform asymptotic stability (which is actually exponential) is verified if all of the eigenvalues of $A$ have real parts less than zero. A well known theorem regarding the stability of linear systems, due to Lyapunov, is now stated.
\begin{thm}[Lyapunov] \label{thm:lyap} The eigenvalues of a real matrix $A$ have all real parts less than zero if and only if there exists a $P=P^{\mathsf{T}}>0$ such that $A^{\mathsf{T}}P+PA<0$.\end{thm}
\noindent A stronger version of Thereom \ref{thm:lyap} will  be needed when discussing the stability of Lotka-Volterra dynamics. 
\begin{defn}\label{def:d}
If there exists a diagonal positive matrix $P$ such that $A^{\mathsf{T}}P+PA<0$ then $A$ is said to be {\em Diagonally Stable}.
\end{defn}



\subsection{Stability of Generalized Lotka Volterra Dynamics}\label{sec:lstab}

Consider dynamics of the form 
\be\label{s:eq:lvd}
\dot x_i(t) = r_i x_i(t) +  x_i(t)\sum_{j=1}^n a_{ij} x_j(t), \quad i=1,\ldots,n
\ee
where $t\in[t_0,\infty)$ is time  with $t_0$ the initial time. The state vector is denoted $x\in \Re^n$ and defined as  $x=[ x_1,\, x_2,\, \ldots,\,  x_n ]^{\mathsf{T}}$. The linear terms are collected in the column vector $r=[ r_1,\, x_2,\, \ldots,\,  r_n ]^{\mathsf{T}}$ and  $A=(a_{ij})_{1 \geq i,j\geq n}$ captures the pair-wise interactions in the generalized Lotka-Volterra dynamics presented in \eqref{s:eq:lvd}. The dynamics in \eqref{s:eq:lvd} can be compactly represented as 
\be \label{s:eq:lvd2}
\dot x(t) = \diag(x(t)) (r+ A x(t)). 
\ee

A discussion regarding the invertability of $A$ is in order. This will become important in determining whether the system in \eqref{s:eq:lvd} has a unique non-trivial steady state. As discussed in the main text, the existence of the Verhulst terms $a_{ii}x_i^2$ increases the likelihood that $A$ is full rank. First consider the case where all $a_{ij}=0$ for all $i \neq j$, then a necessary and sufficient condition for $A$ to be full rank is that all diagonal elements are non-zero.

If $A$ is invertible then there exists a unique non-trivial steady state solution of the dynamics in \eqref{s:eq:lvd}, denoted as $x^* = -A^{-1}r$ \cite{goh1977global}. We are interested in answering the following question. Given an ecological system of $n$ species, is it possible to introduce another species and drive the system to any non-trivial steady state of our choosing? We will show that the answer is yes, and furthermore, if the original $n$-dimensional ecological system satisfies a diagonal stability condition, then we can design the interaction strengths for the introduced species so that the new $n+1$ dimensional ecological system is asymptotically stable for all initial conditions in the positive orthant. We then discuss equivalent results regarding steady state shift and stability when an arbitrary number of species is added.

Consider the following set of assumptions.
\begin{assm}\label{ass1}
 $A$ is invertible.
\end{assm}
\begin{assm}\label{ass2}
For the dynamics in \eqref{s:eq:lvd} the steady state solution ${x^*\in \Re^n_{>0}}$.
\end{assm}
\begin{assm}\label{ass3}
The matrix $A$ is diagonally stable, see Definition \ref{def:d}.
\end{assm}

\begin{thm}[\cite{goh1977global}*{Theorem 1}]\label{thm1} If the system in \eqref{s:eq:lvd} satisfies Assumptions \ref{ass1}-\ref{ass3}, then the steady state $x^*$ is uniformly asymptotically stable for all initial conditions $x_0\in \Re^{n}_{>0}$.
\end{thm}
\begin{proof}
Let $V(x,t)= 2 \sum_{i=1}^n p_i(x_i-x^*_i - x^*_i \log(x_i/x^*_i))$ be the Lyapunov candidate where $p_i$ is the $i$-th diagonal element of a diagonal positive matrix $P$ such that $A^{\mathsf T}P+PA<0$. Differentiating the Lyapunov candidate it follows that
\ben\begin{split}
\dot V(x,t) &= 2 \sum_{i=1}^n p_i\left(\dot x_i - x^*_i \frac{\dot x_i}{x_i}\right) \\
&=  2\sum_{i=1}^n p_i(x_i - x^*_i) \frac{\dot x_i}{x_i}  \\
&=  2\sum_{i=1}^n p_i(x_i - x^*_i) \left(b_i+\sum_{j=1}^n a_{ij} x_j \right)  \\
&=  2\sum_{i=1}^n p_i(x_i - x^*_i) \sum_{j=1}^n a_{ij} (x_j-x_j^*) \\
&=  (x - x^*)^{\mathsf T} (A^{\mathsf T}P+ PA) (x - x^*).
\end{split} \een
Thus the Lyapunov candidate is positive definite in $x-x^*$ and its derivative is negative definite in $x-x^*$.
\end{proof}

\begin{rem} Note that in the original work of Goh \cite{goh1977global}, stability in the positive orthant is denoted, but in fact what he proved is {\em uniform asymptotic} stability in the positive orthant. It is important to distinguish the two as stability only implies boundedness of trajectories and uniform asymptotic stability implies convergence to the equilibrium as well as robustness to bounded persistent disturbances. 
\end{rem}

\subsection{Stability in the presence of new species}\label{sec:stab_add}

\subsubsection{Adding one new species}\label{sec:add1}
For the following examples we are interested in the  $m=n+1$ dimensional dynamics
\be \label{s:eq:lvd3}
\dot z(t) = \diag(z(t))(g + F z(t) )
\ee
where 
\ben
g = \bb r \\ s \eb,\quad  \text{and} \quad  F = \bb A & b \\ c^{\mathsf{T}} & d\eb,
\een
with $A$ and $r$ are as defined in \eqref{s:eq:lvd}, and we have introduced the new elements $s,d\in \Re$, and  $b,c\in \Re^{n}$. These dynamics represent the addition of one new species to the ecological system in \eqref{s:eq:lvd}.

We now introduce an important property of diagonally stable matrices.

\begin{thm}[\cite{shorten2009theorem}*{Theorem 3.1}]\label{thm2} Let $A\in\Re^{m\times m}$ be partitioned as
\ben
A = \bb A_{11} & A_{12} \\ A_{21} & A_{22} \eb
\een
where $A_{11}\in \Re^{(m-1)\times(m-1)}$, $A_{12},A_{21}^{\mathsf{T}}\in \Re^{m-1}$, and $A_{22}<0$. Then $A$ is diagonally stable if and only if $A_{11}$ and the quantity $(A_{11}-A_{12}A_{21}/A_{22})$ have a common diagonal Lyapunov function, i.e., there exists a positive diagonal $P$ such that $A_{11}^{\mathsf{T}}P + PA_{11}<0$ and $(A_{11}-A_{12}A_{21}/A_{22})^{\mathsf{T}}P+P(A_{11}-A_{12}A_{21}/A_{22})<0$.
\end{thm}

\begin{lem}\label{lem1} For any steady state solution $x^*$ of  \eqref{s:eq:lvd} satisfying ${-A x^*=r}$ there exists $b,c,d,s$ such that any $z^*\in \Re^m_{>0}$ can be made to be a steady state solution of \eqref{s:eq:lvd3}.
\end{lem}
\begin{proof}
Any steady solution  of \eqref{s:eq:lvd3} satisfies the relation $g=-Fz^*$, which when expanded denotes the following relation
\ben
\bb
r \\ s 
\eb =-\bb A & b \\  c^{\mathsf{T}}&  d \eb  \bb z^*_{1:n} \\z^*_m \eb.
\een
There are $2n+2$ degrees of freedom in the variables $b,c,d,s$ and there are only $n+1$ constraints. The variable $b$ is fixed by the top row of the above equation and can be expressed in closed form as 
\be\label{eq:l11}
b=-\frac{r+Az^*_{1:n}}{z^*_m}.
\ee
Then for any $c$ and $d$, $s$ can be chosen as 
\be\label{eq:l12}
s = -c^{\mathsf{T}}z^*_{1:n} - d z_m^*. \qedhere
\ee
\end{proof}

\begin{cor}
For any steady state solution $x^*$ of  \eqref{s:eq:lvd} satisfying $-A x^*=r$, and given any $c\in\Re^n$ and $d\in\Re$ there exists $b,s$ such that any ${z^*\in \Re^m_{>0}}$ can be made to be a steady state solution of \eqref{s:eq:lvd3}.
\end{cor}

\begin{thm}\label{thm3}For the dynamics in \eqref{s:eq:lvd} with $p=1$ satisfying Assumption \ref{ass3} there exists $b,c,d,s$ such that any ${z^*\in \Re^m_{>0}}$ can be made to be asymptotically stable for all initial conditions in the positive orthant.\end{thm}
\begin{proof}
From Lemma \ref{lem1} it follows that for any $z^*,c$ and $d$ the elements $s,b$ are fixed. Now we will show that for any $c$ there exists a $d$ such that $z^*$ is asymptotically stable. We begin by assuming that $b$ and $s$ are fixed by \eqref{eq:l11} and \eqref{eq:l12}. From Assumption \ref{ass3} we know that $A$ is diagonally stable, and thus there exists an $\epsilon>0$ and diagonal matrix $P>0$ such that  $A^{\mathsf{T}}P+PA\leq-\epsilon I$, where $I$ is an appropriately dimensioned identity matrix. Choosing $d<0$ and $\abs{d} > {2\lambda_\textrm{max}(P)\norms{bc^{\mathsf{T}}}}/\epsilon$, and noting that
\ben
(A-bc^{\mathsf{T}}/d)^{\mathsf{T}}P + P(A-bc^{\mathsf{T}}/d) \leq -\epsilon I + 2\lambda_\textrm{max}(P)\norms{bc^{\mathsf{T}}}/d
\een
we can deduce that $(A-bc^{\mathsf{T}}/d)^{\mathsf{T}}P + P(A-bc^{\mathsf{T}}/d) <0$. Thus $(A-bc^{\mathsf{T}}/d)$ and $A$ have a common diagonal Lyapunov function. This, in addition with the fact that $d<0$, the conditions of Theorem \ref{thm2} are satisfied and thus $F$ is diagonally stable. Now, applying Theorem \ref{thm1} to the dynamics in \eqref{s:eq:lvd3} we deduce that $z(t)$ can be made to be asymptotically stable for all $z(t_0)\in \Re^m_{>0}$.
\end{proof}

\subsubsection{Adding an arbitrary number of species}\label{sec:addm}

For the following examples we are interested in the  $m=n+p$ dimensional dynamics
\be \label{s:eq:lvd4}
\dot z(t) = \diag(z(t))(g + F z(t) )
\ee
where 
\ben
g = \bb r \\ s \eb,\quad  \text{and} \quad  F = \bb A & B \\ C^{\mathsf{T}} & D\eb,
\een
with $A$ and $r$ are as defined in \eqref{s:eq:lvd}, and we have introduced the new elements $s\in \Re^{p}$, $D\in\Re^{p\times p}$ and  $B,C\in \Re^{n\times p}$. These dynamics represent the addition of $p$ new species to the ecological system in \eqref{s:eq:lvd}. 


\begin{thm}\label{thm9} For any steady state solution $x^*$ of  \eqref{s:eq:lvd} satisfying $-A x^*=r$ there exists $B,C,D,s$ such that any $z^*\in \Re^m_{>0}$ can be made to be a steady state solution of \eqref{s:eq:lvd4}. Furthermore, if $A$ is diagonally stable, then $B,C,D,s$ can be chosen such that the system in \eqref{s:eq:lvd4}  is uniformly asymptotically stable in the positive orthant. 
\end{thm}
\begin{proof}
Any steady solution of  of \eqref{s:eq:lvd4} satisfies the relation $g=-Fz^*$ which when expanded denotes the following relation
\ben
\bb
r \\ s 
\eb =-\bb A & B \\  C^\mathsf{T}&  D \eb  \bb z^*_{1:n} \\z^*_{(n+1):m} \eb.
\een
There are $2np+p^2+p$ degrees of freedom in the variables $B,C,D,s$. The variable $B$ is fixed by the top row of the above equation and must satisfy the following relation 
\ben
B z^*_{(n+1):m} =-(r+Az^*_{1:n}).
\een
There are $n\times p$ degrees of freedom in the selection of $B$ and only $n$ constraints in the above equation. Thus such a $B$ always exists for any $r$, $A$, and $z^*$.
For any $C$ and $D$, $s$ can be chosen as 
\ben
s = -C^{\mathsf{T}}z^*_{1:n} - D z^*_{(n+1):m}. 
\een
Finally, we show that with the extra degrees of freedom in $C$ and $D$ there always exists a diagonal $P_1>0$ and $P_2>0$ such 
\be\label{s:eq:AP}
\bb A & B \\  C^\mathsf{T}&  D \eb^\mathsf{T}  \bb P_1 & 0 \\  0 &  P_2 \eb + \bb P_1 & 0 \\  0 &  P_2 \eb\bb A & B \\  C^\mathsf{T}&  D \eb <0.
\ee
Given that by assumption there exists a diagonal $P_1>0$ such $\tilde A \triangleq A^{\mathsf T}P_1+P_1 A$, then by the Schur complement the inequality in \eqref{s:eq:AP} holds if an only if \ben D^{\mathsf{T}} P_2 + 2 P_2 D - (B^{\mathsf{T}} P_1 + P_2 C^{\mathsf{T}}) \tilde A^{-1} (C P_2 + P_1 B)<0.\een
Given any $A, B, C$ and positive diagonal $P_1,P_2$ there always exists a $D$ such that the above inequality holds.
\end{proof}


\subsubsection{Removing an arbitrary number of species.}\label{sec:remove}
Diagonally stable matrices have a very special property that every principle minor is also diagonally stable, giving us the following definition and theorem.
\begin{defn}
Let $L$ be a proper subset of $N\triangleq \{1,2,\ldots,n\}$ then a {\em principle minor} of $A\in\Re^{n\times n}$ is the matrix obtained by omitting the columns and rows of $A$ whose index appear in $L$.
\end{defn}
\begin{thm}[{\cite[Theorem 1]{Cross:1978aa}}]\label{thm:cross}\label{thm:remove}
If $A\in\Re^{n\times n}$ is diagonally stable, then all principle minors of $A$ are also diagonally stable.
\end{thm}
\noindent In the context of Lotka-Volterra dynamics this implies that if a given systems is diagonally stable, then even if an arbitrary numbers of species are removed from the system, the resulting system is still uniformly asymptotically stable in the positive orthant.

\begin{cor}\label{cor:remove}
A necessary condition that $A$ is diagonally stable is that all of its diagonal elements must be strictly negative.
\end{cor}

\subsection{Robustness to disturbances.}\label{sec:a:robustness}
We now consider three classes of Lotka-Volterra dynamics in the presence of disturbances, the first two are determinstic and the second is stochastic.

\subsubsection{Deterministic Dynamics}
In this section we will analyze the following two dynamical systems
\be\label{s:eq:d1}
\dot x(t) =  \diag(x(t))(r+ A x(t)+w(t)) 
\ee 
and 
\be \label{s:eq:d2}
\dot x(t) = d(t) + \diag(x(t)) (r+ A x(t)) 
\ee
where $d(t)$ and $w(t)$ are known to be a priori bounded. In \eqref{s:eq:d1} the term $w(t)$ represents uncertainty in the growth rates of the species. While in \eqref{s:eq:d1}  the term $w$ is deterministic, in the since that it is bounded, in terms of the ecology literature this term is sometimes referred to as the stochastic effect. Note that we will treat this term in a purely stochastic since shortly, complete with It\^o calculus. The term $d(t)$ is a migration term. The following theorems show that in the presence of these disturbances, the systems are bounded for all initial conditions (in the positive orthant).

\begin{thm}\label{thm:d1} If the system in \eqref{s:eq:d1} satisfies Assumptions \ref{ass1}-\ref{ass3} with $\norm{d(t)}\leq \alpha$ then the state $x$ is uniformly bounded for all initial conditions $x_0\in \Re^{n}_{>0}$.
\end{thm}
\begin{proof}
{
The proof is given for two different scenarios. The first scenario (when the disturbance is small) uses the same Lyapunov function as was used in the analysis of the disturbance free dynamics. In the second scenario a new Lyapunov candidate is introduced. A few definitions are needed before the different scenarios are analyzed. Let $P$ be a diagonal positive solution to the Lyapunov equation $A^{\mathsf T}P+PA=-Q$, where $Q=Q^{\mathsf T}>0$, and $p_i$ denotes the $i$-th diagonal element of $P$, just as in the disturbance free case. Let $q_\text{min}$ denote the minimum eigenvalue of $Q$ and $p_\text{max}$ denote the maximum diagonal element in $P$. Scenario: (a) the compact set $$\{x: \norm{x -x^*} \leq 2 p_\text{max} \alpha / q_\text{min}\} $$ does not intersect any of the $n$-axes; and (b) when the above compact set does intersect at least one of the $n$-axes. 
}

We now address the stability of Scenario (a). Let 
\be\label{eq:a:v}
V(x)= 2 \sum_{i=1}^n p_i(x_i-x^*_i - x^*_i \log(x_i/x^*_i))
\ee be a Lyapunov candidate. Differentiating the Lyapunov candidate it follows that
\ben\begin{split}
\dot V(x) &= 2 \sum_{i=1}^n p_i\left(\dot x_i - x^*_i \frac{\dot x_i}{x_i}\right) \\
&=  2\sum_{i=1}^n p_i(x_i - x^*_i) \frac{\dot x_i}{x_i}  \\
&=  2\sum_{i=1}^n p_i(x_i - x^*_i) \left(b_i+d_i(t)+\sum_{j=1}^n a_{ij} x_j \right)  \\
&=  2\sum_{i=1}^n p_i(x_i - x^*_i) \left ( d_i(t) +\sum_{j=1}^n a_{ij} (x_j-x_j^*) \right) \\
&=  (x - x^*)^{\mathsf T} (A^TP+ PA) (x - x^*) + 2(x-x^*)^{\mathsf T}P d.
\end{split} \een
Recall that $q_\text{min}$ denotes the minimum eigenvalue of $Q$ and $p_\text{max}$ denotes the maximum diagonal element in $P$, then it follows that 
\ben
\dot V \leq -q_\text{min} \norm{x-x^*}^2 + 2 p_\text{max} \norm {x-x^*} \alpha. 
\een
Thus for all $\norm{x-x^*} > 2 p_\text{max} \alpha / q_\text{min}$, $\dot V <0$ and thus $V(x(t))<\infty$ for all $t\geq t_0$. 
It follows that for all $x_0 \in \Re^n_{>0}$ the state $x(t)$ is bounded. 
Note that $x(t)$ can never be negative due to the fact that the disturbance appears as $x_i(t)d_i(t)$. {
This completes the proof for Scenario (a). 

Scenario (b) has to be treated differently do to the fact that if any of the $x_i=0$, then it follows that for the Lyapunov candidate in \eqref{eq:a:v} $V=\infty$. This was not possible in Scenario (a), but is possible in Scenario (b). We define a new Lyapunov candidate
\be\label{eq:a:v2}
V(x) = 2 \sum_{i=1}^n v_i(x_i)
\ee
where 
\ben
v_i(x) = \begin{cases} p_i(x_i-x^*_i - x^*_i \log(x_i/x^*_i)) & x^*_i \leq x_i \\
              0 & 0 \leq x_i < x^*_i.
\end{cases}
\een
For a general set of dynamics the above candidate function could not be a Lyapunov candidate. However, do to the special form of the Lotka-Volterra dynamics $x(t) \in R^n_{\geq 0}$ for all time regardless of the specific coefficients. Any population dynamic model should have this property, as negative abundances would make no sense. Differentiating $V$ in \eqref{eq:a:v2} we have that
\ben
\dot V(x) = 2 \sum_{i=1}^n \dot v_i(x_i)
\een
where 
\ben
\dot v_i(x) = \begin{cases}p_i(x_i - x^*_i) \left ( d_i(t) +\sum_{j=1}^n a_{ij} (x_j-x_j^*) \right)& x^*_i \leq x_i \\
              0 & 0 \leq x_i < x^*_i,
\end{cases}
\een
which is continuous in $x$, note that $p_i(x_i - x^*_i) \left ( d_i(t) +\sum_{j=1}^n a_{ij} (x_j-x_j^*) \right) = 0$ when $x_i=x_i^*$.

We now define the index set for all species as 
\be\label{eq:a:i}
\mathcal I = \{i : 0\leq i \leq n, i \in \mathbb N \}
\ee 
and the set of indices for species abundances that are greater than $x_i^*$ is defined as 
\be\label{eq:a:s}
\mathcal S = \{ i : x_i > x_i^*\} \subset \mathcal I.
\ee
Using this notation we can write the derivative of the Lyapunov function as 
\ben
\dot V(x) =  2\sum_{i\in S} p_i(x_i - x^*_i) \left ( d_i +\sum_{j\in\mathcal S} a_{ij} (x_j-x_j^* ) +\sum_{j\in\mathcal S^\mathsf c} a_{ij} (x_j-x_j^*) \right).
\een
The above equation can be rewritten in a more compact matrix-vector form as 
\be\begin{split}\label{eq:a:stept1}
\dot V(x) = & \left(x_\mathcal S - x^*_\mathcal S\right)^{\mathsf T} \left( A_{\mathcal S,\mathcal S}^{\mathsf T} P_{\mathcal S,\mathcal S}+ P_{\mathcal S,\mathcal S}A_{\mathcal S,\mathcal S} \right)\left(x_\mathcal S - x^*_\mathcal S\right)+ 2(x_{\mathcal S}-x_{\mathcal S}^*)^{\mathsf T}P_{\mathcal S,\mathcal S} d_{\mathcal S} \\ & +2 
\left(x_\mathcal S - x^*_\mathcal S\right)^{\mathsf T}  P_{\mathcal S,\mathcal S}A_{\mathcal S,\mathcal S^\mathsf c}\left(x_{\mathcal S^\mathsf c} - x^*_{\mathcal S^\mathsf c}\right).
\end{split}\ee
Using Theorem \ref{thm:cross} we can exploit a very unique property of diagonal Lyapunov functions, and that is the diagonal stability of all principle submatrices. It is also easy to confirm that the same $P$ matrix can be used for the sub-diagonal Lyapunov equations. Therefore, it follows that $A_{\mathcal S,\mathcal S}^{\mathsf T} P_{\mathcal S,\mathcal S}+ P_{\mathcal S,\mathcal S}A_{\mathcal S,\mathcal S}=-Q_{\mathcal S,\mathcal S}<0$ for any $\mathcal S$.
Using the following definitions
\be\label{eq:a:bounds1}
\begin{split}
q_\mathrm{min}^\prime(\mathcal S) &= \min_i \lambda_i (Q_{\mathcal S,\mathcal S})\\
p_\mathrm{max}^\prime(\mathcal S) &= \max_i \lambda_i (P_{\mathcal S,\mathcal S})\\
\beta^\prime(\mathcal S) &= \norm{A_{\mathcal S,\mathcal S^\mathsf c}},
\end{split}
\ee
the equality in \eqref{eq:a:stept1} can be bounded as
\ben
\dot V(x) \leq  -q_\mathrm{min}^\prime(\mathcal S)  \norm{x_\mathcal S - x^*_\mathcal S}^2+ 2 p_\mathrm{max}^\prime(\mathcal S)  \norm{x_\mathcal S - x^*_\mathcal S} \left( \alpha   + \beta^\prime(\mathcal S) \norm{x_{\mathcal S^\mathsf c}-x^*_{\mathcal S^\mathsf c}}\right).
\een
Given our definition of $\mathcal S$ it follows that 
$\norm{x_{\mathcal S^\mathsf c}-x^*_{\mathcal S^\mathsf c}}\leq \norm{x^*_{\mathcal S^\mathsf c}}\leq \norm{x^*}$. This allows the above inequality to be further simplified as 
\be\label{eq:a:t3}
\dot V(x) \leq  -q_\mathrm{min}^\prime(\mathcal S)  \norm{x_\mathcal S - x^*_\mathcal S}^2+ 2 p_\mathrm{max}^\prime(\mathcal S)  \norm{x_\mathcal S - x^*_\mathcal S} \left( \alpha   + \beta^\prime(\mathcal S) \norm{x^*}\right).
\ee
The bound in \eqref{eq:a:t3} is still not sufficient to address stability as the bounds depend on the set of indices  in $\mathcal S$. In a two step process we obtain uniform constants and then introduce a distance function that is independent of $\mathcal S$. We will now give uniform analogs to the bounds $q_\mathrm{min}^\prime(\mathcal S), 
p_\mathrm{max}^\prime(\mathcal S),
\beta^\prime(\mathcal S)$ by taking the appropriate supremum or infimum over the powerset $2^\mathcal I$ of all subsets $\mathcal S\subset \mathcal I$. The powerset of interest is finite and thus the following are well defined
\be\label{eq:a:bbounds}
\begin{split}
\bar q &= \inf_{\mathcal S \subset \mathcal I \setminus \emptyset} q_\mathrm{min}^\prime(\mathcal S) \\
\bar p &=  \sup_{\mathcal S \subset \mathcal I}p_\mathrm{max}^\prime(\mathcal S) \\
\bar \beta &=  \sup_{\mathcal S \subset \mathcal I} \beta^\prime(\mathcal S).
\end{split}
\ee
Using the uniform bounds above the inequality in \eqref{eq:a:t3} can be replaced by
\be\label{eq:a:t4}
\dot V(x) \leq  -\bar q  \norm{x_\mathcal S - x^*_\mathcal S}^2+ 2 \bar p  \norm{x_\mathcal S - x^*_\mathcal S} \left( \alpha   + \bar \beta \norm{x^*}\right).
\ee

We now  define a distance metric between an element $y\in \Re^n$ and a compact set $\mathcal A \in \Re^n$ as
\be\label{eq:a:dist}
d(y,\mathcal A) \triangleq \inf\{\norm{y-z}: z \in \mathcal A\}.
\ee
Given that $\mathcal A$ is compact, such a $z\in \mathcal A$ always exists. Our compact set of interest is defined as 
\be\label{eq:A:setX}
\mathcal X = \{ x : 0\leq x_i \leq x_i^*\}.
\ee
Using the compact set defined just above, the inequality in \eqref{eq:a:t4} can be equivalently rewritten as 
\ben\label{eq:a:t5}
\dot V(x) \leq  -\bar q d(x,\mathcal X)^2+ 2 \bar p d(x,\mathcal X) \left( \alpha   + \bar \beta \norm{x^*}\right).
\een
Thus for all $d(x,\mathcal X)> 2 \bar p \left( \alpha   + \bar \beta \norm{x^*}\right)/\bar q $, $\dot V<0$. Therefore $V(x(t))$ is bounded for all $t\geq t_0$.}
\end{proof}

\begin{thm}\label{thm:d2} If the system in \eqref{s:eq:d2} satisfies Assumptions \ref{ass1}-\ref{ass3} with $\norm{d(t)}\leq \alpha$ and furthermore we exclude the possibility for the disturbance to generate negative state values\footnote{This is not a restrictive assumption in the context of ecology as a disturbance can never result in the creation of a negative population.}, then the state $x$ is uniformly bounded for all initial conditions $x_0\in \Re^{n}_{>0}$.
\end{thm}
{
\begin{proof}
Consider the Lyapunov candidate in \eqref{eq:a:v2} and differentiating along the system dynamics in \eqref{s:eq:d2} it follows that  
\ben
\dot V(x) =  2\sum_{i\in S} p_i(x_i - x^*_i) \left ( d_i/x_i +\sum_{j\in\mathcal S} a_{ij} (x_j-x_j^* ) +\sum_{j\in\mathcal S^\mathsf c} a_{ij} (x_j-x_j^*) \right).
\een
where $\mathcal S\subset I$ was defined in \eqref{eq:a:i} and \eqref{eq:a:s}. Rearranging terms slightly with regard to $d_i/x_i$ it follows that 
\ben
\dot V(x) =  2\sum_{i\in S} p_i(x_i - x^*_i) \left( \sum_{j\in\mathcal S} a_{ij} (x_j-x_j^* ) +\sum_{j\in\mathcal S^\mathsf c} a_{ij} (x_j-x_j^*) \right) + 2\sum_{i\in S} d_i p_i(1- x^*_i/x_i).
\een
 Given our definition of $\mathcal S$ it follows that $(1- x^*_i/x_i) \leq 1$ when $i\in \mathcal S$. Therefore it follows that 
 \ben
\dot V(x) \leq  2\sum_{i\in S} p_i(x_i - x^*_i) \left( \sum_{j\in\mathcal S} a_{ij} (x_j-x_j^* ) +\sum_{j\in\mathcal S^\mathsf c} a_{ij} (x_j-x_j^*) \right) + 2\sum_{i\in S} \abs{d_i} p_i.
\een
The above inequality can be rewritten in a more compact matrix-vector form as 
\ben
\begin{split}
\dot V(x) \leq & \left(x_\mathcal S - x^*_\mathcal S\right)^{\mathsf T} \left( A_{\mathcal S,\mathcal S}^{\mathsf T} P_{\mathcal S,\mathcal S}+ P_{\mathcal S,\mathcal S}A_{\mathcal S,\mathcal S} \right)\left(x_\mathcal S - x^*_\mathcal S\right)+ 2 \norm{P_{\mathcal S,\mathcal S}}\norm{ d_{\mathcal S}} \\ & +2 
\left(x_\mathcal S - x^*_\mathcal S\right)^{\mathsf T}  P_{\mathcal S,\mathcal S}A_{\mathcal S,\mathcal S^\mathsf c}\left(x_{\mathcal S^\mathsf c} - x^*_{\mathcal S^\mathsf c}\right).
\end{split}
\een
Using the bounds in \eqref{eq:a:bounds1} the above inequality can be reduced to 
\ben
\dot V(x) \leq  -q_\mathrm{min}^\prime(\mathcal S)  \norm{x_\mathcal S - x^*_\mathcal S}^2+ 2 p_\mathrm{max}^\prime(\mathcal S)  \norm{x_\mathcal S - x^*_\mathcal S} \beta^\prime(\mathcal S) \norm{x_{\mathcal S^\mathsf c}-x^*_{\mathcal S^\mathsf c}} + 2 p_\mathrm{max}^\prime(\mathcal S)  \alpha  .
\een
Given our definition of $\mathcal S$ it follows that 
$\norm{x_{\mathcal S^\mathsf c}-x^*_{\mathcal S^\mathsf c}}\leq \norm{x^*_{\mathcal S^\mathsf c}}\leq \norm{x^*}$. This allows the above inequality to be further simplified as 
\be\label{eq:a:t3:2}
\dot V(x) \leq  -q_\mathrm{min}^\prime(\mathcal S)  \norm{x_\mathcal S - x^*_\mathcal S}^2+ 2 p_\mathrm{max}^\prime(\mathcal S)  \norm{x_\mathcal S - x^*_\mathcal S} \beta^\prime(\mathcal S) \norm{x^*} + 2 p_\mathrm{max}^\prime(\mathcal S) \alpha.
\ee
Using the definitions in \eqref{eq:a:bbounds} and the set distance function in \eqref{eq:a:dist} with the compact set $\mathcal X$ defined in \eqref{eq:A:setX} the above inequality can be written as 
\ben
\dot V(x) \leq  -\bar q d(x,\mathcal X)^2+ 2 \bar p d(x,\mathcal X) \bar \beta \norm{x^*}+ 2 \bar p \alpha
\een
Rewriting the above inequality as
\ben
\dot V(x) \leq  -\frac{\bar q}{2} d(x,\mathcal X)^2 -\frac{\bar q}{2} \left(d(x,\mathcal X)^2- 4 \frac{\bar p}{\bar q} d(x,\mathcal X) \bar \beta \norm{x^*} \right)+ 2 \bar p \alpha
\een 
and completing the square with respect to the middle term $\frac{\bar q}{2} \left(d(x,\mathcal X)^2- 4 \frac{\bar p}{\bar q} d(x,\mathcal X) \bar \beta \norm{x^*} \right) $ the following inequality holds
\ben
\dot V(x) \leq  -\frac{\bar q}{2} d(x,\mathcal X)^2 -\frac{\bar q}{2} \left(d(x,\mathcal X)- 2 \frac{\bar p}{\bar q}\bar \beta \norm{x^*} \right)^2+ 2 \bar p \alpha + 2\frac{\bar p^2 \bar\beta^2 \norm{x^*}^2}{\bar q}
\een
Noting that the term $-\frac{\bar q}{2} \left(d(x,\mathcal X)- 2 \frac{\bar p}{\bar q}\bar \beta \norm{x^*} \right)^2\leq 0$ it follows that 
\be\label{eq:a:fb:2}
\dot V(x) \leq  -\frac{\bar q}{2} d(x,\mathcal X)^2 + 2 \bar p \alpha+ 2\frac{\bar p^2 \bar\beta^2 \norm{x^*}^2}{\bar q}
\ee
From the inequality in \eqref{eq:a:fb:2} it follows that \ben d(x,\mathcal X)>2\sqrt{\frac{\bar p \alpha}{\bar q} + \frac{\bar p^2  \bar \beta^2 \norm{x^*}^2}{\bar q^2}} \een implies  $\dot V<0$. Therefore, $x(t)$ is uniformly bounded and asymptotically converges to the compact set \ben d(x,\mathcal X)\leq2\sqrt{\frac{\bar p \alpha}{\bar q} + \frac{\bar p^2  \bar \beta^2 \norm{x^*}^2}{\bar q^2}}.\qedhere\een
\end{proof} }


{
\subsubsection{Stochastic Dynamics}
In this section we wish to analyze a stochastic differential equation that is similar to \eqref{s:eq:d1}, but we no longer make the assumption that the disturbance is bounded, instead we assume that the disturbance appears as a Brownian motion $w(t) \in {\Re}^n$, resulting in the following differential equation,
\be\label{eq:sde}
\mathrm d x = \diag(x)(r+Ax)\, \mathrm dt + c\,\diag(x)\,\mathrm dw.
\ee
The variable $c\in \Re$ will be a constant used to scale the square root of the variance of the brownian motion. It will be shown that $x(t)$ converges in a probabilistic since to a compact set which is proportional to $c$. The analysis of stochastic differentials is significantly more challenging than their deterministic counterparts. Before stating the stochastic version of the stability result, we first need to formally define filtration, Brownian motion, and give a key result of It\^o. In this work the linear ordered set of interest will always be time $t\in [t_0,\infty)$. All of the following definitions are given in a less general context compared to their original presentation \cite{doob2012classical}.

\begin{defn}[Stochastic Process, \cite{doob2012classical}*{Part 2 \S I.8}]
Let $(\Omega, \mathcal F,\mathbf P)$ be a probability space. A {\em stochastic process} on $(\Omega, \mathcal F,\mathbf P,[t_0,\infty))$ is a family of random variables $\{y(t),t\in[t_0,\infty)\}$ with map $(t,\omega) \mapsto y(t,\omega)$ from $[t_0,\infty) \times \Omega$ to $R$, the variable space of interest. Given that each instance in time $y(t)$ is a random variable, the variable space  $R=(R,\mathcal R)$ is necessarily a measurable space.
\end{defn}

\begin{defn}[Filtration, \cite{doob2012classical}*{Part 2 \S I.1}]
Let $(\Omega, \mathcal F)$ be a measurable space and $t\in[t_0,\infty)$ is time, then a {\em filtration} of that measurable space  is a map $t\mapsto \mathcal F(t)$ of increasing sub $\sigma$ algebras such that $\mathcal F(s)\subset \mathcal F(t) \subset \mathcal F$ when $s\leq t$.
\end{defn}

\begin{defn}[Adaptation, \cite{doob2012classical}*{Part 2 \S I.1}]
Let $\{y(t),t\in[t_0,\infty)\}$ be a stochastic process from a {\em filtered measure space} $\{\Omega, \mathcal F, \mathcal F(\cdot) \}$ into the measurable space $\{R,\mathcal R\}$. The process is {\em adapted} to $\mathcal F(\cdot)$ if for each $t$, $y(t):(\Omega, \mathcal F(t) ) \to (R,\mathcal R)$ is measurable.
\end{defn}

\begin{defn}[Progressively Measurable, \cite{doob2012classical}*{Part 2 \S I.2}]
Let $\{y(t),\mathcal F(t), t\in[t_0,\infty)\}$ be an adapted stochastic process. The function $y: [t_0,\infty)\times \Omega \to R$, where $(R,\mathcal R)$ is a measurable space, is deemed {\em progressively measurable} if 
\ben
y:( [t_0,t] \times \Omega, \mathsf{Borel}([t_0,t]) \times \mathcal F(t) ) \to (R,\mathcal R)
\een  is measurable for any $t\geq t_0$.
\end{defn}


Let us pause now and discuss these definitions. Defining a stochastic processes is rather obvious. We extended the definition of a random variable to incorporate time. At each fixed instance a stochastic process is nothing but a random variable. The extra definitions and the progression from filtrations, adaptations, and progressive measurability, are needed so that we can better understand what is needed to perform the following integration \cite{stroock2010probability}*{Remark 7.1.1}. Let $\{y(\cdot),\mathcal F(\cdot)\}$ be an adapted stochastic process defined as before where $y:[t_0,\infty) \times \Omega \to R$. Let $f: R \to \Re$ be bounded and $\mathcal R$ measurable, then the map $(t,\omega)\mapsto z = \int_{t_0}^t f(y(\tau,\omega)) \, \mathrm d\tau$ need not be adapted. However if in addition to being adapted the stochastic process is progressively measurable, then $z$ will be progressively measurable as well. This process is a key necessary ingredient when studying stochastic differential equations as it can be inherited. Next we formally define a Brownian motion and give a sufficient condition for the adaptation so that the progressive measurability condition is satisfied.

\begin{defn}[Brownian,  \cite{doob2012classical}*{Part 2 \S VII.2 }]
A {\em Brownian} motion is an adapted stochastic process $\{w(\cdot),\mathcal F(\cdot)\}$ from the filtered probability space $(\Omega,\mathcal F, \mathbf P, \mathcal F(t), t\in[t_0,\infty))$ into the measurable space $(\Re^n,\mathsf{Borel}(\Re^n))$ 
\begin{itemize}
\item that is  {\em Markovian}, i.e. when $s<t$ and $\mathcal A \in \mathsf{Borel}(\Re^n)$ it follows that \ben\mathbf P (w(t)\in \mathcal A | \mathcal F(s)) =\mathbf P (w(t)\in \mathcal A |  w(s)) \een almost surely.
\item with a stationary stochastic transition function (the transition function is independent of the current time of the process) with density $\rho(t,\psi -\xi)$ defined on $\Re\times\Re^n$ relative to the $n$-dimensional Lebesgue measure, where $\rho$ is defined as
\ben
\rho(t,\psi;\sigma,t_0) = \begin{cases} (2\pi\sigma^2t)^{-n/2} \mathrm{exp}\frac{-\abs{\psi}^2}{2\sigma t} & \text{if } t>t_0 \\
0 & \text{if } t\leq t_0,\end{cases}
\een
where $t\in \Re$ is time and $\psi\in\Re^n$ is the space variable.
\item that is almost surely continuous
\end{itemize}
\end{defn}

We are almost ready to address the problem at hand, however we still need to show how we can ensure that the Brownian motion is progressively measurable. Unfortunately we need  a few more definitions and then we will show that Brownian motions can be naturally adapted so as to have the progressive measurability property.

\begin{defn}[Right Continuous Filtration,  \cite{doob2012classical}*{Part 2 \S I.1 }]
Let $(\Omega, \mathcal F, \mathcal F(t), t\in[t_0,\infty))$ be a filtered measurable space, and define $\mathcal F^+(t)=\bigcap_{s>t} \mathcal F(s)$ for all $t\in [t_0,\infty)$. The filtration is {\em right-continuous} if $\mathcal F(\cdot) = \mathcal F^+(\cdot)$.
\end{defn}

\begin{thm}
Let $\{y(\cdot), \mathcal F(\cdot)\}$ be a Brownian motion into $\Re^n$ with respect to time $t\in[t_0,\infty)$, and if $\mathcal F(t)$  is generated by the null sets and $\sigma(y(s);s\leq t)$, the smallest sigma algebra for which all $y(s$) are measureable for all $s\in[t_0,t]$, then $\mathcal F(\cdot)=\mathcal F^+(\cdot)$
\end{thm}
\begin{proof} see   \cite{doob2012classical}*{Part 2, \S VI, Theorem 8}
\end{proof}

This theorem implies that one can assume without loss of generality that the filtration is right-continuous. That is, there is a natural way to construct them for any brownian motion.

\begin{thm}
Let $\{y(\cdot), \mathcal F(\cdot)\}$ be a Brownian motion with right-continuous filtration containing the null sets, then the Brownian motion is progressively measurable.
\end{thm}
\begin{proof} see \cite{meyer1966probability}*{Part A Theorem 47}
\end{proof}

As stated before, this progressive measurability is necessary when discussing the solutions to stochastic differential equations as it allows for integrals containing stochastic processes to inherit the progressively measurable property.
We now state a classic result do to It\^o \cite{Ito:1944aa}. Our version follows from   \cite{doob2012classical}*{Part 2 \S VIII.12}.

\begin{lem}\label{lem:ito}
Let $\{w(\cdot),\mathcal F(\cdot)\}$ be a Brownian motion into $\Re^n$ with a right-continuous filtration containing the null sets. Consider the dynamics $x(t)\in \Re^n$ given by
\ben
\mathrm dx = \mu \mathrm d t + \sigma \mathrm d w
\een
where $t\in[t_0,\infty)$, $\mu(x)\in\Re^n$ is locally Lipschitz in $x$, and $\sigma(x)\in\Re^{n\times n}$ is globally Lipschitz in $x$. Assuming $(x,t)\mapsto f(x,t)\in \Re$  is twice differentiable and continuous in terms of $x$ $($class $C^2$ with respect to $x)$ and once differentiable and continuous in terms of $t$ $($class $C^1$ with respect to $t)$, then
 $$\mathrm df = \frac{\partial f}{\partial t} \mathrm d t+  (\nabla_x f)^\mathsf T \mu \,\mathrm dt  + \frac{1}{2} \trace( \sigma \hess_x(f) \sigma)\, \mathrm d t +  (\nabla_x f)^\mathsf T \sigma \, \mathrm dw.$$
\end{lem}

\begin{rem}
In most constructions it is assumed that $\mu$ is globally Lipschitz. Indeed this assumption was made in one of It\^o's original papers \cite{ito1946stochastic}. A detailed discussion regarding the existence of solutions when only locally Lipschitz conditions are assumed can be found in \cite{protter2013stochastic}. Exploiting the stability of our dynamics the existence and uniqueness of solutions can be deduced directly from the original work of It\^o however. In our analysis we will show that for the dynamics of interest the state is uniformly bounded almost surely, a formal definition is given just below this remark. Therefore, we can define equivalent dynamics 
\ben
\mathrm d \bar x = \bar \mu \mathrm d t + \sigma \mathrm d w
\een
where $\bar\mu=\mu$ on the compact set of interest, and zero outside this compact set. Given that $\mu$ is locally Lipschitz it follows that the function $\bar \mu$ with compact support is by definition globally Lipschitz. Solutions $\bar x$ exist, are unique, and are uniformly bounded almost surely. This then implies the existence and uniqueness of $x$ for the dynamics $
\mathrm dx = \mu \mathrm d t + \sigma \mathrm d w$ on the same compact set which $x(\cdot)$ remains in, almost surely. 
\end{rem}


\begin{defn}\label{defn:boundedas}
The state  $x(t;t_0,x_0)$  as a solution to the difference equation in \eqref{eq:sde} is {\em uniformly bounded with probability one} if for ever  $r>0$ there exists a $B(r)>0$ such that $\norm{x(t_0)}\leq r$ implies 
\ben
\mathbf P\left(\sup_t \norm{x(t)} \leq B(r)\right)=1
\een
for all $t\geq t_0$. An equivalent statements would be $x(t)$ is {\em uniformly bounded almost surely}.
\end{defn}


\begin{thm}\label{thm:stochastic_stab}
Consider the dynamics in \eqref{eq:sde} for $x(t)\in \Re^n$. Let  assumptions \ref{ass1}-\ref{ass3} hold with $\{w(\cdot),\mathcal F(\cdot)\}$ a Brownian motion into $\Re^n$ with a right-continuous filtration containing the null set. The following then hold
\begin{enumerate}
\item The state variable $x(\cdot)$ is uniformly bounded in expectation, 
\item $x(\cdot)$ is uniformly bounded with probability one,  and
\item $x(\cdot)$ asymptotically converges to the compact set
\be\label{a:eq:setd}
\mathcal D \triangleq \left\{x: \norm{x-x^*} \leq c \sqrt{\frac{n x^*_\mathrm{max}p_\mathrm{max}}{q_\mathrm{min}}} \right\}\ee
 with probability one. Stated more precisely,  
\be\label{thm:sder}
\mathbf P \left(\lim_{t\to\infty} x(t) \in \mathcal D \right) =1.
\ee
\end{enumerate}
\end{thm}
\begin{proof}[Sketch of the proof]
This proof only outlines the analysis when the compact set $\mathcal D$ does not intersect any of the $n$-axes. The more complicated scenario when this is not the case can be handled just as it was in the proof of Theorem \ref{thm:d1}. Consider the Lyapunov candidate ${V(x)= 2 \sum_{i=1}^n p_i(x_i-x^*_i - x^*_i \log(x_i/x^*_i))}$. Taking the differential along the lines of Lemma \ref{lem:ito} (It\^o)  results in
\begin{multline}
\mathrm dV= (\nabla_x V)^\mathsf T{\diag}(x)(r+Ax) \,\mathrm dt  + \frac{c^2}{2} \trace(\diag(x) \hess_x(V) \diag(x))\, \mathrm d t \\ + (\nabla_x V)^\mathsf T c\, \diag (x)\, \mathrm d w  
\end{multline} 
Recall from the steps in the proof of Theorem \ref{thm:d1} that 
$$
(\nabla_x V)^\mathsf T\diag(x)(r+Ax) = -(x-x^*)^\mathsf T Q (x-x^*)
$$ 
where $A^\mathsf T P + PA = -Q$. Thus giving
\begin{multline}
\mathrm dV=  -(x-x^*)^\mathsf T Q (x-x^*) \,\mathrm dt  + \frac{c^2}{2} \trace(\diag(x) \hess_x(V) \diag(x))\, \mathrm d t \\ +(\nabla_x V)^\mathsf T c\, \diag (x)\, \mathrm d w  
\end{multline}
Next, note that $[\hess_x(V)]_{ii} = 2  \frac{p_ix_i^*}{x_i^2} $ and  $[\hess_x(V)]_{ij} = 0$ when $i\neq j$. Substitution into the above equation results in 
\be \label{eq:fatouapply}
\mathrm dV=  -(x-x^*)^\mathsf T Q (x-x^*) \,\mathrm dt  + c^2 \sum_i p_i \, \mathrm d t + (\nabla_x V)^\mathsf Tc\, \diag (x)\, \mathrm d w  
\ee
Note that $x(t)$ and $\mathrm dw(t)$ are independent, $x(t)$ is only dependent on $\mathrm dw(s)$ when $s<t$. Therefore $\mathbf E( (\nabla_x V)^\mathsf Tc\, \diag (x) \, \mathrm d w)=0 $, given that for a Brownian motion $\mathbf E\, \mathrm d w=0$.\footnote{In order to address this rigorously we would rewrite the expression in \eqref{eq:fatouapply} in integral form. Then we would integrate only up to a stopping time which we designed so that $(\nabla_x V)^\mathsf Tc\, \diag (x) <\infty$ up to the time of interest. Then we would use the fact that $\mathbf E g w =0$ for any bounded $g$. The bound used to generate the stopping time would then be relaxed and the desired result would be obtained as $t\to \infty$ using Fatou's Lemma and monotone convergence. A similar procedure is carried out in \cite{deng2001stabilization}*{(4.3)-(4.5)}}  Therefore
\be\label{eq:a:edv}
\frac{\mathbf{E}\, \mathrm d V}{\mathrm dt} \leq  -q_\mathrm{min}  \norm{x-x^*}^2 \, \mathrm + c^2n x_\mathrm{max}^* p_\mathrm{max}
\ee
where $q_\mathrm{min}$ is the minimum eigenvalue of $Q$, $p_\mathrm{max}$ is the maximum diagonal element in $P$, and $x^*_\mathrm{max}$ is the value in the vector $x^*$. Therefore,
\be\label{eq:a:signvcompact}
\norm{x-x^*}> c \sqrt{\frac{n x^*_\mathrm{max}p_\mathrm{max}}{q_\mathrm{min}}}  \implies \frac{\mathbf{E}\, \mathrm d V}{\mathrm dt} < 0.
\ee
Given that $\mathbf E \mathrm d V/\mathrm dt \leq 0$ outside a compact set it follows that $\mathbf E V$ is uniformly bounded in terms of the initial condition $x(t_0)$ \cite{khasminskii2011stochastic}*{Lemma 5.4}. Given that $V(x)\geq 0$ is continuous in $x$ and convex it follows from Jensen's inequality that $\mathbf E x$ is uniformly bounded as well \cite{jensen1906fonctions}. Jensen's inequality will be used without reference from this point forward.  Claim 1 has been proven. 

We now move on to the second claim, namely that the dynamics are uniformly bounded with probability one. 
We address this claim with sub scenarios: (a) $x(t_0)\notin \mathcal D$, and (b) $x(t_0) \in \mathcal D$. Under scenario (a) we will now show that $\mathbf P(\sup V(x(t)) - V(x(t_0)  \leq \delta_1 ) = 1$ for any $\delta_1\in(0,\infty)$. The previous statement is equivalent to $\mathbf P(\sup V(x(t)) - V(x(t_0)  > \delta_1 ) = 0$. This statement will be proved by contradiction. Assume that there exists a $\delta_2\in (0,1]$ such that $\mathbf P(\sup V(x(t)) - V(x(t_0)  > \delta_1 ) \geq \delta_2$. Using Markov's inequality it follows that 
 \ben
  \mathbf E \sup V(x(t)) - V(x(t_0)) \geq \delta_1\delta_2.
 \een
This statement however contradicts \eqref{eq:a:signvcompact} which states that outside the compact set of interest the expected value of the Lyapunov candidate is strictly decreasing. Therefore it follows that $\mathbf P(\sup V(x(t)) - V(x(t_0)  > \delta_1 ) = 0$ which is equivalent to the claim that $\mathbf P(\sup V(x(t)) - V(x(t_0)  \leq \delta_1 ) = 1$. Application of Jensen's inequality then implies that $x$ is uniformly bounded with probability one for sub-scenario (a).

We now address sub-scenario (b) where it is assumed that $x(t_0)\in \mathcal D$. If $x(\cdot)$ remains in $\mathcal D$ for all time, then we are done and the state is uniformly bounded. Therefore, assume at some time $t_1$ the state $x(t_1)\notin \mathcal D$. This sub-scenario is now equivalent to sub-scenario (a) with time shifted. Therefore, it follows that $x$ is uniformly bounded with probability 1 in sub-scenario (b) as well. This completes the proof of claim 2.
 

We approach claim 3 following a method proposed in the proof of \cite{deng2001stabilization}*{Theorem 2.1}. Consider the probability of three mutually exclusive scenarios, just as in \cite{deng2001stabilization}*{Theorem 2.1}
\begin{align*}
p_1 &= \mathbf P  \left(\limsup_{t\to\infty} d(x(t),\mathcal D)=0 \right)\\
p_2 &= \mathbf P  \left(\liminf_{t\to\infty} d(x(t),\mathcal D)>0 \right)\\
p_3 &= \mathbf P  \left(\liminf_{t\to\infty} d(x(t),\mathcal D)=0 \ \text{and}\ \limsup_{t\to\infty} d(x(t),\mathcal D)>0 \right)
\end{align*}
We wish to prove that $p_1=1$ and $p_2,p_3=0$.

We first prove that $p_2=0$. This will be proved by contradiction. Assume $p_2 = \epsilon_1$ where $\epsilon_1\in(0,1]$, then there exists an $\epsilon_2$ such that $\mathbf P  \left(\liminf_{t\to\infty} d(x(t),\mathcal D)\geq\epsilon_2 \right)=\epsilon_1$. Using Markov's inequality it follows that 
\ben
\frac{1}{\epsilon_2}\mathbf E \liminf_{t\to\infty} d(x(t),\mathcal D) \geq \epsilon_1.
\een
Multiplying both sides by $\epsilon_2$ it follows that 
\ben
\mathbf E \liminf_{t\to\infty} d(x(t),\mathcal D) \geq \epsilon_1 \epsilon_2.
\een
From the definition of $\liminf$ it follows that for any $\epsilon_3\in(0,\epsilon_1\epsilon_2)$ there exist a finite $T_1\in[t_0,\infty)$ such that \be\label{a:eq:c1}
\mathbf E\,  d(x(t),\mathcal D) \geq \epsilon_3 \ \text{for all }t \geq T_1.
\ee
Therefore, for $t\geq T_1$ it follows that $\mathbf E \mathrm d V / \mathrm d t <0$. Given that $V(x^*)=0$ and $V(x)$ is strictly increasing away from $x=x^*$ and tends to infinity for large $x$ in the positive orthant it follows that $\mathbf E \norm{x-x^*}$ is strictly decreasing in time. Given that $\mathcal D$ is centered at $x^*$, there exists $T_2$ such that 
$$
\mathbf E\,  d(x(t),\mathcal D) < \epsilon_3 \ \text{for all }t \geq T_2.
$$
which contradicts \eqref{a:eq:c1}. Therefore $p_2=0$. That is $$\mathbf P  \left(\liminf_{t\to\infty} d(x(t),\mathcal D)>0 \right)=0.$$ 

We now establish that $p_3=0$. As before we achieve this by contradiction. For any $\epsilon_5>0$ assume 
\be\label{eq:a:cont3}
\mathbf P  \left(\liminf_{t\to\infty} d(x(t),\mathcal D)=0 \ \text{and}\ \limsup_{t\to\infty} d(x(t),\mathcal D)\geq \epsilon_5 \right)\neq 0.
\ee
For any $\epsilon_6$ and $\epsilon_7$  such that $0<\epsilon_7<\epsilon_6 <\epsilon_5$, consider the {\em stopping times}\footnote{For more detail on stopping times see  \cite{doob2012classical}*{Part 2, Chapter II}.} $T_i^{\prime} : \Omega \to [t_0,\infty)$, $T_i^{\prime\prime}:\Omega\to [t_0,\infty)$ with $i\in \mathbb N_{>0}$, where \begin{align*} T_i^\prime &= \inf \left\{ t \geq T_{i-1}^{\prime\prime} : d(x(t),\mathcal D) \leq \epsilon_7 \right\} \\ T_i^{\prime\prime} &= \inf \left\{ t \geq T_{i}^{\prime} : d(x(t),\mathcal D) \geq \epsilon_6 \right\}.\end{align*} 
From \eqref{eq:a:cont3} and the stopping times defined above, it follows that
$$
\lim_{i\to \infty} T_i^\prime = \infty \text{ and } \lim_{i\to\infty} T_{i}^{\prime\prime} = \infty
$$
In order to continue with this proof by contradiction we need to obtain a lower bound on the expectation $T_{i+1}^{\prime}-T_{i}^{\prime\prime} $. Indeed if this can be done, then we expect the solutions to the differential equation to spend a non negligible amount of time within a domain of the state space where the Lyapunov function will be decreasing. Thus we can see how this may lead to a contradiction, for how can trajectories be expected to spend an infinite amount of time in a domain where the Lyapunov function is always decreasing. Following the same procedures as in the proof of the bound in \cite{deng2001stabilization}*{(2.27)} it can be shown that there exists an $\epsilon_8>0$ such that
\ben
\mathbf E\left(T_{i+1}^{\prime}-T_{i}^{\prime\prime} \  |\  \mathcal F(T_i^{\prime\prime})\right)\geq \epsilon_8.
\een


Using the bound above and the definition of $\mathcal D$ in \eqref{a:eq:setd} it follows that
\begin{multline*}\label{eq:d123}
\mathbf E  \left(\int_{T_i^{\prime\prime}}^{T^{\prime}_{i+1}} q_\mathrm{min}\norm{x(\tau)-x^*}^2 - c^2n x_\mathrm{max}^* p_\mathrm{max} \, \mathrm{d} \tau \ \Big|\ \mathcal F(T_i^{\prime\prime})\right) \\  \geq \left(q_\mathrm{min} c \sqrt{\frac{n x_\mathrm{max}^* p_\mathrm{max}}{q_\mathrm{min}}}\epsilon_6 + q_\mathrm{min} \epsilon_6^2   \right) \epsilon_8  
\end{multline*}
From the bound in \eqref{eq:a:edv} it follows that
\ben\begin{split}
V(x(t_0)) \geq & \mathbf E \int_{t_0}^\infty  q_\mathrm{min}\norm{x(\tau)-x^*}^2 - c^2n x_\mathrm{max}^* p_\mathrm{max} \, \mathrm{d} \tau \\
\geq & \sum_{i=1}^\infty  \left(q_\mathrm{min} c \sqrt{\frac{n x_\mathrm{max}^* p_\mathrm{max}}{q_\mathrm{min}}}\epsilon_6 + q_\mathrm{min} \epsilon_6^2   \right) \epsilon_8  \mathbf P(T_i^{\prime\prime} <\infty).
\end{split}\een
Noting that $V(x(t_0)) <\infty$, it then follows from the Borel-Cantelli Lemma \cite{doob1953stochastic}*{Chapter III, Theorem 1.2} that  $\mathbf P( \exists N<\infty\ \text{s.t.}\ \forall i\geq N, \quad T_i^{\prime\prime} <\infty)=0$ \cite{deng2001stabilization}*{(2.28)-(2.31)}. This implies that $\mathbf P  \left(\limsup_{t\to\infty} d(x(t),\mathcal D)\geq \epsilon_5 \right) = 0$ which contradicts \eqref{eq:a:cont3}. Therefore $p_3=0$, that is 
\ben
\mathbf P  \left(\liminf_{t\to\infty} d(x(t),\mathcal D)=0 \ \text{and}\ \limsup_{t\to\infty} d(x(t),\mathcal D(0))>0 \right)=0.
\een 
Finally, given that the events associated with $p_1$, $p_2$, and $p_3$ are nonintersecting, it follows that $p_1=1$. Summarizing, we have shown that 
\begin{align*}
p_1 &= \mathbf P  \left(\limsup_{t\to\infty} d(x(t),\mathcal D)=0 \right)=1\\
p_2 &= \mathbf P  \left(\liminf_{t\to\infty} d(x(t),\mathcal D)>0 \right)=0\\
p_3 &= \mathbf P  \left(\liminf_{t\to\infty} d(x(t),\mathcal D)=0 \ \text{and}\ \limsup_{t\to\infty} d(x(t),\mathcal D(0))>0 \right)=0,
\end{align*}
and thus claim 2 of the theorem has been proven. We wish to reiterate the fact that this sketch barrows heavily from \cite{deng2001stabilization}. In most instances intermediate steps where stopping times are needed were glossed over. It might be worthwhile to revisit this analysis in the future, seeing as there does not seem to be much literature along this direction other than the references here in.
\end{proof}

\begin{rem}
The definition given for uniformly bounded in probability in this work is stronger than any other definitions that could be found in the literature and follows the classic definition, see Definition \ref{defn:a:stab} and compare it to Definition \ref{defn:boundedas}. A trajectory is uniformly bounded in probability, if for all $r>0$ there exists a $B(r)>0$ such that $\norm{x(t_0)}\leq r$ implies 
\ben
\mathbf P \left( \sup_t \norm{x(t)} \leq B(r)\right)=1
\een
for all $t\geq t_0$. In \cite{deng2001stabilization}*{Definition 2.2}  the following definition is given, for all $r>0$ and any $\epsilon>0$ there exists a $B(r,\epsilon)>0$ such that $\norm{x(t_0)}\leq r$ implies \ben \mathbf P \left( \sup_t \norm{x(t)} \leq B(r)\right)=1-\epsilon. \een
In  \cite{khasminskii2011stochastic}*{\S1.4} the definition of uniformly bounded is defined as follows, 
\ben
\sup_t \mathbf P(\norm{x(t)}>R)\to 0 \text{ as }  R\to\infty.
\een
Our definition of asymptotic attractivity follows that of Deng and Kristi\'c \cite{deng2001stabilization} as it is allready as strong as the classic definition. Again however the often cited work by Khasminskii \cite{khasminskii2011stochastic} gives  a much weaker definition. Compare the following definitions for asymptotically attracting
\begin{align}
\mathbf P \left( \lim_{t\to\infty}\norm{x(t)}=0 \right)=1 \tag{\cite{deng2001stabilization}*{Definition 2.2}} \\
\lim_{x_0 \to 0}\mathbf P \left( \lim_{t\to\infty}\norm{x(t)}=0 \right)=1. \tag{\cite{khasminskii2011stochastic}*{Equation (5.15)}}
\end{align}
It is imperative when discussing stability that uniform bounds are achieved.
\end{rem}
}

\subsection{Diagonal Stability of Random Matrices}\label{sec:diag_random}

\begin{thm}\label{thm:asym_stab}
If $A_n\in\Re^{n\times n}$ is chosen as \ben[A_n]_{ij}\sim\frac{1}{\sqrt{(2+\delta)n}}\mathcal N(0,1),\quad i\neq j\een  for any $\delta>0$, and $[A_n]_{ii}=-1$, then $A_n$ is asymptotically almost surely diagonally stable.
\end{thm}
\begin{proof}
Consider the random matrix $\bar A_n\in\Re^{n\times n}$ defined as \ben[\bar A_n]_{ij}\sim\frac{1}{\sqrt{(2+\delta)n}}\mathcal N(0,1),\quad i \neq j,\een and $[\bar A_n]_{ii}=0$. Then it follows from \eqref{eq:var} that $\mathbf{Var}[\bar A_n]_{ij}=\frac{1}{(2+\delta)n}$ for $i\neq j$. Let $\bar B_n= \bar A_n + \bar A_n^ \mathsf{T}$, then it follows that $\mathbf{Var}[\bar B_n]_{ij}=\frac{2}{(2+\delta)n}$ when $i\neq j$ and $0$ otherwise. From Theorem \ref{thm:spec_bound} we have that \ben \begin{split}\sup_{1\leq i \leq n} \lambda_i (\bar B_n) &= 2\sqrt {\frac{2n}{(2+\delta)n }}(1+o(1)) \\ &< 2(1+o(1)) \end{split}\een as $n\to\infty$ asymptotically almost surely.
Noting that $A_n^{\mathsf T}+A_n \equiv \bar B_n -2 I_{n\times n}$ it follows that $A_n^{\mathsf T}+A_n <0$ asymptotically almost surely.
\end{proof}

\begin{cor}\label{cor:asym_stab}
If $A_n\in\Re^{n\times n}$ i is chosen as \ben[A_n]_{ij}\sim \mathcal N\left(0,\frac{1}{(2+\delta)n}\right),\quad i\neq j,\een  for any $\delta>0$, and $[A_n]_{ii}=-1$, then $A_n$ is asymptotically almost surely diagonally stable.
\end{cor}
%


%% file: clustering3.tex
\section{Clustering Analysis and Ordination Methods}\label{sec:clust}
In the following section clustering techniques are explored in detail as well as ordination techniques for visualizing data.

\subsection{Distance and Metrics}\label{sec:dm}

Let $x = [x_1,\, x_2,\, \ldots , x_n]^\mathsf{T} \in \Re^n$, then the Euclidian norm is defined as
\ben
\norm{x} = \left( \sum_{i=1}^{n}  x_i^2\right)^{1/2}.
\een
For a row vector the euclidian norm is similary defined. If no subscript is given with $\norm{\cdot }$  then we assume the Euclidian norm is being used. The following is the Euclidian distance function $d(x,y)=\norm{x-y}$.
%
%

A common distance metric used in ecology is the {\em Jensen-Shannon Distance} (JSD) metric   \cite{kullback1951information,61115,Jeffreys:1998aa}
\be\label{eq:jsd}
\mathrm{JSD}(y,z) \triangleq \sqrt{\tfrac{1}{2}\mathrm{JS}(y,z)},
\ee
where the Jensen-Shanon divergence, $
\mathrm{JS}(y,z) \triangleq \mathrm{KL}(y,\tfrac{1}{2}(y+z))+\mathrm{KL}(z,\tfrac{1}{2}(z+y))
$, is simply the symetrized version of the Kulback-Liebler directed divergence $
\mathrm{KL}(y,z) \triangleq \sum_{i=1}^p y_i\log \frac{y_i}{z_i} 
$. So as to not divide by zero, a pseudo count of 1e-10 is added to zero elements before performing the JSD.

\subsection{$k$-Medoids}\label{sec:medoids}

Assume that one has a collection of samples $X\in \Re^{n\times p}$ where $n$ is the  total number of samples and $p$ is the dimension of each sample. Let $X_i\in\Re^{1\times p}$, $i = 1,2,\ldots, n$, be the row vectors of $X$ as defined below
\be\label{methods:data}
X = \bb X_1 \\ X_2 \\ \vdots \\ X_n \eb.
\ee

We would like each $X_j$ to belong to a unique cluster within a collection of $k$ possible clusters ${\mathcal C}_1,{\mathcal C}_2,\ldots,{\mathcal C}_k$. The unique cluster that contains sample $X_j$ is denoted ${\mathcal C}(j;k)$. If sample $X_j$ is contained in cluster $3$ then ${\mathcal C}(j)={\mathcal C}_3$. If one is given an a-priori number of clusters then it is possible to  perform this task using the popular paradigm of $k$-medoids. The paradigm works as follows. Initially, $k$ samples are chosen at random as representative {\em medoids}. $k$ clusters are then constructed by associating other samples to the nearest medoid. Within each cluster all elements are tested so as to see if a different sample has a smaller within cluster sum of distances. The element with the smallest within cluster sum of distances is chosen as the new medoid for that cluster. This process is performed for each cluster. New clusters are constructed with the $k$ new medoids and the algorithm repeats again. The MATLAB command \verb*:kmedoids: performs $k$-medoids with a variety of different methods depending on the number of samples. When the number of samples is less than 3000 MATLAB implements $k$-medoids using the {\em Partitioning Around Medoids} (PAM) algorithm \cite{kaufman1987clustering}.

\subsection{Silhouette Value}\label{sec:sil}
 For each sample $X_j$ there is a corresponding silhouette value $s_j\in[-1,1]$ which is defined as follows
\ben
s_j(k) \triangleq \frac{b_j(k)-a_j(k)}{\max\{a_j(k),b_j(k)\}}
\een
where $a_j$ is the average dissimilarity between sample $j$ and all other samples within its own cluster, $b_j$ is the average dissimilarity between $X_j$ and the elements of the nearest cluster, and $k$ is the total number of apriori designated clusters. These two quantities are now formally defined as follows
\ben
a_j(k) \triangleq \frac{1}{\abs{{\mathcal C}(j;k)}-1}\sum_{\substack{X_i\in{\mathcal C}(j;k) \\ i\neq j}} d(X_i,X_j).
\een
Note that we have used $\abs{{\mathcal C}(j)}$ to denote the cardinality of ${\mathcal C}(j)$, the total number of samples in ${\mathcal C}(j)$. We similarly define the $b_j$ as 
\ben
b_j(k) \triangleq  \min_{{\mathcal C}_m\neq{\mathcal C}(j;k)}   \frac{1}{\abs{{\mathcal C}_m}}\sum_{ X_i\in{\mathcal C}_m}  d(X_i,X_j).
\een
For a given sample set and a given number of clusters $k\geq 2$ there is a corresponding $s_j(k)$. The Silhouette Index for a sample data set is then the maximum of the mean silhouette value for each total number of clusters.
\ben
\mathrm{SI}(X) \triangleq  \max_k \frac{1}{n} \sum_{j=1}^n s_j(k)
\een
The optimal number of clusters is then $\argmax_k \frac{1}{n} \sum_{j=1}^n  s_j(k)$. Silhouette analysis is performed in MATALAB using the command \verb*:silhouette:.

\subsection{Variance Ratio Criterion and the  Cali\'nski-Harabasz Index }\label{sec:vrc}
We now define the {\em Variance Ratio Criterion} {(VRC)} which holds for any distance function. When the Euclidian metric is used  the VRC is referred to as the {\em Cali\'nski-Harabasz} (CH) index \cite{Calinski:1974aa}.
As before, assume that a collection of samples has already been grouped into $k$ clusters. 
The VRC is defined as
\ben
\mathrm{VRC}(k) \triangleq \frac{\mathrm{BG}(k)}{\mathrm{WG}(k)}\frac{n-k}{k-1}
\een
where BG is the {\em Between Group} variance and WG is the {\em Within Group} variance defined below,
\ben\begin{split}
\mathrm{BG}(k) & \triangleq \sum_{j=1}^k \frac{1}{ \abs{\mathcal C_j}\sum_{m>j}^k \abs{\mathcal C_m}} \sum_{ X_i\in{\mathcal C}_j}  \sum_{\substack{X_\ell \in{\mathcal C}_m \\ m> j}} d(X_i,X_\ell)^2 \\
\mathrm{WG}(k) & \triangleq \sum_{j=1}^k  \frac{2}{\abs{\mathcal C_j}(\abs{\mathcal C_j}-1)} \sum_{\substack{ i=1 \\ X_i\in{\mathcal C}_j}}^n  \sum_{\substack{ \ell>i \\ X_\ell\in{\mathcal C}_j}}^n d(X_i,X_\ell)^2. 
\end{split}
\een

\subsection{Principle Coordinate Analysis}

The purpose of  {\em Principle Coordinates Analysis} (PCoA) is to represent a collection of high dimensional data in a lower dimension. Once again
assume that one has a collection of samples $X\in \Re^{n\times p}$ where $n$ is total number of samples and $p$ is the dimension of each sample. Let $X_i\in\Re^{1\times p}$, $i = 1,2,\ldots, n$ be the row vectors of $X$ as defined in \eqref{methods:data}. The question  answered in this section is how one obtains a $Y\in \Re^{n\times k}$, $k\leq n$  with $Y_i\in\Re^{1\times k}$, $i = 1,2,\ldots, n$  defined as follows
\ben
Y = \bb Y_1 \\ Y_2 \\ \vdots \\ Y_n \eb
\een
that is a faithful representation of $X$. We begin by defining the dissimilarity between samples $i$ and $j$ as 
$
\delta_{ij} = d(X_i,X_j).
$
Then the goal of this method is to find $Y$ such that $d(Y_i,Y_j)$ is similar to $d(X_i,X_j)$ for the distance measure of interest. Let $D = - \frac{1}{2}(\delta^2_{ij})_{1\leq i,j\leq n} $ and 
\ben
B = (I_{n\times n} - {n}^{-1}\mathbf 1_n \mathbf 1_n^\mathsf{T})D (I_{n\times n} - {n}^{-1}\mathbf 1_n \mathbf 1_n^\mathsf{T}).
\een
where $\mathbf 1_n$ is an $n$-dimensional column vector with each entry equal to 1.
The $n\times k$ dimensional representation of the $n\times p$ sample data is then \cite[Chapter 5]{seber}
\ben
Y = [  q_1\sqrt{\lambda_1},\, q_2\sqrt{\lambda_2},\,     \dotsc, \, q_k\sqrt{\lambda_k}].
\een
where $B = Q \Lambda Q^{-1}$ is the eigenvalue decomposition with eigenvalues $\lambda_i\in\Re$, and normalized eigenvectors $q_i\in \Re^{n}$ for $i=1,2,\ldots,n$ with 
$Q= [ q_1,\, q_2,\, \ldots,\, q_n]$ and $[\Lambda]_{ii} = \lambda_i$ the diagonal eigenvalue matrix. Due to the fact that $B$ is symmetric all eigenvalues and eigenvectors will be real valued. It is furthermore assumed that the eigenvalues are arranged such that $\lambda_1 \geq \lambda_2 \geq \cdots \geq \lambda_n$. When the Euclidian distance function is used $B$ is always positive semi-definite. However, it may be possible to have negative eigenvalues when other distance functions are used and thus it is necessary to check and make sure that that $\lambda_i$ for each $i \in \{1,2,\ldots,k\} $ is positive for the $k$-dimensional space of interest. The command that performs this task in MATLAB is \verb:cmdscale:. MATLAB also rotates the data, but that has no affect on the dissimilarity measure, as rotation is distance preserving in Euclidian space.

\subsection{Principle Component Analysis}\label{sec:pca}
{\em Principle Component Analysis} (PCA) is similar in spirit to PCoA in that one wishes to represent a set of data in a lower dimension.
Assume that one has a collection of samples $X\in \Re^{n\times p}$ where $n$ is total number of samples and $p$ is the dimension of each sample. As before the goal is to find a  $Y\in \Re^{n\times k}$, $k\leq p$  that is representative of the original data. Assume without loss of generality that the columns of $X$ have mean zero. Let
\ben
 X = U \Sigma W^T
\een
be the singular value decomposition of $X$ where $\Sigma\in \Re^{n\times p}$ is a diagonal matrix with the singular values $\sigma_i = [\Sigma]_{ii}$ where $\sigma_i \geq \sigma_j$ for $i< j$.  The columns of $U \in \Re^{n\times n}$ are the left singular vectors and the columns of $W\in \Re^{p\times p}$ are the right singular vectors. The reduced order representation is then defined as 
\ben
Y = X W_{1:n,1:k}
\een
where $W_{1:n,1:k}$ contains only the first $k$ columns of $W$. Note that $W_{1:n,1:k}$ can be thought of as a right hand side projection operator and can be used to project subsequent measurements into pre-existing principle components. Also, note that when the Euclidean distance is used in PCoA it is equivalent to PCA. The command that performs this task in MATLAB is \verb:pca:.

%% file: model3.tex
\section{Modeling}

\begin{figure}[b!]

\psfrag{t1}[cl][cl][1]{\bf Universal Dynamics}
\psfrag{t2}[cl][cl][1]{\bf Multiple Sets of Dynamics }

\psfrag{a1}[cc][cc][1]{$\mathbf S,\mathbf A,\mathbf r$}
\psfrag{a2}[cc][cc][.8]{$S^{[1]},A^{[1]},r^{[1]}$}
\psfrag{a3}[cc][cc][.8]{$S^{[2]},A^{[2]},r^{[2]}$}
\psfrag{a4}[cc][cc][.8]{$S^{[3]},A^{[3]},r^{[3]}$}
\psfrag{a5}[cc][cc][.8]{$S^{[q]},A^{[q]},r^{[q]}$}
\psfrag{a6}[cc][cc][1]{$\mathrm{LC}_{1}$}
\psfrag{a7}[cc][cc][1]{$\mathrm{LC}_{2}$}
\psfrag{a8}[cc][cc][1]{$\mathrm{LC}_{3}$}
\psfrag{a9}[cc][cc][1]{$\mathrm{LC}_{q}$}

\psfrag{b1}[cc][cc][1]{$\mathbf S,\mathbf A_a,\mathbf r_a$}
\psfrag{b2}[cc][cc][.8]{$S^{[1]},A^{[1]},r^{[1]}$}
\psfrag{b3}[cc][cc][.8]{$S^{[2]},A^{[2]},r^{[2]}$}
\psfrag{b4}[cc][cc][.8]{$S^{[3]},A^{[3]},r^{[3]}$}
\psfrag{b5}[cc][cc][.8]{$S^{[q]},A^{[q]},r^{[q]}$}
\psfrag{b6}[cc][cc][1]{$\mathrm{LC}_{1}$}
\psfrag{b7}[cc][cc][1]{$\mathrm{LC}_{2}$}
\psfrag{b8}[cc][cc][1]{$\mathrm{LC}_{3}$}
\psfrag{b9}[cc][cc][1]{$\mathrm{LC}_{q}$}

\psfrag{c0}[cc][cc][1]{$\mathbf S$}
\psfrag{c1}[cc][cc][1]{$\mathbf A_1,\mathbf r_1$}
\psfrag{c2}[cc][cc][1]{$\mathbf A_2,\mathbf r_2$}
\psfrag{c3}[cc][cc][1]{$\vdots$}
\psfrag{c4}[cc][cc][1]{$\mathbf A_\ell,\mathbf r_\ell$}
\psfrag{c5}[cc][cc][1]{$\mathbf A_5,\mathbf r_5$}
\psfrag{c6}[cc][cc][1]{$\mathrm{LC}_{1}$}
\psfrag{c7}[cc][cc][1]{$\mathrm{LC}_{2}$}
\psfrag{c8}[cc][cc][1]{$\mathrm{LC}_{3}$}
\psfrag{c9}[cc][cc][1]{$\mathrm{LC}_{q}$}

\centering
  \includegraphics[width=3in]{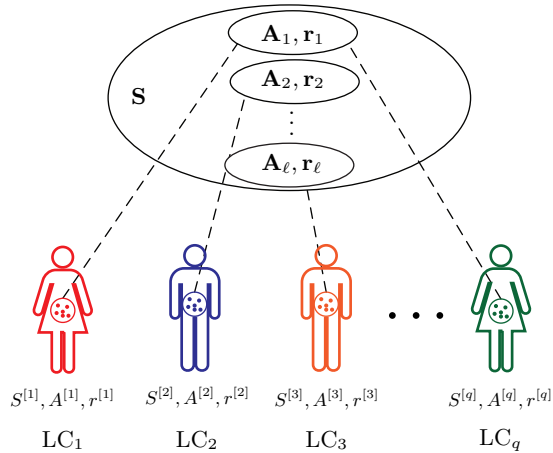}
  \caption[Modeling Paradigms]{{\em Modeling Paradigms}.  Two different modeling paradigms are presented. In the {\em universal} paradigm there is a universal list of species $\mathbf S$, a universal interaction matrix $\mathbf A$, containing all the pairwise interaction strengths, and a universal growth rate vector $\mathbf r$. Then the dynamics of each {Local Community} LC is determined by the collection of species in that community. In the {\em multiple set} paradigm the dynamics are not only determined by the collection of species present but the dynamics are not universal and come from a collection of possible dynamics. In the example above $\mathrm{LC}_1$ dynamics are derived from the first dynamic pair $(\mathbf A_1, \mathbf r_1)$, $\mathrm{LC}_2$ dynamics are derived from $(\mathbf A_2, \mathbf r_2)$, $\mathrm{LC}_3$ dynamics are derived from $(\mathbf A_\ell, \mathbf r_\ell)$, and  $\mathrm{LC}_q$ dynamics  are derived from the dynamic pair $(\mathbf A_1, \mathbf r_1)$.} 
  \label{fig:modeling}
\end{figure}

Two approaches for modeling individual human microbial communities are now presented. The first model is the same as that presented in the main text and from this point forward is referred to as the {\em universal model}. This model assumes that all species interact in a universal manner independent of the host. The second model allows for there to be multiple possible interaction strengths and growth rates for species. This model will be referred to as the  {\em multiple set model}. We use Figure \ref{fig:modeling} as a visual reference when discussing the modeling paradigms.

\subsection{Universal Model}\label{sec:u}
Consider a universal species pool indexed by a set of integers $\mathbf S = \{1,\ \ldots, \ n\}$, a global ${n\times n}$ matrix $\mathbf A$ representing all possible pairwise interactions between species, and a universal vector $\mathbf r$ of size $n$ containing the growth rates for all the  $n$ species. The global variables for our ecological system are completely defined by the triple $(\mathbf S, \mathbf A, \mathbf r)$.
Then $q$ {\em Local Communities} (LCs)  are defined by sets $S^{[\nu]}$, which are subsets of $\mathbf S$ denoting the specific microbes present in  $\mathrm{LC}_\nu$, $\nu=1,\ldots, q$. For simplicity we assume that each LC contains only $p$ species ($p\leq n$), randomly selected from the universal species pool. The GLV dynamics for each LC is given by 
\be\label{eq:glv_environment}
\mathrm{LC}_\nu: \quad \dot x^{[\nu]}(t) = \diag\left(x^{[\nu]}(t)\right) \left(r^{[\nu]}+A^{[\nu]}x^{[\nu]}(t)\right)\ee
where the LC specific interaction matrix and growth vector are defined as $\aa \triangleq \mathbf A_{S^{[\nu]},S^{[\nu]}}$ and $\rr \triangleq \mathbf r_{S^{[\nu]}}$, respectively.
That is, $\aa$ is obtained from $\mathbf A$ by only taking the rows and columns of $\mathbf A$ that are contained in the set $S^{[\nu]}$. A similar procedure is performed in order to obtain $\rr$. Finally there is a map $m_\nu$ that takes the abundances $\xx$ in the index $\ss$ and carries them to the universal  index in $\mathbf S$ giving the vector $\mathbf x^{[\nu]} = m_\nu\left(\xx\right)$ which results in $\mathbf x^{[\nu]}_{S^{[\nu]}_j} = x_j^{[\nu]}$ for $j=1,\ldots, p$ and $\mathbf x^{[\nu]}_j=0$ for $j$ in $\mathbf S \setminus S^{[\nu]}$. 
This modeling procedure is inspired by \cite{Costello08062012}. A toy example of how this model is used to construct a single local community is now given

\begin{exmp}
Consider a global set of 4 species and thus $\mathbf S = \{1,2,3,4\}$. Let 
\ben
\mathbf A = 
\begin{bmatrix}
 	{-1} & { 0.1} & 0.4 & -0.1 \\
  	{ 0.7} & -1 & 0  & {0.4} \\
   	 -0.1 & 0.7 & -1 & 0 \\
  	-0.8 & {-0.2} & {0.4} & -1\end{bmatrix} \quad \text{and} \quad
\mathbf r = 
	\bb 0.2 \\ 0.4 \\ 0.5 \\0.4 \eb.  \een 
Then, if the first local community has the species list $S^{[1]}=\{1,3,4\}$ then  it follows that
\ben
 A^{[1]} = 
\begin{bmatrix}
 	{-1} &  0.4 & -0.1 \\
   	 -0.1 &  -1 & 0 \\
  	-0.8 &   {0.4} & -1\end{bmatrix} \quad \text{and} \quad
 r^{[1]}  = 
	\bb 0.2 \\ 0.5 \\0.4 \eb.  \een 
\end{exmp}

\subsection{Mutiple Set Model}\label{sec:mm}
As before there is a universal set of species $\mathbf S = \{1, \ldots, n\}$, but now there are $\ell$ pairs of possible global dynamics $\{( \mathbf A_1, \mathbf r_1), ( \mathbf A_2, \mathbf r_2), \ldots, ( \mathbf A_\ell, \mathbf r_\ell)\}$. The $q$ LCs  are defined by sets $S^{[\nu]}$, $\nu= 1, \ldots, q $  and a map $w:  \{1, 2, \ldots, q \}\to \{1, 2, \ldots, \ell\}$ which determines which model pair the LC is derived from. The GLV dynamics for each LC are given by 
\eqref{eq:glv_environment}
where  $\aa \triangleq [\mathbf A_{w(\nu)}]_{S^{[\nu]},S^{[\nu]}}$ and $\rr \triangleq [\mathbf r_{w(\nu)}]_{S^{[\nu]}}$, respectively.
That is, $\aa$ is obtained from $\mathbf A_{w(\nu)}$ by only taking the rows and columns of $\mathbf A_{w(\nu)}$ that are contained in the set $S^{[\nu]}$. A similar procedure is performed in order to obtain $\rr$. The map $m_\nu$ that takes the abundances $\xx$ in the index $\ss$ and carries them to the universal  index  $\mathbf S$ is defined as before $\mathbf x^{[\nu]} = m_\nu(\xx )$. It is worth noting that results from this modeling paradigm are not presented in the main text, because the emergence of community types in this case is trivial (see \S\ref{sec:mmodelsim}) \cite{Koren:2013aa}.



%% file: simulations3.tex
\section{Simulation Results and Analysis}
We are finally in a position to address the main topic of this work. Revealing properties that allow for community types to arise.
Four case studies are now presented. The first case study will use the multiple set model paradigm in Seciton \ref{sec:mm}, and it will be shown that under this paradigm that clustering of the steady states occurs trivially. The subsequent three case studies use the universal model from Section \ref{sec:u}. The case studies explore  heterogeneity (in terms of interaction strength and network topology), mean degree of the network, and community overlap and how these affect the existence of community types for our dynamics of interest.

\subsection{Multiple Set Models}
\label{sec:mmodelsim}
\begin{figure}[b!]
\centering
  \includegraphics[width=3.5in]{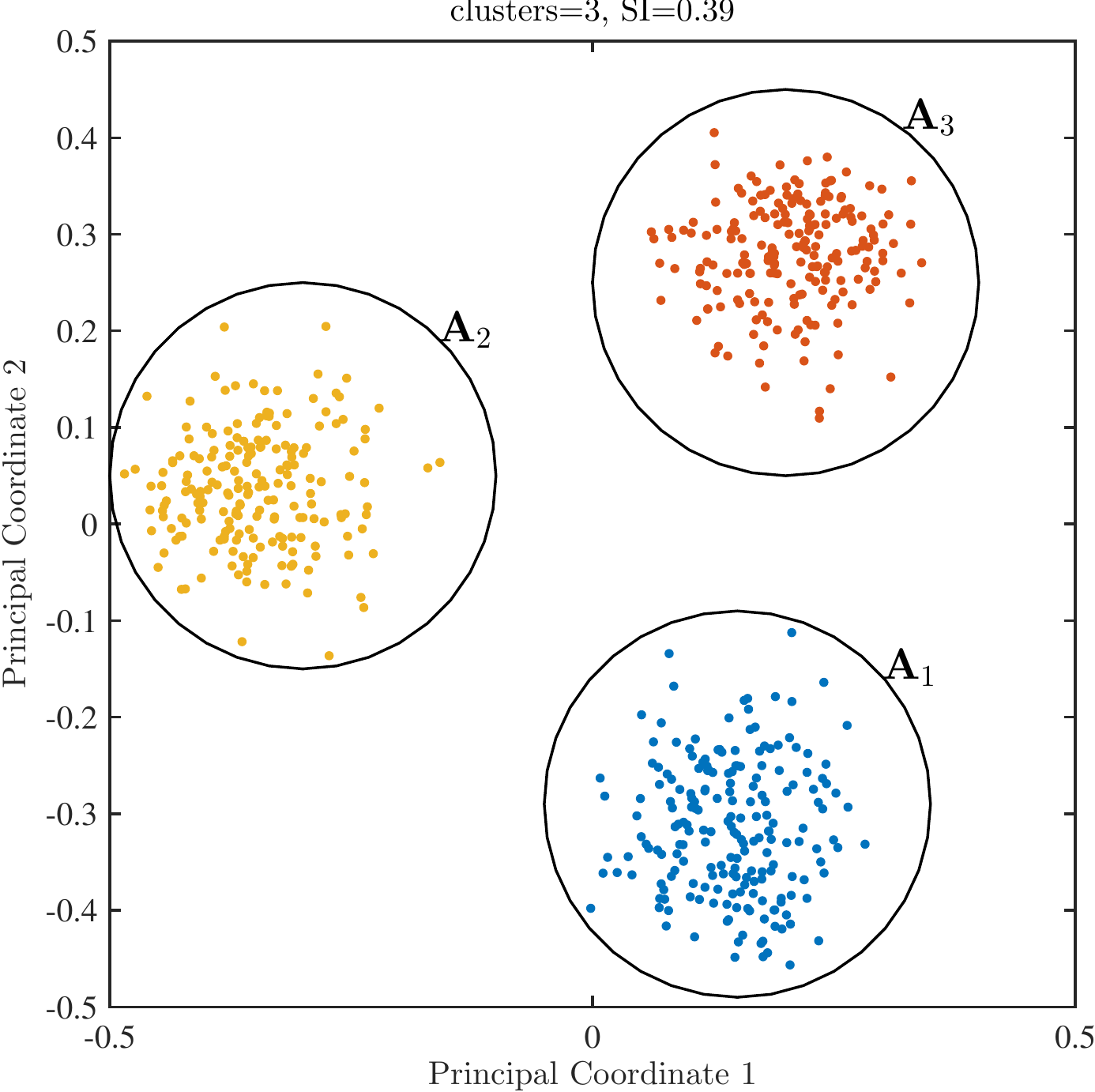}
  \caption[Principle coordinates for multiple set model case study]{Principle coordinates for multiple model case study.} 
  \label{fig:multiple_model}
\end{figure}

For the multiple set model study three universal pairs $\{( \mathbf A_1, \mathbf r_1), ( \mathbf A_2, \mathbf r_2),  ( \mathbf A_3, \mathbf r_3)\}$ were considered. For this study there was a total of 100 global species. Each $\mathbf A_i$ was a 100 by 100 matrix with values drawn from a normal distribution with mean 0 and variance $0.0049$. Then the diagonal elements of each matrix are set to -1, that is $[\mathbf A_i]_{jj}=-1$, $i=1,2,3$,  $j = 1,2,\ldots,100$. Note that with the above selection, the variance of the off diagonal elements of $\mathbf A_i$ satisfy the following bound $0.0049 < \frac{1}{2\cdot 100}$, which from Corollary \ref{cor:asym_stab} is the sufficient condition on the variance for asymptotic almost sure diagonal stability when $n=100$. Also, note the interaction network is implicitly a complete graph without any structural heterogeneity. Finally, each $\mathbf r_i$ had elements drawn from a uniform distribution between 0 and 1.

From the three pairs above 600 local communities were constructed, 200 from each universal pair. Each of the local communities contained 80 species. The dynamics were then simulated for 100 seconds with initial conditions drawn from $\mathcal U(0,1)$. The abundances of the species in the communities were then normalized, and the relative abundances of the 600 local communities were clustered using $k$-medoids from Section \ref{sec:medoids} and silhouette indexed as defined in Section \ref{sec:sil},  each using the Jensen-Shannon metric in \eqref{eq:jsd}. A principle coordinate plot of the results is given in Figure \ref{fig:multiple_model} with the elements colored according to cluster assignment from $k$-medoids. It is clear that there are three clusters. Furthermore each cluster exactly coincides with the universal $\mathbf A_i$ that the local community dynamics were obtained from.

\subsection{Universal Model}\label{sec:smodel}

For the three universal case studies the global interaction matrix is defined as \ben \mathbf A =  \mathbf N  \mathbf H\circ  \mathbf G s,\een
which contains four components. 
\begin{enumerate}[label=(\roman*)]
\item $ \mathbf N\in\Re^{n\times n}$ is the nominal component where each element is sampled from either a normal distribution or a uniform distribution. 
\item The matrix $\mathbf H$ is a diagonal matrix that captures the overall interaction heterogeneity of different species. When interaction strength heterogeneity is employed  the diagonal elements of $ \mathbf H$ are drawn from a power-law distribution \eqref{eq:pldist} with exponent $-\alpha$, $[ \mathbf H]_{ii}\sim \mathcal P(\alpha)$, which are subsequently normalized so that the mean of the diagonal is equal to 1. Without interaction heterogeneity $\mathbf H$ is simply the identity matrix.
\item The matrix $ \mathbf G$ is the adjacency matrix  of the underlying ecological network: $[\mathbf G]_{ij}=1$ if species $i$ is affected by the presence of species $j$ and $0$ otherwise. When the underlying network is a complete digraph all elements in $\mathbf G$ are equal to 1. For details on the construction of $\mathbf G$ when the underlying network is Erd\H{o}s-R\'enyi or a power-law digraph see Sections \ref{sec:ER} and \ref{sec:powerdigraph} respectively.
\item The last element $s$ is simply a scaling factor between 0 and 1.  \end{enumerate}
As before we set $[\mathbf A]_{ii} = -1$  to ensure one of the necessary conditions for diagonal stability of a matrix is satisfied (Theorem \ref{thm:remove} and Corollary \ref{cor:remove}). Finally, the elements in the global growth rate vector are defined as $$\mathbf r=\mathbf h \circ \mathbf n$$ where $\mathbf n$ is the nominal component taken from a uniform distribution, and $\mathbf h$ captures the growth rate heterogeneity. When there is growth rate heterogeneity $\mathbf h$ is drawn from a power-law distribution with exponent $-\alpha$ and subsequently normalized to have a mean of 1. Without growth rate heterogeneity $\mathbf h$ is simply a column vector of ones. Note that $\mathbf h$ is not included in the main text, and is only used in one scenario of one case study within this section.

\subsubsection{Universal Model: Heterogeneity Study}\label{sec:hetstudy}
Next is a systematic study of heterogeneity and its effects on the dynamics in the universal model. Eight different scenarios were tested in this study. For each of the following scenarios the number of global species is $n=100$. Table \ref{tab:parameters1} outlines the differences between the scenarios.

\begin{enumerate}[label=Scenario \arabic*.,leftmargin=*, labelindent=\parindent]
\item $[\mathbf N]_{ij}\sim \mathcal U(-0.5,0.5)$, $[ \mathbf H]_{ii}\sim \mathcal P(\alpha)$ where $\alpha \in [1.2, 7]$ and subsequently normalized to have a mean of 1, $\mathbf G$ is the adjacency matrix for a complete digraph and thus all entries are equal to 1, and the scaling factor is set to  $s=0.07$. There is no growth rate heterogeneity and thus the column vector $\mathbf h$ is all ones and finally $[\mathbf n]_i\sim\mathcal U(0,1)$.
\item The same as Scenario 1 but with $\mathbf [\mathbf N]_{ij}\sim \mathcal N(0,1)$ and $s=0.07$.
\item $[\mathbf N]_{ij}\sim \mathcal U(-0.5,0.5)$, $[ \mathbf H]_{ii}\sim \mathcal P(\alpha)$ where $\alpha \in [1.2, 7]$ and subsequently normalized to have a mean of 1, $\mathbf G$ is the adjacency matrix for an Erd\H{o}s-R\'enyi digraph with a mean out-degree of $10$, and the scaling factor is set to  $s=0.5$. There is no growth rate heterogeneity and thus the column vector $\mathbf h$ is all ones and finally $[\mathbf n]_i\sim\mathcal U(0,1)$. For more details on Erd\H{o}s-R\'enyi digraphs see \S \ref{sec:ER}.
\item The same as Scenario 3 but with $[\mathbf N]_{ij}\sim \mathcal N(0,1)$ and $s=0.1$.
\item $[\mathbf N]_{ij}\sim \mathcal N(0,1)$, $ \mathbf H$ is the identity matrix, $\mathbf G$ is the adjacency matrix for a digraph with the out-degree drawn from a power-law distribution $\mathcal P(\alpha)$ with a mean out-degree of 10 where $\alpha \in [1.2, 7]$, and the scaling factor is set to  $s=0.2$. There is no growth rate heterogeneity and thus the column vector $\mathbf h$ is all ones and  $[\mathbf n]_i\sim\mathcal U(0,1)$.  For more details on power-law out-degree digraphs see \S \ref{sec:powerdigraph}.
\item $[\mathbf N]_{ij}\sim \mathcal N(0,1)$, $[ \mathbf H]_{ii}\sim \mathcal P(\alpha)$ where $\alpha \in [1.2, 7]$ and subsequently normalized to have a mean of 1, $\mathbf G$ is the adjacency matrix for a digraph with the out-degree drawn from a power-law distribution $\mathcal P(\alpha)$ with a mean out-degree of 10 where $\alpha \in [1.2, 7]$, and the scaling factor is set to  $s=0.1$. The power-law degree distribution and interaction strength heterogeneity are uncoupled. There is no growth rate heterogeneity and thus the column vector $\mathbf h$ is all ones and  $[\mathbf n]_i\sim\mathcal U(0,1)$. For more details on power-law out-degree digraphs see \S \ref{sec:powerdigraph}.
\item $[\mathbf N]_{ij}\sim \mathcal N(0,1)$, $[ \mathbf H]_{ii}\sim \mathcal P(\alpha)$ where $\alpha \in [1.2, 7]$ and subsequently normalized to have a mean of 1, $\mathbf G$ is the adjacency matrix for a digraph with the out-degree drawn from a power-law distribution $\mathcal P(\alpha)$ with a mean out-degree of 10 where $\alpha \in [1.2, 7]$, and the scaling factor is set to  $s=0.02(\alpha+1)$. The power-law degree distribution and interaction strength heterogeneity are coupled so that the node with the highest out-degree is also the node with the largest interaction strength scaling. There is no growth rate heterogeneity and thus the column vector $\mathbf h$ is all ones and  $[\mathbf n]_i\sim\mathcal U(0,1)$. For more details on power-law out-degree digraphs see \S \ref{sec:powerdigraph}.
\item $[\mathbf N]_{ij}\sim \mathcal N(0,1)$, $ \mathbf H$ is the identity matrix, $\mathbf G$ is the adjacency matrix for an Erd\H{o}s-R\'enyi digraph with a mean out-degree of $10$, and the scaling factor is set to  $s=0.1$. There is growth rate heterogeneity and thus the column vector $\mathbf h$ has elements drawn from a power-law distribution $\mathcal P(\alpha)$ and subsequently normalized to have a mean of  1  and  $[\mathbf n]_i\sim\mathcal U(0,1)$. For more details on Erd\H{o}s-R\'enyi digraphs see \S \ref{sec:ER}.
\end{enumerate}

For each of the eight scenarios above and for every $\alpha$, ${q=500}$ local communities were generated each with $p=80$ species selected at random from the ${n=100}$ global species. The dynamics as described above following the modeling paradigm in Section \ref{sec:u} were then simulated for 100 seconds with initial conditions drawn from $\mathcal U(0,1)$. If any of the 500 simulations crashed due to instability or if the norm of the terminal discrete time derivative was greater than $0.01$ then that local community was excluded from the rest of the study. Those simulations that finished without crashing and with small terminal discrete time derivative were deemed steady. Less than $1\%$ of simulations were deemed unstable. The abundances of the species in the communities were then normalized, and the relative abundances of the 500 local communities were clustered using $k$-medoids from Section \ref{sec:medoids} and silhouette indexed as defined in Section \ref{sec:sil},  each using the Jensen-Shannon distance metric in \eqref{eq:jsd}.

The results from the above scenarios are presented in Figures \ref{fig:parameter1}-\ref{fig:parameter8}. The first plot is a comprehensive clustering analysis of the steady state values obtained from the simulations. The $x$-axis denotes the heterogeneity value $\alpha$. The box plots are the silhouette values pertaining to the number of clusters for which the silhouette index was defined. The total number of clusters pertaining to the silhouette index is denoted on the top $x$-axis. The second row is a principle coordinate analysis of the  steady state values obtained at three different heterogeneity values. Clusters are color coded to match the optimal clustering from $k$-medoids. The third row of the figure plots 
\ben \max_{i,j} \mathrm{real}\left(\lambda_i (A^{[j]})\right)\een
as a function of $\alpha$. The fourth row is a plot of $\lambda_i (A^{[j]})$ for $i\in \{1,2,\ldots, 80\}$ and  $j\in \{1,2,\ldots, 500\}$  at three values of $\alpha$.

Figures \ref{fig:parameter1} and \ref{fig:parameter2} show that regardless of whether the nominal component is drawn from a uniform or a normal distribution the increase in interaction strength heterogeneity (decreasing $\alpha$) leads to steady state clustering in the data. Figures \ref{fig:parameter3} and \ref{fig:parameter4} illustrate that the same phenomenon also holds for Erd\H{o}s-R\'enyi digraphs as well. When the underlying graph topology follows a power-law degree distribution, but there is no interaction strength heterogeneity, clustering of steady states is not observed, see Figure \ref{fig:parameter5}. Figure \ref{fig:parameter6} illustrates the fact that when there is interaction strength heterogeneity and network degree heterogeneity it is possible to have clustering of steady states when $\alpha$ is in the range $[3,1]$. However the trend is not smooth and is inconsistent, as compared to Figures \ref{fig:parameter1}-\ref{fig:parameter4}. This is due to the fact that when the underlying interaction network is being constructed, there is no guarantee that the high-degree node will also have a large interaction strength. For instance, if a node with no out edges is randomly selected to have high interaction strength scaling, then the impact of that node on the rest of the nodes is still zero. When the interaction strength heterogeneity is coupled with the out-degree for a power-law out-degree digraph then the trends from Figures \ref{fig:parameter1}-\ref{fig:parameter4} are recovered. Finally, if the growth rates of the species are derived from a power-law distribution, then clustering of steady states also occurs as $\alpha$ decreases.

Rows three and four illustrate the spectrum of  $A^{[j]}$, and are included so that we can infer the asymptotic stability of the system for certain paradigms. Note that regardless of whether the nominal interactions are drawn from a normal distribution or a uniform distribution when $\alpha$ is large the spectrum represents a uniform disk in the complex plain as predicted by Theorem \ref{thm:circ}. Furthermore, for scenario 2 with $s=0.07$ and $[\mathbf N]_{ij}\sim \mathcal N(0,1)$ it follows that $\mathbf{Var} [\mathbf A]_{ij} < \frac{1}{\sqrt{2n}}$ for $n=100$. From Theorem \ref{thm:asym_stab} it follows that $\mathbf A$ is diagonally stable, in a probabilistic sense. Then invoking Theorem \ref{thm:remove} we know that any principle minor of $\mathbf A$ is diagonally stable as well. Therefore, in Scenario 2 for large $\alpha$ each $A^{[j]}$ is diagonally stable, in a probabilistic sense. 

Also, for all of the scenarios with interaction heterogeneity all of the eigenvalues of $\mathbf A$,  and consequently $A^{[j]}$, converge to $-1$ as $\alpha$ tends to 1. In the limit of $\alpha$ tending to 1, only one of the columns of $\mathbf A$ has non-zero values off the diagonal. Therefore, in the limit of $\alpha$ tending to 1  the following inequality holds $\mathbf A^\mathsf T P+ P\mathbf A<0$ where $P=[1, \epsilon, \ldots, \epsilon]^\mathsf T$ and $\epsilon$ is sufficiently small.\footnote{Without loss of generality we have assumed that the first column of $\mathbf A$ is the highly weighted column.}  Thus $\mathbf A$ is diagonally stable in the limit of $\alpha$ tending to 1. Therefore, for low interaction strength heterogeneity and for high interaction strength heterogeneity one of the necessary conditions for uniform asymptotic stability in the positive orthant is satisfied, see Theorem \ref{thm1}. Note that even when the $A^{[j]}$ are not diagonally stable it does not imply that the state trajectory $x^{[j]}(t)$ is unstable.


\subsubsection{Universal Model: Sparsity Study}\label{sec:sparse_study}

Next is a study where the mean degree of the Erd\H{o}s-R\'enyi digraph along with the interaction strength heterogeneity is varied. Table \ref{tab:sparse} outlines the differences between the scenarios. The details of the study are as follows: $[\mathbf N]_{ij}\sim \mathcal U(-0.5,0.5)$, $[ \mathbf H]_{ii}\sim \mathcal P(\alpha)$ where $\alpha \in [1.2, 7]$ and subsequently normalized to have a mean of 1, $\mathbf G$ is the adjacency matrix for an Erd\H{o}s-R\'enyi digraph with a mean out-degree of \ben d \in\{1,\, 3,\, 5,\, 7,\, 9,\, 11,\, 13,\, 15 ,\, 17,\, 19\},
\een and the scaling factor is set to  $s=1/\sqrt{d}$. There is no growth heterogeneity and thus the column vector $\mathbf h$ is all ones and finally $[\mathbf n]_i\sim\mathcal U(0,1)$. For each of the ten scenarios above and for every $\alpha$ the same procedure as in \S \ref{sec:hetstudy} was  carried out and the results are shown in Figures \ref{fig:sparse1}-\ref{fig:sparse10}. 

From Figures \ref{fig:sparse1}-\ref{fig:sparse10} it can be concluded that so long as the mean out-degree of the ER digraph is greater than 2 the steady states of the GLV model increase in SI as $\alpha$ decreases. That is, the same trends as observed in the previous study hold, so long as the ER digraph is connected. When the mean degree is 1 for an ER graph it is very unlikely that the graph will be connected \cite{erdos}. Thus for a digraph with mean out-degree 1, it is even more unlikely that it will be connected. When the underlying digraph $\mathbf G$ has many isolated nodes then the scaling of interaction strengths does not influence as many other species in the GLV system and thus clustering is not observed.

\subsubsection{Universal Model: Community Size Study}\label{sec:com_study}

In all previous studies each local community contained 80 species. For this study the size of the local communities take on values in the following set 
\ben\label{eq:com_size}
p \in\{ 100,\ 99,\ 95,\ 90,\ 80,\ 70,\ 60,\ 50\}.
\een
Details of the study are as follows: $[\mathbf N]_{ij}\sim \mathcal U(-0.5,0.5)$, $[ \mathbf H]_{ii}\sim \mathcal P(\alpha)$ where $\alpha \in [1.2, 7]$ and subsequently normalized to have a mean of 1, $\mathbf G$ is the adjacency matrix for an Erd\H{o}s-R\'enyi digraph with a mean out-degree of $10$, and the scaling factor is set to  $s=0.5$. There is no growth heterogeneity and thus the column vector $\mathbf h$ is all ones and finally $[\mathbf n]_i\sim\mathcal U(0,1)$. Table \ref{tab:remove} outlines the differences between the scenarios. For each of the eight scenarios above and for every $\alpha$ the same procedure as in \S \ref{sec:hetstudy} was  carried out and the results are shown in Figures \ref{fig:remove1}-\ref{fig:remove8}.

These results show that the trends observed in the earlier studies still hold when the community sizes are varied between 95 and 50 species, Figures \ref{fig:remove3}-\ref{fig:remove8}. As the community sizes approach 50 the trend of increased clustering with increased heterogeneity is less significant. When the number of species in the LCs approaches the number of species in the meta-community, 100, the results do not follow the same trends as before and the Silhouette Indices are near 1, independent of the interaction strength heterogeneity. In Figure \ref{fig:remove1} all the simulations are identical, each LC has the same 100 species but with different initial conditions. Due to asymptotic stability all of the simulations converge to the same steady state (only those simulations for $\alpha$ between 2.2 and 3 have the potential to be unstable). Even though the Silhouette Indices are near 1 across the heterogeneity spectrum, all of the steady state values are within  $10^{-3}$ to $10^{-2}$ in terms of the first two principle coordinates. When all of the LCs differ by only one species, Figure \ref{fig:remove2}, once again the Silhouette Indices are near 1, yet the PCoA illustrates that most of the samples are very close together with just a few outliers. Figures \ref{fig:remove1} and \ref{fig:remove2} illustrate the sometimes confounding results when clustering analysis and  PCoA are performed \cite{milligan1987methodology,Jain:1999:DCR:331499.331504}.


%% file: tables3.tex
\clearpage

\newcounter{case_num}

\begin{sidewaystable}[b!]
\caption{Parameter settings for heterogeneity case study.}\label{tab:parameters1}
\centering

\begin{tabular}{l | c c c c c c c}
Scenario &   \begin{tabular}{c}Growth Rate\\ Distribution  \end{tabular}& \begin{tabular}{c}Nominal Interaction\\ Distribution \end{tabular}&\begin{tabular}{c}Interaction \\Structure  \end{tabular}& \begin{tabular}{c}Mean \\  Degree  \end{tabular} & \begin{tabular}{c}Interaction Strength \\Heterogeneity  \end{tabular} & \begin{tabular}{c}Network Structure \\Heterogeneity  \end{tabular} &  \begin{tabular}{c}Growth Rate\\Heterogeneity  \end{tabular}\\
\hline
1 & uniform & uniform & complete  & 100 & $\alpha\in[1.2,\ 7]$ & none & none\\
2 & uniform & normal & complete &  100 & $\alpha\in[1.2,\ 7]$ & none& none\\
3 & uniform & uniform &  Erd\H{o}s-R\'enyi  &  10 &  $\alpha\in[1.2,\ 7]$& none& none\\
4 & uniform & normal &  Erd\H{o}s-R\'enyi &  10 &  ${\alpha\in[1.2,\ 7]}$ & none&none\\
5 & uniform & normal & power-law &  10 & none& ${\alpha\in[1.2,\ 7]}$ &none\\
6 & uniform & normal & power-law&  10 &  $\alpha\in[1.2,\ 7]$&  ${\alpha\in[1.2,\ 7]}$&none\\
7 & uniform & normal & power-law&  10 &  $\alpha\in[1.2,\ 7]$&  ${\alpha\in[1.2,\ 7]}$(coupled)&none\\
8 & uniform & normal & Erd\H{o}s-R\'enyi  & 100 & none & none & $\alpha\in[1.2,\ 7]$ \\
\end{tabular}
\end{sidewaystable}%

\begin{sidewaystable}[b!]
\caption{Parameter settings for sparsity study.}\label{tab:sparse}
\centering
\begin{tabular}{l | c c c c c c c}
Scenario &   \begin{tabular}{c}Growth Rate\\ Distribution  \end{tabular}& \begin{tabular}{c}Nominal Interaction \\ Distribution \end{tabular}&\begin{tabular}{c}Interaction \\Structure  \end{tabular}& \begin{tabular}{c}Mean \\  Degree  \end{tabular} & \begin{tabular}{c}Interaction Strength \\Heterogeneity  \end{tabular} & \begin{tabular}{c}Network Structure \\Heterogeneity  \end{tabular} &  \begin{tabular}{c}Growth Rate \\Heterogeneity  \end{tabular}\\
\hline
1 & uniform & uniform &  Erd\H{o}s-R\'enyi  &  1 &  $\alpha\in[1.2,\ 7]$& none& none\\
2 & uniform & uniform &  Erd\H{o}s-R\'enyi  & 3 &  $\alpha\in[1.2,\ 7]$& none& none\\
3 & uniform & uniform &  Erd\H{o}s-R\'enyi  &  5 &  $\alpha\in[1.2,\ 7]$& none& none\\
4 & uniform & uniform &  Erd\H{o}s-R\'enyi &   7 &  ${\alpha\in[1.2,\ 7]}$ & none&none\\
5 & uniform & uniform &  Erd\H{o}s-R\'enyi  &  9 &  $\alpha\in[1.2,\ 7]$& none& none\\
6 & uniform & uniform &  Erd\H{o}s-R\'enyi  &  11 &  $\alpha\in[1.2,\ 7]$& none& none\\
7 & uniform & uniform &  Erd\H{o}s-R\'enyi  &  13 &  $\alpha\in[1.2,\ 7]$& none& none\\
8 & uniform & uniform &  Erd\H{o}s-R\'enyi  &  15 &  $\alpha\in[1.2,\ 7]$& none& none\\
9 & uniform & uniform &  Erd\H{o}s-R\'enyi &   17 &  ${\alpha\in[1.2,\ 7]}$ & none&none\\
10 & uniform & uniform &  Erd\H{o}s-R\'enyi  &  19 &  $\alpha\in[1.2,\ 7]$& none& none\\
\end{tabular}
\end{sidewaystable}%

\begin{sidewaystable}[b!]
\caption{Parameter settings for community size study.}
\label{tab:remove}
\centering
\begin{tabular}{l | c c c c c c c}
Scenario &   \begin{tabular}{c}Growth Rate\\ Distribution  \end{tabular}& \begin{tabular}{c}Nominal Interaction\\ Distribution \end{tabular}&\begin{tabular}{c}Interaction \\Structure  \end{tabular}& \begin{tabular}{c}Mean \\  Degree  \end{tabular} & \begin{tabular}{c}Interaction Strength \\Heterogeneity  \end{tabular} & \begin{tabular}{c}Network/Growth \\Heterogeneity  \end{tabular} &  \begin{tabular}{c}Community \\Size  \end{tabular}\\
\hline
1 & uniform & uniform &  Erd\H{o}s-R\'enyi  &  10 &  $\alpha\in[1.2,\ 7]$& none& 100\\
2 & uniform & uniform &  Erd\H{o}s-R\'enyi  &  10 &  $\alpha\in[1.2,\ 7]$& none& 99\\
3 & uniform & uniform &  Erd\H{o}s-R\'enyi  & 10 &  $\alpha\in[1.2,\ 7]$& none& 95\\
4 & uniform & uniform &  Erd\H{o}s-R\'enyi  &  10 &  $\alpha\in[1.2,\ 7]$& none& 90\\
5 & uniform & uniform &  Erd\H{o}s-R\'enyi &   10 &  ${\alpha\in[1.2,\ 7]}$ & none& 80\\
6 & uniform & uniform &  Erd\H{o}s-R\'enyi  &  10 &  $\alpha\in[1.2,\ 7]$& none& 70\\
7 & uniform & uniform &  Erd\H{o}s-R\'enyi  &  10 &  $\alpha\in[1.2,\ 7]$& none& 60\\
8 & uniform & uniform &  Erd\H{o}s-R\'enyi  &  10 &  $\alpha\in[1.2,\ 7]$& none& 50\\
\end{tabular}
\end{sidewaystable}

%% file: figures3.tex
\setcounter{case_num}{1}
\forloop{case_num}{1}{\value{case_num} < 9}{
\begin{figure}[htbp]
\centering
  \includegraphics[width=5in]{codev6/figure/fig_SI_\arabic{case_num}-eps-converted-to.eps}\\
  \includegraphics[width=5in]{codev6/figure/fig_pca_growth_rate_\arabic{case_num}-eps-converted-to.eps}\\
  \includegraphics[width=5in]{codev6/figure/fig_eig_max_\arabic{case_num}-eps-converted-to.eps}\\
  \includegraphics[width=5in]{codev6/figure/eig_detail_\arabic{case_num}-eps-converted-to.eps}
\caption[Universal Model Heterogeneity Study Scenario \arabic{case_num}]{Universal Model Heterogeneity Study Scenario \arabic{case_num} in Table \ref{tab:parameters1}. The first plot is a comprehensive clustering analysis of the steady state values obtained from the Lotka-Volterra simulations. The $x$-axis denotes the heterogeneity value $\alpha$. The box plots are the silhouette values pertaining to the number of clusters for which the silhouette index (maximum over the number of clusters of the mean silhouette value for each given total number of cluster) was defined. The total number of clusters pertaining to the silhouette index is denoted on the top $x$-axis. The second row is a principle coordinate analysis of the  steady state values obtained at three different heterogeneity values. Clusters are color coded to match the optimal clustering from $k$-medoids. The third row of the figure plots $\max_{i,j} \mathrm{real}(\lambda_i (A^{[j]}))$ as a function of $\alpha$. The fourth row is a plot of $\lambda_i (A^{[j]})$for $i\in \{1,2,\ldots, 80\}$ and  $j\in \{1,2,\ldots, 500\}$  at three values of $\alpha$.}
\label{fig:parameter\arabic{case_num}}
\end{figure}
\clearpage
}

\setcounter{case_num}{1}
\forloop{case_num}{1}{\value{case_num} < 11}{
\begin{figure}[htbp]
\centering
  \includegraphics[width=5in]{codev6/figure/sparse_fig_SI_\arabic{case_num}-eps-converted-to.eps}
\\
  \includegraphics[width=5in]{codev6/figure/sparse_fig_pca_growth_rate_\arabic{case_num}-eps-converted-to.eps}
  \\
  \includegraphics[width=5in]{codev6/figure/sparse_fig_eig_max_\arabic{case_num}-eps-converted-to.eps}\\
  \includegraphics[width=5in]{codev6/figure/sparse_eig_detail_\arabic{case_num}-eps-converted-to.eps}
\caption[Universal Model Sparsity Study Scenario \arabic{case_num}]{Universal Model Sparsity Study Scenario \arabic{case_num} in Table \ref{tab:sparse}. The first plot is a comprehensive clustering analysis of the steady state values obtained from the Lotka-Volterra simulations. The $x$-axis denotes the heterogeneity value $\alpha$. The box plots are the silhouette values pertaining to the number of clusters for which the silhouette index (maximum over the number of clusters of the mean silhouette value for each given total number of cluster) was defined. The total number of clusters pertaining to the silhouette index is denoted on the top $x$-axis. The second row is a principle coordinate analysis of the  steady state values obtained at three different heterogeneity values. Clusters are color coded to match the optimal clustering from $k$-medoids. The third row of the figure plots $\max_{i,j}  \mathrm{real}(\lambda_i (A^{[j]}))$ as a function of $\alpha$. The fourth row is a plot of $\lambda_i (A^{[j]})$ for $i\in \{1,2,\ldots, 80\}$ and  $j\in \{1,2,\ldots, 500\}$  at three values of $\alpha$.}
\label{fig:sparse\arabic{case_num}}
\end{figure}
\clearpage
}

\clearpage

\setcounter{case_num}{1}
\forloop{case_num}{1}{\value{case_num} < 9}{
\begin{figure}[htbp]
\centering
  \includegraphics[width=5in]{codev6/figure/remove_fig_SI_\arabic{case_num}-eps-converted-to.eps}
\\
  \includegraphics[width=5in]{codev6/figure/remove_fig_pca_growth_rate_\arabic{case_num}-eps-converted-to.eps}
  \\
  \includegraphics[width=5in]{codev6/figure/remove_fig_eig_max_\arabic{case_num}-eps-converted-to.eps}\\
  \includegraphics[width=5in]{codev6/figure/remove_eig_detail_\arabic{case_num}-eps-converted-to.eps}
\caption[Universal Model Community Size Overlap Study Scenario \arabic{case_num}]{Universal Model Community Size Overlap Study Scenario \arabic{case_num} in Table \ref{tab:remove}. The first plot is a comprehensive clustering analysis of the steady state values obtained from the Lotka-Volterra simulations. The $x$-axis denotes the heterogeneity value $\alpha$. The box plots are the silhouette values pertaining to the number of clusters for which the silhouette index (maximum over the number of clusters of the mean silhouette value for each given total number of cluster) was defined. The total number of clusters pertaining to the silhouette index is denoted on the top $x$-axis. The second row is a principle coordinate analysis of the  steady state values obtained at three different heterogeneity values. Clusters are color coded to match the optimal clustering from $k$-medoids. The third row of the figure plots $\max_{i,j}  \mathrm{real}(\lambda_i (A^{[j]}))$ as a function of $\alpha$. The fourth row is a plot of $\lambda_i (A^{[j]})$ for $i\in \{1,2,\ldots, p\}$ and  $j\in \{1,2,\ldots, 500\}$  at three values of $\alpha$.}
\label{fig:remove\arabic{case_num}}
\end{figure}
\clearpage
}